 \documentclass[a4paper,12pt]{article}
\usepackage{xspace}
\usepackage{graphicx}
\usepackage{pslatex}
\usepackage{amsmath}
\usepackage{bbold}
\usepackage{amssymb}
\usepackage[latin1]{inputenc}
\usepackage{txfonts}
\usepackage{color}

 \newcommand{\ecm}{e\,{\rm cm}}
 
 \newcommand{\pp}{\boldsymbol{\rm p}}
 \newcommand{\kk}{\boldsymbol{\rm k}}

 \newcommand{\hp}{\boldsymbol{\hat{\rm p}}}
 \newcommand{\hk}{\boldsymbol{\hat{\rm k}}}
 
 \newcommand{\qp}{\boldsymbol{\rm q}_+}
 \newcommand{\qm}{\boldsymbol{\rm q}_-}
 
 \newcommand{\bq}{\boldsymbol{\rm q}}
 \newcommand{\hbq}{\boldsymbol{\hat{\rm q}}}
 
 \newcommand{\hqp}{\boldsymbol{\hat{\rm q}}_+}
 \newcommand{\hqm}{\boldsymbol{\hat{\rm q}}_-}

 \newcommand{\sip}{\boldsymbol{\sigma}_+}
 \newcommand{\ssim}{\boldsymbol{\sigma}_-}
 \newcommand{\ssig}{\boldsymbol{\sigma}}

\def\nn{\nonumber}
\def\one{{1\!\!\mbox{l}}}
\newcommand{\cO}{\mathcal O}
\def\v{{\varv}}

\def\Re{{\rm Re}}
\def\Im{{\rm Im}}

\def\GeV{{\rm GeV}}

\newcommand{\be} {\begin{equation}}
\newcommand{\ee} {\end{equation}}
\newcommand{\bma} {\begin{math}}
\newcommand{\ema} {\end{math}}
\newcommand{\beqa} {\begin{eqnarray}}
\newcommand{\eeqa} {\end{eqnarray}}

\newcommand{\bc} {\begin{center}}
\newcommand{\ec} {\end{center}}

\newcommand{\simgt}{\hbox{ \raise3pt\hbox to 0pt{$>$}
    \raise-3pt\hbox{$\sim$} }}
\newcommand{\simsm}{\hbox{ \raise3pt\hbox to 0pt{$<$}
    \raise-3pt\hbox{$\sim$} }}
    
\begin{document}

\begin{titlepage}
  \begin{flushright}
    TTK-21-03 \\
  \end{flushright}
  \vspace{0.01cm}
  
  \begin{center}
    {\LARGE \bf Electric dipole moment of the tau lepton revisited} \\
    \vspace{1.5cm}
    {\bf Werner Bernreuther}\,$^{a,}$\footnote{\tt
      breuther@physik.rwth-aachen.de}, 
    {\bf Long Chen}\,$^{a,}$\footnote{\tt longchen@physik.rwth-aachen.de}
    {\bf and  Otto Nachtmann}\,$^{b,}$\footnote{\tt o.nachtmann@thphys.uni-heidelberg.de}
    \par\vspace{1cm}
    $^a$Institut f\"ur Theoretische Teilchenphysik und Kosmologie, \\
    RWTH Aachen University,  52056 Aachen, Germany\\
    $^b$ Institut f{\"u}r Theoretische Physik, Universit{\"a}t Heidelberg, 69120 Heidelberg, Germany
    \par\vspace{1cm}
    {\bf Abstract}\\
    \parbox[t]{\textwidth}
    {\small{ We reconsider the issue of the search for a nonzero electric dipole form factor (EDM) $d_\tau(s)$ using optimal observables 
     in $\tau^+\tau^-$ production 
     by $e^+ e^-$ collisions in the center-of-mass energy range from the $\tau$-pair threshold to about $\sqrt{s} \sim 15$ GeV.
       We discuss the general formalism of optimal observables and apply it
  to two $CP$-odd observables that are sensitive to the real and imaginary part of $d_\tau(s)$, respectively. 
  We compute the  expectation values and covariances of these optimal $CP$ observables for 
  $\tau$-pair production   at $\sqrt{s}=10.58$ GeV
 with subsequent decays of $\tau^\pm$ into major leptonic 
   or semihadronic modes. For the $\tau$ decays to two pions and three charged pions we take the full kinematic information of the hadronic system into account.
   Assuming that the Belle II experiment at the KEKB accelerator will eventually 
   analyze data corresponding to an integrated luminosity of 50 ab$^{-1}$
   and applying acceptance cuts on the final-state pions we find that 
   1~s.d. sensitivities $\delta \Re d_\tau = 6.8 \times 10^{-20} \ecm$ and $\delta \Im d_\tau = 4.0 \times 10^{-20} \ecm$ can be obtained 
    with events where both $\tau$'s decay semihadronically.
     We consider also the ideal case that no cuts on the final-state particles are applied. With  50 ab$^{-1}$ at $\sqrt{s}=10.58$ GeV corresponding to 
     $4.5 \times 10^{10}$  $\tau^+ \tau^-$ events  
    we find  the 1 s.d. sensitivities $\delta \Re d_\tau = 5.8 \times 10^{-20} \ecm$ and $\delta \Im d_\tau = 3.2 \times 10^{-20} \ecm$, again for events
     where both $\tau$ leptons decay semihadronically.
    Furthermore, we analyze the potential magnitude of the $\tau$ EDM form factor 
    in the type-II two-Higgs doublet extension and in two scalar leptoquark extensions of the Standard Model, 
    taking into account phenomenological constraints.}}
    
  \end{center}
  \vspace*{0.7cm}

\end{titlepage}

\setcounter{footnote}{0}
\renewcommand{\thefootnote}{\arabic{footnote}}
\setcounter{page}{1}

\section{Introduction} 
\label{sec:intro}
The search for electric dipole moments (EDMs) of fundamental fermions is an important 
aspect of experimental investigations hunting for physics beyond the Standard Model (SM) 
of particle physics, in particular for $CP$ violation beyond the Kobayashi-Maskawa mechanism.
So far only upper bounds for these EDMs exist~\cite{Zyla:2020zbs}. For the electron an impressive 
 upper limit was obtained rather recently by the ACME Collaboration~\cite{Andreev:2018ayy}. The best muon EDM limit to date
  was set by the Muon $(g-2)$ Collaboration~\cite{Bennett:2008dy}. These limits are 
\begin{align}\label{Eq.01.01}
|d_{e}| &< 1.1 \times 10^{-29} \, \ecm \, \text{
at }\, 90\% \, {\rm C.L.}  \, , \\ \label{Eq.01.02}
|d_{\mu}|& < 1.8\phantom{1}\times 10^{-19}\, \ecm \, \text{
at }\, 95\% \, {\rm C.L.} \, .
\end{align}

 The lifetime of the $\tau$ lepton is too short to allow for the measurement of its 
  static moments. Instead information on the nonstatic $\tau$ EDM form factor\footnote{In this paper we use the acronym EDM  for both the 
  static moment and the form factor at $q^2\neq 0$.} can be retrieved, for 
   instance, from the measurement
   of $CP$-violating correlations in $\tau$-pair production by $e^+ e^-$ collisions. 
   The $\tau$ EDM form factor can be a complex quantity 
    for timelike momentum transfer. The best limits to date on its real and imaginary parts were obtained 
    by the Belle I Collaboration~\cite{Inami:2002ah} at $q^{2}=(10.58~{\rm GeV})^{2}$:
\begin{align}    \label{Eq.01.03}
-2.2 \times 10^{-17} \, {\ecm} &<\Re~d_{\tau}(q^{2})<4.5\times10^{-17} {\ecm} \, \text{
at }\, 95\% \,{\rm C.L.} \, , \notag \\
-2.5 \times 10^{-17} \, {\ecm} &<\Im~d_{\tau}(q^{2})<0.8\times10^{-17} {\ecm} \, \text{
at }\, 95\% \, {\rm C.L.} \,.
\end{align}

In a series of articles where two of the authors of this paper were involved, 
ways of searching for $CP$-violating effects in $e^{+}e^{-}$ collisions, in particular for a nonzero $\tau$ EDM, 
 were proposed~\cite{Bernreuther:1988jr,Bernreuther:1989kc,Korner:1990zk,Bernreuther:1991xe,Bernreuther:1993nd}. 
The observables and results of \cite{Bernreuther:1989kc,Bernreuther:1993nd} were used 
in the experimental searches for an EDM form factor of the $\tau$ lepton by~\cite{Inami:2002ah} 
 and earlier by the  ARGUS Collaboration~\cite{Albrecht:2000yg} that obtained the results
\begin{align}\label{Eq.01.05}
|\Re~d_{\tau}(q^{2})|&<4.6\times10^{-16} {\ecm} \, \text{
at }\, 95\% \,{\rm C.L.}\, , \notag \\ 
|\Im~d_{\tau}(q^{2})|&<1.8\times10^{-16} {\ecm} \, \text{
at }\, 95\% \,{\rm C.L.} 
\end{align}
at a c.m. energy  $\sqrt{s}=\sqrt{q^{2}}=10~{\rm GeV}$ of the reaction $e^+e^- \to \tau^+\tau^-$. 
 For reviews of the search results for the $\tau$ EDM and its weak dipole
 form factor (the analogue of the EDM for the coupling of the Z boson to fermions); 
 see, for instance, \cite{Stahl:2000aq,Lohmann:2005im}.
 Further discussions of possible measurements of the anomalous magnetic
 moment and the EDM of the $\tau$ lepton can be found in \cite{Bernabeu:2006wf,Bernabeu:2007rr,Eidelman:2016aih,Chen:2018cxt,Dyndal:2020yen}
  and references therein. 

The experimentation at Belle II \cite{Abe:2010gxa} which  started recently at the KEKB accelerator 
 offers new possibilities for 
measuring the $\tau$ EDM form factor, in particular, because a huge number of 
recorded $\tau$-pair events are expected at the end of data taking~\cite{Kou:2018nap}.
Also the BES III experiment, where $e^+e^-$ collisions at a center-of-mass (c.m.) energy $\sqrt{s} \sim 4$ GeV are studied,
expects to collect and analyze a large number of $\tau^+\tau^-$ pairs \cite{Ablikim:2019hff}.
Therefore, we reconsider the issue with  particular emphasis on using optimal 
observables~\cite{Atwood:1991ka,Davier:1992nw,Diehl:1993br}  for tracing the $\tau$ EDM form factor
 in $\tau$-pair production at c.m. energies from threshold up to about 15 GeV where the contribution from $Z$-boson exchange
  is negligible. 
  In our numerical analysis we consider $\tau$-pair production at $\sqrt{s}=10.58$ GeV.
 Moreover, we analyze this form factor in a few SM extensions
 that can induce a potentially sizable $\tau$ EDM \cite{Bernreuther:1996dr}.
 
Our paper is organized as follows. In Section~\ref{Sec:02} we recall the form factor
decomposition of the $\gamma\tau\tau$ vertex and in particular the definition of the $\tau$ EDM
form factor.
In section~\ref{Sec:03} we discuss the production and decay matrices for the 
process $e^{+}e^{-}\rightarrow\tau^{+}\tau^{-}$ with the $\tau$'s decaying into
one, two, or three  particles that are measured in an experiment. Section~\ref{Sec:04} deals with simple and 
optimal observables \cite{Atwood:1991ka,Davier:1992nw,Diehl:1993br} 
for tracing the EDM of the $\tau$ lepton.
 Section~\ref{sec:results}  contains our numerical results, in particular 
 our estimates of the sensitivities with which the real and the imaginary parts of 
  the $\tau$ EDM form factor can be measured in various $\tau$ decay channels. 
  In Section~\ref{sec:FFBSM} we consider the $\tau$ EDM form factor in a type-II two-Higgs doublet extension and in two 
  leptoquark extensions of the SM and analyze the potential magnitude of the $\tau$ EDM taking into account
   experimental constraints. 
  Moreover, we show that within these models $CP$-violating box contributions to the $S$-matrix element of $e^+e^-\to \tau^+ \tau^-$ 
  are negligible as compared to that of the $\tau$ EDM form factor.
   We conclude in Section~\ref{sec:concl}. In Appendix~\ref{app:taudec} 
   we list the density matrices for several major decays of polarized $\tau^\pm$ leptons. In particular, we present the explicit
    form of the differential decay density matrices for $\tau \to 2 \pi  \nu_\tau$ and  $\tau \to 3 \pi \nu_\tau$.
   Appendix~\ref{app:exCoCP} contains a detailed analysis of the expectation values and covariances of the $CP$-odd optimal observables 
    used in Sec.~\ref{sec:results} in various $\tau^+\tau^-$ decay channels.
 
\section{Form Factors}
 \label{Sec:02}
 We consider $\tau^+\tau^-$ production in $e^+ e^-$ collisions at c.m. energies $\sqrt{s}$ from threshold up to about 15 GeV,
  with $\tau^-$ and $\tau^+$ decaying into a final state $A$ and ${\overline B}$, respectively,
\begin{equation}\label{Eq.01.04}
e^{+}(p_{+}) + e^{-}(p_{-}) \rightarrow \tau^{+}(k_{+},\alpha) +\tau^{-}(k_{-},\beta) \rightarrow {\overline B} \; + \; A \, .
\end{equation}
The four-momenta and the corresponding three-momenta are denoted in the $e^+e^-$ c.m.
 frame by $p_\pm =(p^0_\pm, {\pp}_\pm)^T$, $k_\pm =(k^0_\pm, {\kk}_\pm)^T$.
We consider unpolarized electrons and positrons and neglect their masses; the labels $\alpha,\beta\in \{\pm 1/2\}$ 
denote the spin indices of the tau leptons. In the c.m. frame we have ${\pp}_+ + {\pp}_- = {\kk}_+ + {\kk}_- = 0$.

\begin{figure}[h!]
\begin{center}
{\includegraphics[width=0.89\textwidth]{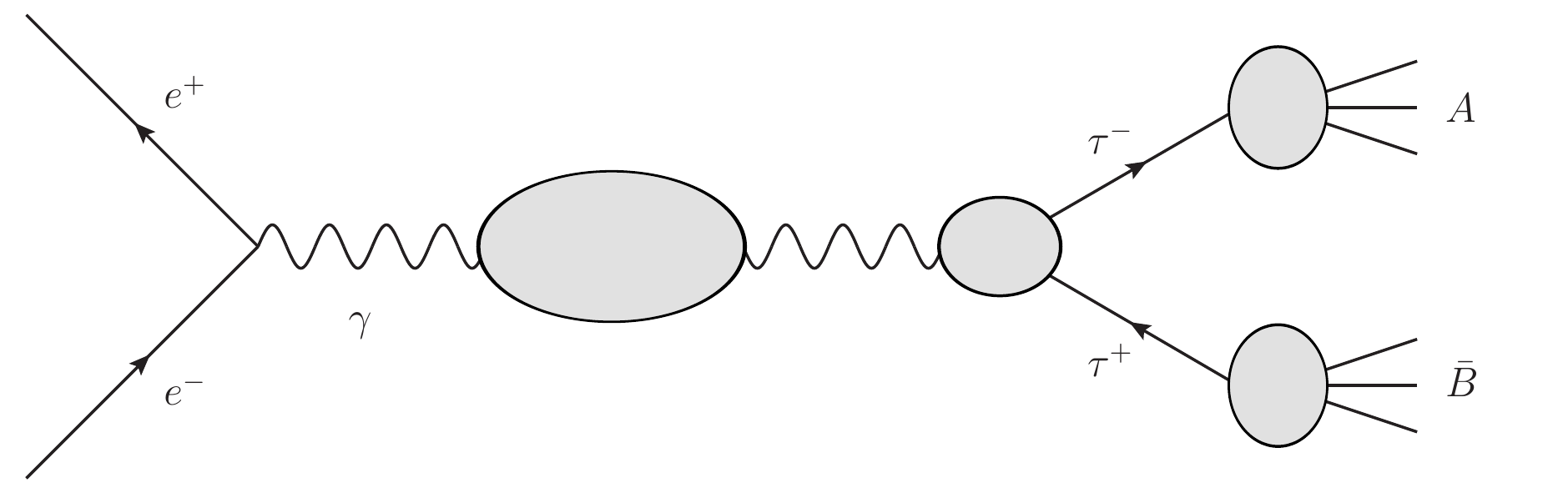}}
\caption{The reaction \eqref{Eq.01.04} in the one-photon-exchange approximation.}
\label{Fig.tauprod}
\end{center}
\end{figure}

For unpolarized $e^{+}$ and $e^{-}$ the initial state
is described by a $CP$-invariant density matrix. 
Thus, any non zero $CP$-odd correlation observed in the final state
indicates a genuine $CP$-violating effect that
 can be located in the production and/or in 
 the decays of the $\tau$'s. We consider tau-pair production  by
  one-photon-exchange only. At the energies considered here $Z$-boson exchange is negligible.
 This will be justified at the end of this section.
  The diagram shown in Fig.~\ref{Fig.tauprod}  exhibits this approximation with the full photon
   propagator  \begin{equation} \label{eq:fullphpr}
    i \Delta_{\mu\nu}^{(\gamma)}(q) = \frac{- i g_{\mu\nu}}{q^2[ 1 + e^2 \Pi_c(q^2)]} \, ,
   \end{equation}
 where $\Pi_c(q^2)$ is the vacuum-polarization function; see e.g. Eq.~(19.45) of \cite{Bjorken:1965zz}.
 For instance, at the mass of the $\Upsilon(4{\rm S})$ resonance, at $\sqrt{q^2}= 10.58$ GeV, this vacuum polarization effect
  produces an enhancement of the cross section. For a detailed discussion of the $\tau$-pair cross section at this
   energy, including radiative corrections, we refer to \cite{Banerjee:2007is}. Below we consider only normalized 
    expectation values of $CP$ observables where such resonance enhancements enter only through the number of events which we 
    take as input from experiment.

\begin{figure}[h!]
\begin{center}
{\includegraphics[width=0.89\textwidth]{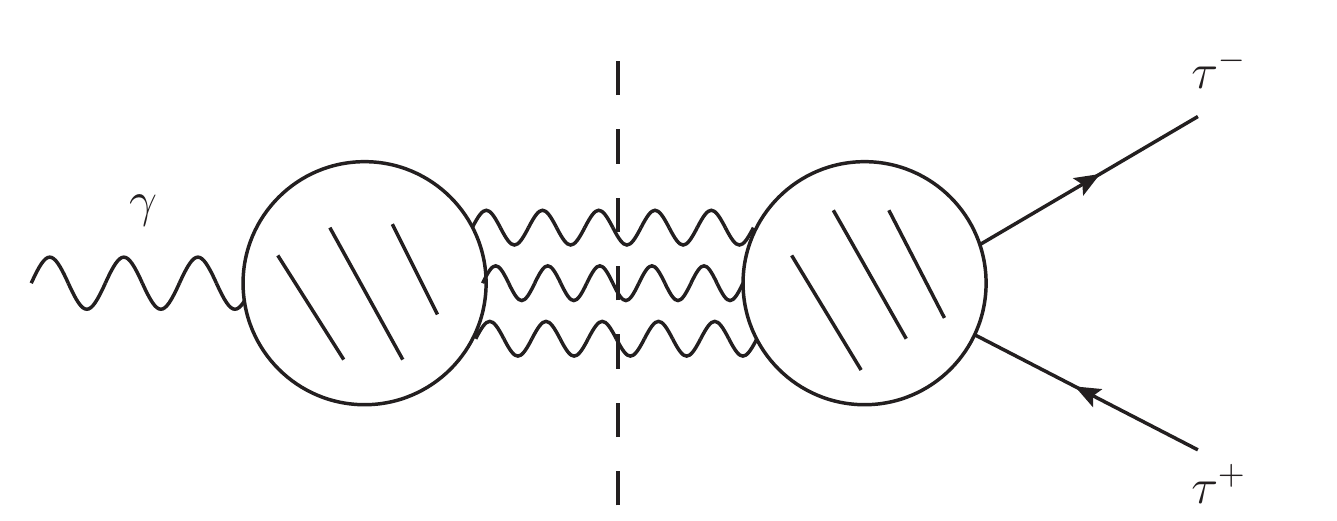}}
\caption{A cut diagram leading to an imaginary part of the form factors in \eqref{Eq.02.03} for $q^{2}>0$.}
\label{Fig.01}
\end{center}
\end{figure}

    In the following we  assume that the only source of $CP$ violation in the diagram of Fig.~\ref{Fig.tauprod} is due to
  a nonzero  EDM form factor in the $\gamma\tau\tau$ vertex.  
  This vertex is given by the following one-particle irreducible (1PI) matrix element
  of the electromagnetic current $J_{\lambda}^{em}$ between the vacuum and the $\tau^+\tau^-$ final state:
\begin{eqnarray}\label{Eq.02.03}
\big{\langle}\tau^{-}(k_{-},\beta), \, \tau^{+}(k_{+},\alpha) \, {\rm out}|J_{\lambda}^{em}(0) |0 \big{\rangle} = & \nonumber \\
 - \overline{u}_{\beta}(k_{-}) \Bigl[ eF_{1}(q^{2})\gamma_{\lambda}+\dfrac{i}{2m_{\tau}}\sigma_{\lambda\mu}q^{\mu} e F_{2}(q^{2}) 
+d_{\tau}(q^{2})\sigma_{\lambda\mu}q^{\mu}\gamma_{5} \Bigr.  & \nonumber \\
 \Bigl. +\dfrac{1}{8\pi}A(q^{2})(q^{2}\gamma_{\lambda}-2m_{\tau}q_{\lambda})\gamma_{5} \Bigr] {\v}_{\alpha}(k_{+})\;, &
\end{eqnarray}
where $q = k_+ + k_-$. The right-hand side of~\eqref{Eq.02.03} represents the most general decomposition of this matrix element taking 
into account the conservation of the
electromagnetic current. Moreover,  $e=\sqrt{4\pi\alpha_{em}}>0$ denotes the $\tau^{+}$ charge and 
we use the $\gamma$-matrix conventions of \cite{Bjorken:1965zz}.
Note that the order of $\tau^{-}$ and $\tau^{+}$ in the matrix element \eqref{Eq.02.03}  matters because we are dealing with fermions. 
The form factors $F_{1,2}(q^2),$ $d_\tau(q^2)$, and $A(q^{2})$ are 
analytic functions of $q^{2}$ in the complex $q^{2}$ plane with a cut on the positive real axis satisfying
\begin{align}\label{Eq.02.04}
F_{i}(q^{2*})^{*}&=F_{i}(q^{2})\, , \quad i=1,2 \,, \notag \\
d_{\tau}(q^{2*})^{*}&=d_{\tau}(q^{2})\,, \notag \\
A(q^{2*})^{*}&=A(q^{2})\,.
\end{align} 
That is, on the real $q^{2}$ axis, the form factors are real functions for $q^{2}<0$ and can have 
imaginary parts for $q^{2}>0$. At higher order in $\alpha_{em}$ these cuts start at $q^{2}=0$ due to cut diagrams of the type
shown in Fig.~\ref{Fig.01} with three photons in the intermediate state.
In the decomposition \eqref{Eq.02.03} we have $q^{2}\geq 4m_{\tau}^{2}$
  and we have to set $q^{2}+i\varepsilon$, that is, to take $q^{2}$  above the cut.

Next we recall the transformation properties of the  $\gamma\tau\tau$ coupling terms associated with the four form factors
in \eqref{Eq.02.03} under charge conjugation ($C$), parity ($P$), and $CP$. Assuming that the interaction is invariant under these
 transformations and using the transformation of $J_{\lambda}^{em}(x)$ under $C$, $P$, and $CP$, one 
 gets the transformation properties  listed in Table~\ref{tab:CPvertex}.

The $e^+e^-\to \tau^+\tau^-$ amplitude can receive also $CP$-odd 1PI box contributions, for instance contributions with
 Lorentz structure $({\bar e} e) ({\bar\tau} i\gamma_5\tau)$. We do not take such contributions into account in the following.
 We discuss a few SM extensions in Section~\ref{sec:FFBSM} that can induce sizable $\tau$ EDM form factors.
 For these models we show in Section~\ref{suse:boxc} that the $CP$-violating box contributions can be neglected 
 as compared to that of the induced $\tau$ EDM form factor.

\vspace{2mm}
\begin{table}[htbp]
\begin{center}
  \caption{Transformation properties of the $\gamma\tau\tau$ coupling terms corresponding to the four form factors
  in the decomposition of the matrix element \eqref{Eq.02.03} of the 
  electromagnetic current.} 
  \vspace{1mm}
\begin{tabular}{c c c c} \hline \hline
            & $C$  & $P$ & $CP$  \\ \hline 
 $F_1(q^2)$ & $+$ & $+$  & $+ $  \\
 $F_2(q^2)$ & $+$ & $+$  & $+ $  \\
 $d_\tau(q^2)$ & $+$ & $-$  & $-$  \\
 $A(q^2)$   & $-$ & $-$  & $+$  \\ \hline \hline
\end{tabular}
\label{tab:CPvertex}
\end{center}
\end{table}

For the matrix elements of the current between $\tau^{-}$ and $\tau^{+}$ states, respectively,
we get, using the standard crossing relations: 
\begin{equation}\label{Eq.02.06}
\big{\langle}\tau^{-}(k',\beta') |J_{\lambda}^{em}(0) |\tau^{-}(k,\beta)  \big{\rangle} 
=-\overline{u}_{\beta'}(k')\Gamma_{\lambda}(q)u_{\beta}(k)\;,
\end{equation}
\begin{equation}\label{Eq.02.07}
\big{\langle}\tau^{+}(k',\alpha') |J_{\lambda}^{em}(0) |\tau^{+}(k,\alpha)  \big{\rangle} 
=\overline{\v}_{\alpha}(k)\Gamma_{\lambda}(q){\v}_{\alpha'}(k')\;,
\end{equation}
where the vertex function $\Gamma_{\lambda}(q)$ is given by the expression in the square brackets of Eq.~\eqref{Eq.02.03}
with $q=k'-k$ and $q^2\leq 0.$

The form factor $F_{1}(q^2)$ is the electric or Dirac form factor with the normalization
\begin{equation}\label{Eq. 02.10}
F_{1}(0)=1\;.
\end{equation}
The magnetic or Pauli form factor $F_{2}(q^2)$ at $q^2=0$ yields the $\tau$ anomalous magnetic moment:
\begin{equation}\label{Eq.02.11}
F_{2}(0)=a_{\tau}=\frac{1}{2}(g_{\tau}-2)\;.
\end{equation}
The $\tau^{-}$ and $\tau^{+}$ electric dipole moments, respectively, are obtained from the EDM form factor  
$d_{\tau}(q^2)$  at $q^2=0$:
\begin{equation}\label{Eq.02.12}
d_{\tau^{-}}=-d_{\tau^{+}}=d_{\tau}(0)\;.
\end{equation}
The form factor $A(q^{2})$ at $q^2=0$ defines the anapole moment \cite{Zeldovich58,Flambaum:1980sb,Flambaum:1984fb,Flambaum:1984fc}
of the $\tau^{-}$:
\begin{equation}\label{Eq.02.13}
A_{\tau^{-}}= A(0)\;.
\end{equation}
For a $\tau^-$ at rest, $k=k_{R}=(m_\tau,\boldsymbol{0})^T$, one has
\begin{equation}\label{Eq.02.14}
\big{\langle}\tau^{-}(k_R,\beta') |(-\pi)\int d^{3}x|\boldsymbol{x}|^2 \boldsymbol{J}^{em}(\boldsymbol{x} , 0)|\tau ^{-}(k_{R},\beta) \big{\rangle}
=\frac{1}{2}{\ssig}_{\beta '\beta} A_{\tau^-}\;.
\end{equation}

 A comment on the gauge invariance of the form-factor decomposition of the vertex function 
 \eqref{Eq.02.03} is in order. Electromagnetic gauge invariance is obvious, because  conservation of the electromagnetic current
  was used in the decomposition of \eqref{Eq.02.03}. As to the invariance with respect to the electroweak gauge group 
  ${\rm SU(2)}\times{\rm U(1)}$: The static moments at $q^2=0$, in particular the anomalous magnetic and electric dipole moment
   and the anapole moment are gauge invariant, as they correspond to terms in the $\tau\to  \tau$ 
   $S$-matrix element in the soft-photon limit. Yet, for obtaining a gauge-invariant amplitude for $e^+e^- \to \tau^+\tau^-$
   one cannot, of course, use \eqref{Eq.02.03} in isolation, but must take into account all contributions (including box 
   contributions at one-loop order and beyond) to the   $S$-matrix element order by order in the electroweak couplings.
   However, in the following we use only the tree-level $\gamma\tau\tau$ vertex supplemented by the $\tau$ EDM form factor.
   The $\tau$ EDM is extremely small in the SM, as will be briefly reviewed at the beginning of Section~\ref{sec:FFBSM}.
   Thus, a sizable value for $d_\tau$ must come from ``beyond the Standard Model'' (BSM) physics. 
   In Section~\ref{sec:FFBSM} we discuss a few BSM extensions that can induce a sizable $\tau$ EDM form factor and compute 
    it at one-loop order. The form factors $d_\tau(q^2)$ given in that section are invariant with respect to the electroweak 
    gauge group.

 As is well-known one may introduce a $\tau$ EDM, together with an analogous $CP$-violating weak dipole moment (WDM)
 $d_\tau^Z$ in the $Z\tau\tau$ vertex, by using a ${\rm SU(3)}\times{\rm SU(2)}\times{\rm U(1)}$ invariant effective 
  Lagrangian approach for BSM couplings. Imposing baryon and lepton number conservation the leading gauge-invariant operators
   have mass dimension 6 \cite{Buchmuller:1985jz} and the relevant effective Lagrangian takes the
   form (see, for instance, \cite{Bernreuther:1991gh}):
 \begin{eqnarray}\label{Eq.Leffinv}
\mathcal{L}_{\textup{eff}}(x) = &-i \frac{c_1}{\Lambda^2}~{\bar\tau_R}(x)\sigma^{\mu\nu}
\phi^\dagger(x)\left[g'\frac{\tau^a}{2}W_{\mu\nu}^a(x) - \frac{g}{2}B_{\mu\nu}(x)\right] L_L(x) \nonumber \\
 & -i \frac{c_2}{\Lambda^2}~{\bar\tau_R}(x)\sigma^{\mu\nu}
\phi^\dagger(x)\left[g\frac{\tau^a}{2}W_{\mu\nu}^a(x) + \frac{g'}{2}B_{\mu\nu}(x)\right] L_L(x)  + {\rm H.c.} \, .
\end{eqnarray}
 Here $c_1$ and $c_2$ are dimensionless real coupling constants, $\Lambda\gg {\rm v}_0$ denotes the energy scale of 
 new physics that is assumed to
 be considerably larger than the electroweak symmetry breaking scale ${\rm v}_0=246$ GeV, $g$ and $g'$ are 
 the ${\rm SU(2)}$ and ${\rm U(1)}$
 gauge couplings, respectively, $W_{\mu\nu}$ and $B_{\mu\nu}$ are the gauge field strength tensors corresponding to these groups, 
 $\phi$ is the Higgs doublet field, and $\tau_R$ and $L_L^T=(\nu_\tau, \tau)_L^T$ are the right-handed singlet 
 and left-handed lepton doublet fields 
 of the third generation. (Our notation follows \cite{Nachtmann:1990ta}.) 
 After spontaneous symmetry breaking the effective Lagrangian \eqref{Eq.Leffinv} contains the EDM interactions
\begin{equation}\label{Eq.EWeff}
\mathcal{L}_{\textup{eff}}(x) \supset -\frac{i}{2}d_\tau~\overline{\tau}(x)\sigma^{\mu\nu}\gamma_{5}\tau(x)F_{\mu\nu}(x) 
-\frac{i}{2}d^Z_\tau~\overline{\tau}(x)\sigma^{\mu\nu}\gamma_{5}\tau(x)Z_{\mu\nu}(x)  \, ,
\end{equation}
where $F_{\mu\nu}=\partial_{\mu}A_{\nu}-\partial_{\nu}A_{\mu}$ and $Z_{\mu\nu}=\partial_{\mu}Z_{\nu}-\partial_{\nu}Z_{\mu}$ 
are, respectively, the Abelian field strength tensors of the photon and $Z$ boson and $d_\tau$ and $d^Z_\tau$ the electric and weak dipole moments
 of the $\tau$ lepton:
 \begin{equation}\label{Eq.dEWeff}
 d_\tau = \frac{{\rm v}_0}{\Lambda^2}\frac{\sqrt{g^2+g'^2}}{\sqrt{2}} c_1 \, , \qquad 
 d^Z_\tau = \frac{{\rm v}_0}{\Lambda^2}\frac{\sqrt{g^2+g'^2}}{\sqrt{2}} c_2 \, .
\end{equation}
This approach constitutes a possibility to introduce the $\tau$ EDM and WDM in a way that respects electroweak 
gauge invariance. 
Using the Hermitian Lagrangian \eqref{Eq.EWeff}
to leading order corresponds to setting
\begin{equation}\label{Eq.02.16}
\Re d_{\tau}(q^{2})=d_{\tau} \,, \quad  \Im d_{\tau}(q^{2})=0 \, 
\end{equation}
and likewise for $d_\tau^Z$. In this framework imaginary parts of $d_{\tau}$ and $d_{\tau}^Z$ will be generated by diagrams 
involving both ${\cal L}_{\textup{eff}}$ of Eq.~\eqref{Eq.EWeff} and SM couplings at higher order.
 We shall take into account in the following that $\Im d_\tau$ can be nonzero for $q^2>0$ but we neglect, as already mentioned above,
the contribution 
from $Z$-boson exchange, in particular the contribution from $d_\tau^Z$.
This can be justified as follows. Eq.~\eqref{Eq.dEWeff} shows that $d_\tau$ and $d^Z_\tau$ will be of the same order of magnitude if the coupling constants
 $c_1$ and $c_2$ are of comparable size. This is the case, for instance, in the BSM models considered in Section~\ref{sec:FFBSM}, 
 as was shown in \cite{Bernreuther:1996dr}. At energies $\sqrt{s}\ll m_Z$ that we consider in this paper, the effects of $d_\tau^Z$ resulting from $Z$-boson exchange
 are then negligible compared to those of $d_\tau$, as is the contribution resulting from the interference of the SM $Z$-boson exchange amplitude with the amplitude
  involving $d_\tau$. One can see this explicitly from the formulas given in 
  \cite{Bernreuther:1993nd} where both $\gamma$ and $Z$-boson exchange were taken into account. From Eq.~(3.10) of \cite{Bernreuther:1993nd} we find that 
  for c.m. energies  $\sqrt{s} \approx 10$ GeV that we are considering $Z$-boson exchange contributions are suppressed by a factor of order
  \begin{equation} \label{eq:suppr}
        s/{m_Z^2} \approx 10^{-2} \, .
  \end{equation}
This holds for $\sqrt{s}$ in the continuum and at the $\Upsilon(4{\rm S})$ resonance where the suppression factor \eqref{eq:suppr} 
is a few percent smaller 
because there  the photon contribution is enhanced as compared to the continuum value; see \cite{Banerjee:2007is}.

%
\section{Matrix elements, production and decay matrices}
\label{Sec:03}

 We are interested in analyzing $CP$-violating effects in $\tau$-pair production 
generated by a nonzero $\tau$ EDM form factor. Therefore we shall analyze the reactions \eqref{Eq.01.04}
by considering on-shell $\tau$-pair production by one-photon exchange,
including the $\tau$ EDM form factor in the $\gamma\tau\tau$ vertex, followed
 by the decays of $\tau^-$ and $\tau^+$ into the final states $A$ and $\overline{B}$, respectively. The $\tau$ spin correlations and
  polarizations will be taken into account. (The reactions \eqref{Eq.01.04} were investigated in \cite{Bernreuther:1993nd} 
   for arbitrary c.m. energies for photon and
  $Z$-boson exchange including besides the EDM also the weak dipole form factor of the $\tau$ lepton.) 

As to the decay channels $A$ and $\overline{B}$, we consider two cases: \\
i) Only one charged particle of $A$ and one of $\overline{B}$ are measured:
\begin{equation}
 \label{eq:1prong}
 \tau^- \rightarrow a(q_-) + X \, , \qquad \tau^+ \rightarrow {\bar b}(q_+) + X' \, ,
\end{equation}
Examples from the main decay modes of $\tau^-$ are 
\begin{align}\label{Eq.03.02}
\tau^{-}  \rightarrow &e^{-}(q_{-})\overline{\nu}_{e}\nu_{\tau}\, ,\quad \,\mu^{-}(q_{-})\overline{\nu}_{\mu}\nu_{\tau} \,,\notag \\
&\pi^{-}(q_{-})\nu_{\tau}\, ,\quad \pi^{-}(q_{-})\pi^{0}\nu_{\tau}  \, ,
\quad \pi^{-}(q_{-})\pi^{0}\pi^{0}\nu_{\tau} \, , \quad \pi^{-}(q_{-})\pi^{+}\pi^{-}\nu_{\tau} \, ,
\end{align}
and the  respective charge-conjugate $\tau^+$ decays. The decay modes \eqref{Eq.03.02} include, in particular, 
 $\tau$ decays to one charged prong. \\
ii) We shall also treat the case where more than one particle from $\tau$ decay is observed, 
specifically the decay to two pions via a $\rho$ and $\rho'$ meson and
 the decay to the $a_{1}$ meson, respectively  to three charged  pions:
\begin{align}
\tau^{-} \rightarrow & \pi^{-}(q_1)\pi^{0}(q_2)  \nu_{\tau}  \, , &
\tau^{+} \rightarrow &\pi^{+}(\bar{q}_1)\pi^{0}(\bar{q}_2)  \bar{\nu}_{\tau} \, ,  \label{eq:tdec2p} \\
\tau^{-} \rightarrow & \pi^{-}(q_1)\pi^{-}(q_2)\pi^{+}(q_3)  \nu_{\tau}     \, ,  &
\tau^{+} \rightarrow &\pi^{+}(\bar{q}_1)\pi^{+}(\bar{q}_2)\pi^{-}(\bar {q}_3)  \bar{\nu}_{\tau} \,.\label{Eq.03.03}
\end{align}

For on-shell $\tau$-pair production and decay the cross section of \eqref{Eq.01.04} can be written as a product of the
 production density matrix $R$ for $e^+ e^-\rightarrow \tau^+ \tau^-$ 
times the  density matrices $\mathcal{D}^{\bar B}_{\alpha' \alpha}$ and
 $\mathcal{D}^A_{\beta'\beta}$ that describe the decays of polarized  $\tau^+ \to \overline{B}$ and $\tau^-\to A$,
 respectively. 
The production density matrix $R$ is  defined as follows:
\begin{eqnarray}
\label{Eq.03.04}
R_{\alpha\alpha'\beta\beta'} = & \dfrac{1}{4}\sum_{\gamma,\delta} \big{\langle}\tau^{+}({k_{+}}, \alpha), 
\tau^{-}({k_{-}}, \beta)|{\cal T}|e^{+}({p_{+}}, \gamma),e^{-}({p_{-}}, \delta)\big{\rangle}  \nonumber \\
& \times~\big{\langle}\tau^{+}({k_{+}}, \alpha'), \tau^{-}({k_{-}}, \beta')|{\cal T}|e^{+}({p_{+}}, \gamma),e^{-}({p_{-}}, 
\delta)\big{\rangle}^*\, \, ,
\end{eqnarray}
where $\gamma,\delta$ are the spin indices of $e^+$ and $e^-$, respectively.
For a decay of $\tau^-$ according to  case i) above the corresponding decay density matrix is given by
\begin{eqnarray}\label{Eq.03.05}
\mathcal{D}^a_{\beta'\beta}\bigl(\tau^{-}(k_{-})\rightarrow a(q_{-})+X\bigr)& =\Gamma^{-1}(\tau^{-}\rightarrow A) \dfrac{1}{2m_{\tau}} 
\int d\Gamma_ {X}(2\pi)^{4}\delta^{(4)}(k_{-}-q_{-}-q_X) \nonumber \\
& \times~ \big{\langle}a(q_{-}), X|{\cal T}|\tau^{-}(k_{-}, \beta)\big{\rangle}
\big{\langle}a(q_{-}), X|{\cal T}|\tau^{-}(k_{-}, \beta')\big{\rangle}^{*} \, .
\end{eqnarray}
Here the normalization is chosen such that
\begin{equation}\label{Eq.03.07}
\int \frac{d^{3}q_{-}}{(2\pi)^{3}2q_{-}^{0}} \,\mathcal{D}^a_{\beta'\beta}\bigl(\tau^{-}(k_{-})\rightarrow a(q_{-})+X)
=\delta_{\beta'\beta}\langle n_{a}\rangle_{A}\,, \\
\end{equation}
where $\langle{n_a}\rangle_{A}$ is the  mean multiplicity of particle $a$ in channel $A$. Formulas analogous to \eqref{Eq.03.05}
and \eqref{Eq.03.07} apply if  decays $\tau^+ \to \bar{b} + X'$ according to case i) are considered.

Thus the cross section for the two-particle inclusive reactions 
\begin{equation} \label{eq:2pincl}
e^+e^- \rightarrow \tau^+\tau^-\rightarrow \bar{B} + A \, ,
\end{equation}
where
\begin{equation} \label{eq:1prdec}
 A = a(q_-)  \, +  \, X \, , \qquad {\bar B} = {\bar b}(q_+)  \, +  \, X' \, ,
\end{equation}
is given in the narrow-width
approximation of the intermediate $\tau$ leptons by
\begin{eqnarray}\label{Eq.03.10}
d\sigma_{a\bar{b}}=\dfrac{\sqrt{1-4m_{\tau}^{2}/s}}{16\pi s} \;
\dfrac{d\Omega_{k_{+}}}{4\pi} \, {\rm Br}(\tau^{-}\rightarrow A) \, {\rm Br}(\tau^{+}\rightarrow \overline{B})  & \nonumber \\
\times ~ R_{\alpha\alpha'\beta\beta'} 
\frac{d^{3}q_{-}}{(2\pi)^{3}2q_{-}^{0}} \mathcal{D}^a_{\beta'\beta}\bigl[\tau^{-}\rightarrow a (q_{-})+X\bigr] 
\frac{d^{3}q_{+}}{(2\pi)^{3}2q_{+}^{0}} \mathcal{D}^{\bar b}_{\alpha'\alpha}\bigl[\tau^{+}\rightarrow \bar{b} (q_{+})+X'\bigr] \, ,  &
\end{eqnarray}
where $s=(p_+ + p_-)^2$, 
the solid angle element $d\Omega_{k_{+}}$ corresponds to the momentum vector
${\kk}_{+}$ in the $e^+e^-$ c.m. frame,
 and ${\rm Br}(\tau^{-}\rightarrow A)$ and ${\rm Br}(\tau^{+}\rightarrow\overline{B})$ denote the branching fractions 
for the decays $\tau^{-}\rightarrow A$ and $\tau^{+}\rightarrow\overline{B}$, respectively.\footnote{Formula (4.3) 
of Ref.~\cite{Bernreuther:1993nd}  contains a typo. These branching fraction factors
 are missing. However, they were taken into account in the numerical results given in that paper. Moreover, the variable $q_0^*$
 on the l.h.s. of Eq.~(4.4) of that reference should be replaced by $|{\bq}^*|/\langle n_A \rangle$.}

For $\tau$ decay to three charged pions whose four-momenta are all measured in an experiment we define the corresponding 
decay density matrix by
\begin{eqnarray}\label{Eq.03.08}
\mathcal{D}^A_{\beta'\beta}\bigl(\tau^{-}(k_{-})\rightarrow \pi^{-}(q_{1})\pi^{-}(q_{2})\pi^{+}(q_{3})  \nu_{\tau}\bigr) = & \nonumber \\
\Gamma^{-1}(\tau^{-} \rightarrow \pi^{-}\pi^{-}\pi^{+}{\nu_{\tau}})\frac{1}{2m_{\tau}} 
\int \frac{d^{3}q_{4}}{(2\pi)^{3}2q_{4}^{0}} (2\pi)^{4} \delta^{(4)} (k_{-} -q_{1}-q_{2}-q_{3}-q_{4}) & \nonumber \\
\times~\big{\langle}\pi^{-}(q_1) \pi^{-}(q_2)  \pi^{+}(q_3)  \nu_{\tau}|{\cal T}|\tau^{-}(k_{-}, \beta)\big{\rangle} 
\big{\langle}\pi^{-} (q_1) \pi^{-}(q_2)  \pi^{+} (q_3) \nu_{\tau}|{\cal T}|\tau^{-}(k_{-}, \beta')\big{\rangle}^* &  \, ,
\end{eqnarray}
where $q_4$ is the four-momentum of $\nu_\tau$, 
and analogously for the decay $\tau^{+}\rightarrow\pi^{+}\pi^{+}\pi^{-}\overline{\nu}_{\tau}$.
 The normalization is 
\begin{equation}\label{Eq.03.09}
\int \prod_{i=1}^{3}    \frac{d^{3}q_{i}}{(2\pi)^{3}2q_{i}^{0}}
\mathcal{D}^A_{\beta'\beta}\bigl(\tau^{-}(k_{-})\rightarrow \pi^{-}(q_{1})\pi^{-}(q_{2})\pi^{+}(q_{3})  \nu_{\tau}\bigr) 
=2\delta_{\beta'\beta} \, ,
\end{equation}
corresponding to the $\pi^{-}$ multiplicity  $2$ in this channel.
   If the analysis is restricted to three pions in a suitably defined invariant mass region around 
the nominal $a_{1}$ mass one has to take into account the corresponding phase-space cuts 
in $\Gamma(\tau^{-}\rightarrow\pi^{-}\pi^{-}\pi^{+}{\nu}_{\tau})$ and in \eqref{Eq.03.09}.

For the $\tau$ decay \eqref{eq:tdec2p} to two pions, where both the charged and the neutral pion are measured, the 
 respective decay density matrix is defined accordingly by integrating the corresponding 
 squared matrix element over the four-momentum  of the neutrino.

In order to get the inclusive cross section for case ii), considering, for instance,
the decay of the $\tau^{-}$ into three observed pions, we have to make in \eqref{Eq.03.10} the replacement 
\begin{equation}\label{Eq. 03.11}
\frac{d^{3}q_{-}}{(2\pi)^{3}2q_{-}^{0}} \mathcal{D}^a_{\beta'\beta}\bigl[\tau^{-}\rightarrow a+X\bigr]\rightarrow
\prod_{i=1}^{3} \frac{d^{3}q_{i}}{(2\pi)^{3}2q_{i}^{0}} \mathcal{D}^A_{\beta'\beta}\bigl[\tau^{-}\rightarrow \pi^{-} \pi^{-} \pi^{+} \nu_\tau\bigr]\;.
\end{equation}
Analogous replacements apply if the decay of $\tau^{+}$ to three observed pions or the decay of $\tau^\mp$ to two observed pions
 are analyzed. 

 The production density matrix $R$  in \eqref{Eq.03.10} is computed in the $e^+ e^-$ c.m. system, see below.
 Instead of calculating the decay density matrices also in this frame we can determine them in 
  the $\tau^{-}$ and $\tau^{+}$ rest systems, respectively, if we use the following:
\begin{itemize}
\item We consider rotation-free Lorentz transformations (boosts) from the c.m. frame to\ the $\tau^{-}$ and $\tau^{+}$ rest systems, respectively.
\item We use standard spinors $u_{\beta}(k)$, ${\v}_{\alpha}(k)$ for the $\tau$'s with $\beta, \alpha$ 
 denoting the spin components in a given $z$ direction (see, e.g., \cite{Nachtmann:1990ta}).
\end{itemize}
As is well known, these spin components are not changed by boost transformations. 
 Let $\Lambda_{\kk}$ be the boost transforming the $\tau^{-}$ momentum $k_{-}$ from the $e^+ e^-$ c.m. system 
to rest,    $\Lambda_{\kk} k_- = k_{-}^{*}$, where $k_{-}^{*} = (m_\tau,\boldsymbol{0})^T.$                        
 We have then with  $\Lambda_{\kk} q_{-}=q_{-}^{*}$; see \eqref{eq:Lorboost} and  \eqref{eq:rebocm},
\begin{equation}\label{Eq.03.18}
\langle a(q_{-}), X|{\cal T}|\tau^{-}(k_{-},\beta)\rangle=\langle a(q_{-}^{*}), X|{\cal T}|\tau^{-}(k_{-}^{*},\beta)\rangle \;.
\end{equation}
Insertion into the decay matrix \eqref{Eq.03.05} proves our statements above. The analogous argumentation applies to  the $\tau^+$
decay density matrices. 

In Appendix~\ref{app:taudec} we give  the explicit forms of the $\tau^\mp$ decay 
density matrices in the respective rest frames for 
 the  decay modes listed in  \eqref{Eq.03.02} -- \eqref{Eq.03.03}.

Finally, using the one-photon-exchange approximation and setting
\begin{equation}\label{eq:setFF}
F_1(q^2) = 1\, , \quad F_2(q^2) = 0  \, , \quad A(q^2) = 0 \, ,
 \end{equation}
the  production density matrix $R$ is given in the $e^+e^-$ c.m. frame by
    \begin{equation} \label{eq:ProRchi}
         R = \frac{\chi}{| 1 + e^2 \Pi_c(s)|^2} \, ,
    \end{equation}
where \cite{Bernreuther:1993nd}
\begin{equation}\label{Eq.03.12}
\chi=\chi_{SM}+ \Re \hat{d}_\tau ~\chi_{CP}^R +\Im\hat{d}_\tau ~\chi_{CP}^I +\chi_{{\hat d}^2} \, ,
\end{equation}
 and
\begin{eqnarray}\label{Eq.03.13}
\chi_{SM} &= & \frac{e^4}{s} \left\{ [k_0^2 + m_\tau^2 +|\kk|^2 (\hk \cdot\hp)^2]\one -({\sip}\cdot{\ssim})|\kk|^2[1-(\hk \cdot \hp)^2] \right. \nonumber \\
& & \left. + 2 (\hk \cdot \sip)(\hk \cdot \ssim)[|\kk|^2 +(k_0-m_\tau)^2 (\hk \cdot \hp)^2]  + 2 k_0^2 (\hp \cdot \sip)(\hp \cdot \ssim) \right. \nonumber\\
& & \left.  - 2 k_0(k_0-m_\tau) (\hk \cdot \hp)[ (\hk \cdot \sip)(\hp \cdot \ssim)
 + (\hk \cdot \ssim)(\hp \cdot \sip)] \right \} \, ,
\end{eqnarray}

\begin{eqnarray}\label{Eq.03.14}
\chi_{CP}^R & = & - 2 e^4  \frac{|\kk|}{s} \left\{ -[m_\tau + (k_0-m_\tau) (\hk \cdot\hp)^2](\sip \times \ssim)\cdot\hk \right. \nonumber \\
 & & \left. + k_0(\hk \cdot\hp) (\sip \times \ssim)\cdot\hp \right\} \, , 
\end{eqnarray}
\begin{eqnarray}\label{Eq.03.14a}
\chi_{CP}^I & = &  2 e^4 \frac{|\kk|}{s} \left\{ -[m_\tau + (k_0-m_\tau) (\hk \cdot\hp)^2](\sip - \ssim)\cdot\hk \right. \nonumber \\
 & & \left. + k_0(\hk \cdot\hp) (\sip - \ssim)\cdot\hp \right\} \, , 
\end{eqnarray}

\begin{equation}\label{Eq.03.15}
\chi_{{\hat d}^{2}}= e^4 [(\Re {\hat d}_\tau)^2 + (\Im {\hat d}_\tau)^2] \frac{ |\kk|^2}{s}[1- (\hk \cdot \hp)^2](\one - \sip \cdot \ssim) \, .
\end{equation}
Compared to Eqs.~(3.8) -- (3.10) of \cite{Bernreuther:1993nd} we neglect here the contributions from $Z$-boson exchange because we restrict 
 ourselves to the kinematic range $s\ll m_Z^2$, but we have included the photon vacuum polarization effects.
 In  \eqref{Eq.03.13} -- \eqref{Eq.03.15} we put $\pp = {\pp}_+,~ \kk = {\kk}_+$, and $\hp$ and $\hk$ denote the respective unit vectors. We have introduced in  \eqref{Eq.03.12}
and \eqref{Eq.03.15} dimensionless EDM form factors defined by
\begin{equation} \label{eq:dimlFF}
    \Re {\hat d}_\tau(s) = \frac{\sqrt{s}}{e} \Re d_\tau(s) \, , \qquad     \Im {\hat d}_\tau(s) = \frac{\sqrt{s}}{e} \Im d_\tau(s) \, .
\end{equation}
Moreover, we use in the equations above the notation \cite{Bernreuther:1993nd}
\begin{eqnarray}\label{Eq.03.16}
\one\equiv(\one\otimes\one)_{\alpha\alpha'\beta\beta'}=\delta_{\alpha\alpha'}\delta_{\beta\beta'}\;, \notag \\
\sip \equiv(\ssig\otimes\one)_{\alpha\alpha'\beta\beta'}={\ssig}_{\alpha\alpha'}\delta_{\beta\beta'} \;, \notag \\
\ssim\equiv(\one\otimes\ssig)_{\alpha\alpha'\beta\beta'}=\delta_{\alpha\alpha'}{\ssig}_{\beta\beta'} \;,
\end{eqnarray}
 where the first and second factors in these tensor products refer to the spin spaces of
  $\tau^+$ and $\tau^-$, respectively. 
  The density matrices $\chi_{\rm SM}$ and $\chi_{{\hat d}^{2}}$ are $CP$-even, whereas $\chi_{CP}^R$ 
 is $CP$- and $T_N$-odd while  $\chi_{CP}^I$ is
 $CP$-odd and $T_N$-even. Here and below $T_N$-even/odd refers to the behavior with respect to the naive
  ``time reversal'' transformation, that is, reflections 
  of three-momenta and spins.

Equation~\eqref{Eq.03.14} shows that a nonzero $\Re {d}_\tau$ generates $CP$-odd $\tau^+ \tau^-$ spin correlations in the $\pp, \kk$ 
scattering plane while a nonzero imaginary part of $d_\tau$ leads to a $CP$-odd asymmetry of the $\tau^+$ and $\tau^-$ polarizations
with projections along $\pp$ and $\kk$, cf. \eqref{Eq.03.14a}. The $\tau$ leptons autoanalyze their spin directions via their 
parity-violating weak decays. In this way these $\tau$ spin correlations and polarization asymmetries induce $CP$-odd angular correlations
 among the $\tau^\pm$ decay products, to which we now turn.

\section{Simple and optimal $CP$ observables}
\label{Sec:04}
In this chapter we discuss simple and optimal observables 
for studying $CP$ violation in the reactions \eqref{Eq.01.04}. Let us first consider the case i) above where only one charged
 particle is measured from $\tau^-$ and $\tau^+$ decay, respectively, i.e., 
  $\tau^- \to a(q_-)+X$ and $\tau^+ \to \bar{b}(q_+)+X'$.
 Simple $CP$ observables for this case were given in Ref.~\cite{Bernreuther:1993nd}.
 Observables sensitive to $\Re d_{\tau}(s)$ are, for instance, the tensors
\begin{equation}\label{Eq.04.01}
\widehat{T}^{ij}=(\hqp-\hqm)^{i}\,\frac{(\hqp\times\hqm)^{j}}{|\hqp\times\hqm|} \, + \, (i\leftrightarrow j) \;,
\end{equation}
\begin{equation}\label{Eq.04.02}
T^{ij}=(\qp-\qm)^{i}\,(\qp\times\qm)^{j}  \, + \, (i\leftrightarrow j) \;.
\end{equation}
Observables sensitive to $\Im  d_{\tau}(s)$ are, for instance,
%
\begin{equation}\label{Eq.04.03}
\widehat{Q}^{ij}=(\hqp+\hqm)^{i}\, (\hqp -\hqm)^{j}+(i\leftrightarrow j) \;,
\end{equation}
\begin{equation}\label{Eq.04.04}
Q^{ij}=(\qp + \qm)^{i}\,(\qp -\qm)^{j}-\dfrac{1}{3}\delta^{ij}(\qp^2 - \qm^2)+(i\leftrightarrow j) \;.
\end{equation}
The momenta ${\bq}_\mp$ in \eqref{Eq.04.01} -- \eqref{Eq.04.04} are defined  in the $e^+ e^-$ c.m. frame, and ${\hbq}_{\pm}={\bq}_{\pm}/|{\bq}_{\pm}|$
and $i,j\in\lbrace1,2,3\rbrace$ are the Cartesian vector indices. 
 These observables, denoted generically by $\cO(\qp,\qm)$, have the property to be odd under $CP$:
\begin{equation}\label{Eq.04.05}
\cO(\qp,\qm)=-\cO(-\qm,-\qp)\;.
\end{equation}
Moreover, Eqs.~\eqref{Eq.04.01} and \eqref{Eq.04.02} are $T_N$-odd while \eqref{Eq.04.03} and \eqref{Eq.04.04} are $T_N$-even.
A nonzero expectation value of any such observable of the form
\begin{eqnarray}\label{Eq.04.06}
\langle\cO\rangle_{ab}& \equiv & \dfrac{1}{2}\bigl{\lbrace}\langle\cO\rangle_{a\bar{b}}+\langle\cO\rangle_{b\overline{a}}\bigr{\rbrace} \nn \\
& = & \dfrac{1}{2}\Bigl{\lbrace}\dfrac{\int d\sigma_{a\bar{b}}\cO}{\int d\sigma_{a\bar{b}}}+\dfrac{\int d\sigma_{b\overline{a}}
\cO}{\int d\sigma_{b\overline{a}}}\Bigr{\rbrace}
\end{eqnarray}
is a genuine signature of $CP$ violation. Here $d\sigma_{a\bar{b}}$ is the cross section \eqref{Eq.03.10} 
of the reaction \eqref{eq:2pincl} and $d\sigma_{b\overline{a}}$ the corresponding one for the charge-conjugate channel.
 We assume that any phase-space  cuts that may be applied are made in a $CP$-symmetric way.

Observables of the type \eqref{Eq.04.01}-\eqref{Eq.04.04} were studied extensively in \cite{Bernreuther:1993nd}.
In Sec.~\ref{sec:results} we give an update of the sensitivities achievable with these observables at the KEKB accelerator with
Belle II. A discussion of the sensitivities achievable with the BES III experiment at the Beijing Electron-Positron Collider II 
is deferred to a future publication.

We shall now turn to optimal observables \cite{Atwood:1991ka,Davier:1992nw,Diehl:1993br} and we follow 
here Ref.~\cite{Diehl:1993br}.
 We denote the measured phase-space variables generically by $\phi$ and the $CP$-transformed ones by $\overline{\phi}$:
\begin{equation}\label{Eq.04.07}
CP:\quad\phi\rightarrow\overline{\phi}\;.
\end{equation}
Phase-space cuts are assumed to be $CP$-symmetric. In the following we denote the dimensionless $CP$-violating
 EDM form factors (cf. Eq.~\eqref{eq:dimlFF}) that are to be measured by
\begin{equation}\label{Eq.04.08}
g_{1}= \Re \,{\hat d}_{\tau} \,,\quad g_{2}= \Im \, {\hat d}_{\tau} \;.
\end{equation}
From experiment we know that these couplings are  small, $|g_{1,2}|\ll 1$. From  \eqref{Eq.01.03}
 we get $|g_{1,2}| \le 2.4 \times 10^{-2}$
 for $\sqrt{s}=10.58$ GeV.
Therefore, we shall work to leading order in these couplings. 
The cross section \eqref{Eq.03.10} can be expanded in the $g_{i}$ as follows, 
neglecting terms of second order in these couplings:
\begin{equation}\label{Eq.04.09}
S^{a\bar{b}}(\phi)=\dfrac{d\sigma_{a\bar{b}}(\phi)}{d \phi}=S_{0}^{a\bar{b}}(\phi)+g_{i}\,S_{1,i}^{a\bar{b}}(\phi)\;.
\end{equation}
Here and in the following we use the summation convention. Moreover, in order not to overload the notation, 
the labels  $a$ and $\bar{b}$ denote in  \eqref{Eq.04.09} and in what follows
 decays of $\tau^-$ and $\tau^+$ to one, two, or three measured particles, respectively.
 The $CP$ properties of $S_{0}$ and $S_{1}$ in \eqref{Eq.04.09} are:
\begin{equation}\label{Eq.04.09a}
 S_{0}^{a\bar{b}}(\phi)= S_{0}^{b\overline{a}}(\overline{\phi}) \;, \qquad
  S_{1,i}^{a\bar{b}}(\phi)=- S_{1,i}^{b\overline{a}}(\overline{\phi})\;.
\end{equation}
We define now the observables
\begin{equation}\label{Eq.04.10}
\cO_{i}^{a\bar{b}}(\phi)=S_{1,i}^{a\bar{b}}(\phi)\big{/}S_{0}^{a\bar{b}}(\phi)\;.
\end{equation}
Their expectation value $E_{0}$ for $g_i=0$ is
\begin{equation}\label{Eq.04.11}
E_{0}(\cO_{i}^{a\bar{b}})=\int d\phi S_{0}^{a\bar{b}}(\phi)\cO_{i}^{a\bar{b}}(\phi)\bigg{/}\int d\phi' S_{0}^{a\bar{b}}(\phi') \;.
\end{equation}
We set
\begin{equation}\label{Eq.04.12}
\cO_{i}^{\prime a\bar{b}}(\phi)=\cO_{i}^{a\bar{b}}(\phi)-E_{0}(\cO_{i}^{a\bar{b}})
\end{equation}
and get for the expectation value of ${\cO_{i}'}^{a\bar{b}}$:    
\begin{equation}\label{Eq.04.13}
E(\cO_{i}^{\prime a\bar{b}})=\int d\phi S^{a\bar{b}}(\phi)\cO_{i}^{\prime a\bar{b}}(\phi)\Bigg{/}\int d\phi'S^{a\bar{b}}(\phi')
=V_{ij}(\cO^{\prime a\bar{b}})g_{j}\;.
\end{equation}
The expression on the right-hand side is obtained by expanding the ratio to first order in the $g_{j}$.
Here $V({\cO'}^{a\bar{b}})=\bigl(V_{ij}({\cO'}^{a\bar{b}})\bigr)$ is the covariance matrix 
of the quantities $\cO'$ for $g_j=0$.
\begin{equation}\label{Eq.04.14}
V_{ij}(\cO^{\prime a\bar{b}})=E_{0}(\cO_{i}^{\prime a\bar{b}}\cO_{j}^{\prime a\bar{b}})
=E_{0}\biggl(\frac{S_{1,i}^{a\bar{b}}}{S_{0}^{a\bar{b}}}\, 
\frac{S_{1,j}^{a\bar{b}}}{S_{0}^{a\bar{b}}}\biggr)-E_{0}
\biggl(\frac{S_{1,i}^{a\bar{b}}}{S_{0}^{a\bar{b}}}\biggr)E_{0}\biggl(\frac{S_{1,j}^{a\bar{b}}}{S_{0}^{a\bar{b}}}\biggr)\;.
\end{equation}
 The covariance matrix $V({\cO'}^{a\bar{b}})$ is positive definite. 
From \eqref{Eq.04.13} we obtain
\begin{equation}\label{Eq.04.15}
g_{i}=V^{-1}_{ij}(\cO^{\prime a\bar{b}})E(\cO_{j}^{\prime a\bar{b}})\;.
\end{equation}

In the remainder of this section we recall from \cite{Diehl:1993br} some general relations for optimal observables  in order to 
make our article self-contained. Also, we shall discuss that in the nondiagonal case $a\neq b$ the theoretically optimal estimators 
 may not always be ``optimal'' from a practical point of view (see the discussion after Eq.~\eqref{Eq.04.33} below). 

We consider first the diagonal case, $a=b$, and assume that $n$ events of this type are analyzed. 
The density function is then
\begin{align}\label{Eq.04.16}
F(\phi_{1},\dots , \phi_{n})=\prod_{k=1}^{n} f(\phi_{k})\;, \nn \\
f(\phi)=S^{a\overline{a}}(\phi)\Bigg{/}\int d\phi'S^{a\overline{a}}(\phi')\;.
\end{align}
The information matrix $I=(I_{ij})$ is defined by
\begin{equation}\label{Eq.04.17}
I_{ij}=E\Biggl{[}\Bigl{(}\frac{\partial}{\partial g_{i}}\ln F\Bigr{)}\Bigl{(}\frac{\partial}{\partial g_{j}}\ln F\Bigr{)}\Biggr{]} \, .
\end{equation}
The optimal estimators for the couplings $g_i$ are in this case:
\begin{equation}\label{Eq.04.18}
\gamma_{i}(\phi)=V^{-1}_{ij}(\cO^{\prime a \overline{a}})\overline{\cO'}_j^{ a \overline{a}}(\phi) \, ,
\end{equation}
where $\overline{\cO_j'}$ denotes the mean value of $\cO_j'$. 
 From Eqs.~\eqref{Eq.04.15} and \eqref{Eq.04.16} we obtain the expectation values
\begin{equation}\label{Eq.04.19}
E(\gamma_{i})=g_{i}
\end{equation}
and  the covariance matrix of the $\gamma_{i}$, evaluated for $g_{i}=0$, is 
\begin{equation}\label{Eq. 04.20}
V_{ij}(\gamma) = E_0(\gamma_{i}\gamma_{j})=\dfrac{1}{n} V^{-1}_{ij}(\cO^{\prime a \overline{a}}) \;.
\end{equation}
We get for the information matrix \eqref{Eq.04.17}:
\begin{equation}\label{Eq. 04.21}
I|_{g=0}=nV(\cO^{\prime a \overline{a}}) \;.
\end{equation}
Therefore, we have here
\begin{equation}\label{Eq.04.22}
V^{-1}(\gamma)=I|_{g=0}
\end{equation}
and the estimators \eqref{Eq.04.18} are optimal  for small $g_j$. That is, the error ellipse obtained with the estimators 
$\gamma_{i}$ in \eqref{Eq.04.18} is given by the one obtained from $I$ which is the smallest one possible.
We note that due to the $CP$ properties  \eqref{Eq.04.09a} of $S_{0}$ and $S_{1,i}$
we have in the diagonal case $a=b$, assuming possible cuts in phase space to be $CP$-symmetric: 
\begin{equation}\label{Eq.04.23}
E_{0}(\cO_{i}^{a \bar{a}})=0 \;, \qquad
\cO_{i}^{\prime a \bar{a}}(\phi)=\cO_{i}^{a \bar{a}}(\phi)\;, \qquad
V(\cO^{\prime a \bar{a}})=V(\cO^{ a \bar{a}})\;,
\end{equation}
and the optimal estimators are
\begin{equation}\label{Eq.04.24}
\gamma_{i}(\phi)=V^{-1}_{ij}(\cO^{a \overline{a}}) \overline{\cO}_{j}^{a\overline{a}}(\phi)\;;
\end{equation}
see \eqref{Eq.04.11}, \eqref{Eq.04.12}, and \eqref{Eq.04.18}.

Finally, we treat the nondiagonal case, $a\neq b$.
We assume that any phase-space cuts made for the channel $a\bar{b}$ are applied to $b\bar{a}$ in a $CP$-conjugate way.
We get then from the $CP$ relations \eqref{Eq.04.09a}:
\begin{equation}\label{Eq.04.25}
\int d\phi S_{0}^{a\bar{b}}(\phi)=\int d\overline{\phi} S_{0}^{b\bar{a}}(\overline{\phi})\;,
\end{equation}
\begin{equation}\label{Eq.04.26}
E_{0}\Biggl{(}\frac{S_{1,i}^{a\bar{b}}}{S_{0}^{a\bar{b}}}\Biggr{)}=-E_{0}\Biggl{(}\dfrac{S_{1,i}^{b\bar{a}}}{S_{0}^{b\bar{a}}}\Biggr{)}\;,
\end{equation}
\begin{equation}\label{Eq.04.27}
V\bigl(\cO^{\prime a \bar{b}}\bigr)=V\bigl(\cO^{\prime b\bar{a}}\bigr)\;.
\end{equation}
We assume that $n_{1}$ events of the type $a\bar{b}$ and $n_{2}$ events $b\bar{a}$ are analyzed. 
The density function is then
\begin{equation}\label{Eq.04.28}
F(\phi_{1}, \dots ,\phi_{n_{1}},\overline{\phi}_{1}, \dots , \overline{\phi}_{n_{2}})
=\prod_{k=1}^{n_{1}} f_{ a \bar{b}}(\phi_{k})\prod_{l=1}^{n_{2}} f_{b \overline{a}}(\overline{\phi}_{l})
\end{equation}
with
\begin{equation}\label{Eq.04.29}
f_{a\bar{b}}(\phi)=S^{a\bar{b}}(\phi)\Big{/}\int d\phi'S^{a\bar{b}}(\phi') \;, \qquad
f_{b\overline{a}}({\overline\phi})=S^{b\overline{a}}(\overline{\phi})\Big{/}\int d\overline{\phi}'S^{b\overline{a}}(\overline{\phi}') \, .
\end{equation}
%
 Here the information matrix $I=(I_{ij})$ is given for $g_{i}=0$ by
\begin{equation}\label{Eq.04.30}
I_{ij}\big|_{g=0}=E\Bigl{[}\Bigl{(}\frac{\partial}{\partial g_{i}}\ln F\Bigr{)}
\Bigl{(}\frac{\partial}{\partial g_{j}}\ln F\Bigr{)}\Bigr{]}\Big|_{g=0}
= n~ V _{ij}(\cO^{\prime a\bar{b}})  \;,
\end{equation}
where $n=n_1 + n_2$. Here it is convenient to use as estimators for 
the couplings $g_i$, with $\cO_{j}$ from \eqref{Eq.04.10}:
\begin{equation}\label{Eq.04.31}
\gamma_{i}(\phi,\overline{\phi})=\dfrac{1}{4}\Bigl[V^{-1}_{ij}(\cO^{\prime a \bar{b}}) + V^{-1}_{ij}(\cO^{\prime b \overline{a}})\Bigr]
\Bigl[\overline{\cO}_j^{ a \bar{b}}(\phi)+\overline{\cO}_j^{ b \overline{a}}(\overline{\phi})\Bigr]\;.
\end{equation}
We have 
\begin{eqnarray} \label{eq:68a}
E\left(\frac{1}{2} \overline{\cO}_i^{ a \bar{b}}+\frac{1}{2}\overline{\cO}_i^{ b \overline{a}} \right) & = &
E\left(\frac{1}{2} {\cO}_i^{ a \bar{b}}+\frac{1}{2}{\cO}_i^{ b \overline{a}} \right) 
 = V_{ij}(\cO'^{ a \bar{b}}) g_j \, , \nonumber \\
 E(\gamma_i) & = &  V^{-1}_{ik}(\cO'^{ a \bar{b}}) V_{kj}(\cO'^{ a \bar{b}}) g_j =  g_i  \, .
\end{eqnarray}
The covariance matrix of these estimators is obtained as
\begin{equation}\label{Eq.04.32}
V(\gamma)=\dfrac{n_{1}+n_{2}}{4n_{1}n_{2}}V^{-1}(\cO^{\prime a \bar{b}}) \, ,
\end{equation}
which implies
\begin{equation}\label{Eq.04.33}
V^{-1}(\gamma)=n\biggl(1-\dfrac{(n_{1}-n_{2})^{2}}{n^{2}}\biggr) V(\cO^{\prime a \bar{b}})
=\biggl(1-\dfrac{(n_{1}-n_{2})^{2}}{n^{2}}\biggr)I\big|_{g=0} \;.
\end{equation}
The $\gamma_i$ in Eq.~\eqref{Eq.04.31} are the optimal estimators for $n_{1}=n_{2}=n/2$. For $n_{1}\neq n_{2}$ 
they are not quite optimal, but for the theoretically optimal estimators one would need in this case the precise knowledge
of $E_{0}(S_{1,i}^{a\bar{b}}/S_{0}^{a\bar{b}})$.
This would introduce an unnecessary source of uncertainty in the measurements.

To conclude this section we remark on the following. A more elaborate description of $\tau$-pair production and decay would take 
higher-order radiative corrections into account. Let us denote the resulting differential cross section by $\tilde{S}^{a\bar{b}}$,
\begin{equation}\label{Eq.04.34}
 \dfrac{d\sigma_{a\bar{b}}}{d\phi}(\phi)=\tilde{S}^{a\bar{b}}(\phi)\; .
 \end{equation}
 If it is $CP$-invariant, we have
 \begin{equation}\label{Eq.04.34a}
 \tilde{S}^{a\bar{b}}(\phi)=\tilde{S}^{b\overline{a}}(\overline{\phi}) \; .
 \end{equation}
Then the corresponding expectation values $\tilde{E}$ of the estimators $\gamma_{i}$    defined in 
 \eqref{Eq.04.24} and \eqref{Eq.04.31} and constructed with the  expressions $S_{0}$, $S_{1,i}$ from~\eqref{Eq.04.09}
 will, of course,  be zero due to \eqref{Eq.04.09a}:
 \begin{equation}\label{Eq.04.35}
 \tilde{E}(\gamma_{i}) = 0 \, .
 \end{equation}
 That is, the observables $\gamma_i$ given in \eqref{Eq.04.24} and \eqref{Eq.04.31} are in all cases genuine $CP$ observables.
 They cannot get nonzero expectation values, neither from $CP$-conserving radiative SM corrections nor from $CP$-conserving interactions
  beyond the SM.

\section{Numerical results at $\sqrt{s} = 10.58$ GeV}
\label{sec:results} 
 We consider now $\tau$-pair production and decay
 at the $\Upsilon(4{\rm S})$ resonance at $\sqrt{s}=10.58~\GeV$ and compute the 
 expectation values of the simple and optimal $CP$ observables discussed in the previous section and estimate the resulting 
    1 s.d. (standard deviation) statistical sensitivities to the EDM form factors $\Re d_\tau$ and $\Im d_\tau$ at this c.m. energy.
  The expectation values of the $CP$ observables are computed to leading order  in the real and imaginary parts of the $\tau$ EDM form factor
   using the expression \eqref{Eq.03.10} for the differential cross section with \eqref{eq:setFF} -- \eqref{Eq.03.14a} and  
    several of the decay density matrices given in Appendix~\ref{app:taudec}.
  First, no phase-space cuts are applied. At the end of this section we analyze also the effects of cuts.

  The expectation values of the observables \eqref{Eq.04.01} -- \eqref{Eq.04.04} at the $\Upsilon(4{\rm S})$ resonance 
   in the decay channels where only one charged particle from $\tau^-$ and one from $\tau^+$ decay is measured (case i) above)
   are of the form:
  \begin{eqnarray}
   \langle T^{ij}\rangle_{ab} = c_{ab}(s) \, \Re {\hat d}_\tau(s)~ s^{ij} \, , &
   \qquad  \langle \widehat{T}^{ij}\rangle_{ab}=   {\tilde c}_{ab}(s)  \, \Re {\hat d}_\tau(s)~ s^{ij}  \, , \label{eq:exsimT} \\
    \langle Q^{ij}\rangle_{ab} = \kappa_{ab}(s)  \, \,\Im {\hat d}_\tau(s)~ s^{ij} \, , &
   \qquad  \langle \widehat{Q}^{ij}\rangle_{ab}=   {\tilde\kappa}_{ab}(s) \, \Im {\hat d}_\tau(s)~ s^{ij}  \, .
   \label{eq:exsimQ}
  \end{eqnarray}
  In the case of nondiagonal decay channels $a\neq b$ the expectation values are calculated as averages defined in \eqref{Eq.04.06}.
  The expectation values of the symmetric traceless tensors \eqref{Eq.04.01} -- \eqref{Eq.04.04} must be proportional to 
  a tensor $s^{ij}$ with the same property. Using the $e^+$ beam direction $\hp$ in the $e^+ e^-$  c.m. frame we have 
  \begin{equation} \label{eq:defsij}
   (s^{ij})= \frac{1}{2} \left({\hat p}^i {\hat p}^j - \frac{1}{3} \delta^{ij} \right)
   = {\rm diag}\left(-\frac{1}{6}, -\frac{1}{6}, \frac{1}{3} \right) \, .   
  \end{equation}
The right-hand side of \eqref{eq:defsij} follows from identifying  $\pp$  with the $z$ axis which we do in the following.
Equation~\eqref{eq:defsij}  is identical to the tensor polarization of the intermediate photon state. 
 Because the diagonal elements
  of the above tensor observables are not independent, we consider only their $3,3$ components that have the largest expectation values.
  Naive ``time reversal" invariance  $T_N$ implies that the 
  expectation values \eqref{eq:exsimT} and \eqref{eq:exsimQ} do not depend on $\Im {\hat d}_\tau$ and $\Re {\hat d}_\tau$, respectively.
  That is, the covariance matrix of the $T$ and $Q$ tensors is diagonal; see Appendix~\ref{app:exCoCP}.
  
  In order to estimate the statistical error in the measurement of the expectation values of the observables $\mathcal{O}$ we compute 
  also the respective standard deviation $\Delta \mathcal{O} =\sqrt{\langle \mathcal{O}^2 \rangle - \langle \mathcal{O} \rangle^2 }$
   of the distribution of $\mathcal{O}$ in the SM for the various decay channels. As discussed
    in Appendix~\ref{app:exCoCP}  the SM expectation values of the tensors 
   $T^{ij}$, ${\widehat T}^{ij}$    
   vanish for the differential cross section as used by us. (Cf. Section~\ref{Sec:03}.) For the tensors $Q^{ij}$, ${\widehat Q}^{ij}$ 
   this is also true in the diagonal case $a = b$. In the nondiagonal case, $a \neq b$, their SM expectation values need not be zero, but
   are found numerically to be negligibly small.
   In 
   Tables~\ref{tab:TThatres} and~\ref{tab:QQhatres}
   we assume that the momenta of
  $\rho^\mp$ mesons can be experimentally determined and we treat them as on-shell particles with the $\tau$-spin analyzing power given in \eqref{eq:spporo}.
   The symbols $\ell$ and $\ell'$ denote either the electron or muon, both are taken to be massless.
   We sum over the diagonal and nondiagonal  $\ell \ell'$ channels  for estimating the respective sensitivity to the real and imaginary parts of the $\tau$ EDM.
    In a diagonal decay channel the number of events  
   is $N_{aa} = N_{\tau\tau} ({\rm Br}(\tau\to a))^2$, while for a nondiagonal channel including its charge-conjugate mode we have
  $N_{ab} = 2 N_{\tau\tau} {\rm Br}(\tau\to a){\rm Br}(\tau\to b)$. The $\tau$ branching ratios are taken from \cite{Zyla:2020zbs}.
  We assume that the Belle II experiment will eventually record $N_{\tau\tau} = 4.5 \times 10^{10}$ $\tau$ pairs \cite{Kou:2018nap}. Considering as an example 
  the measurements of $T_{33}$  and $Q_{33}$ in the decay channels $a\bar{b}$ and $b\bar{a}$ the resulting ideal  1 s.d.
  statistical errors of the dimensionful EDM 
  couplings $\Re d_\tau$  and  $\Im d_\tau$ are given by
  \begin{equation} \label{eq:1sdsenTQ}
  \delta \Re d_\tau(s) = \frac{e}{\sqrt{s}}
   \frac{1}{\sqrt{N_{ab}}} \frac{3 \left[\langle T_{33}^2 \rangle_{ab}\right]^{1/2}}{|c_{ab}|} \, ,  \quad
   \delta \Im d_\tau(s) = \frac{e}{\sqrt{s}}
   \frac{1}{\sqrt{N_{ab}}} \frac{3 \left[\langle Q_{33}^2 \rangle_{ab}\right]^{1/2}}{|\kappa_{ab}|} \, .  
  \end{equation}
Equation~\eqref{eq:1sdsenTQ} yields the absolute value that $\Re d_\tau$ $(\Im d_\tau)$ must have in order that $\langle T_{33} \rangle_{ab}$
 $(\langle Q_{33} \rangle_{ab})$ deviates from its SM prediction, namely zero,  by 1 s.d.
  obtained from the square root of its SM variance. Formulas analogous to \eqref{eq:1sdsenTQ} hold for the
  dimensionless observables $\widehat{T}_{33}$ and $\widehat{Q}_{33}$.
  
\vspace{2mm}
\begin{table}[htbp]
\begin{center}
  \caption{Observables $T^{ij}$ and $\widehat{T}^{ij}$ at $\sqrt{s}=10.58\, \GeV$ $(N_{\tau\tau} = 4.5 \times 10^{10})$.  } 
  \vspace{1mm}
  {\renewcommand{\arraystretch}{1.2}
\renewcommand{\tabcolsep}{0.2cm}
\begin{tabular}{c c c c c c c c}  \hline \hline
 $\tau^-\to$ & $\tau^+\to$ & $c_{ab}$ & $\sqrt{\langle T_{33}^2 \rangle_{ab}}$ & $\delta \Re d_\tau$ & ${\tilde c}_{ab}$ & $\sqrt{\langle \widehat{T}_{33}^2 \rangle_{ab}}$ &  $\delta \Re d_\tau$  \\ 
              &               & $[\GeV^3]$& $  [\GeV^3]$                          & $(\times \, 10^{-19} \ecm)$ &         &    & \ $(\times \, 10^{-19} \ecm)$ \\ \hline\vspace{0.4mm}
$\pi^-\nu$ & $\pi^+ \bar\nu$ &    $4.46$   &       $11.34$  &  $6.21$     &     $0.332$    &   $1.02$    &            $7.50$                        \\ 
$\rho^-\nu$ & $\rho^+ \bar\nu$ & $0.71 $       & $10.07$     &  $14.7$    &   $0.043$      &  $1.06$  &                $25.5$                   \\
$\pi^-\nu$ & $\rho^+ \bar\nu$  & $1.79$       & $ 10.71 $ &     $6.74$      &   $0.110$      & $1.03$ &                 $10.5$                  \\ 
$\ell^-\nu\bar\nu$ & ${\ell'}^+ {\bar\nu}\nu$ & $ 0.36 $ & $4.68$  &  $9.86$   &    $0.037$     & $0.98$ &               $19.9$                    \\ 
$\ell^-\nu\bar\nu$ & $\pi^+ \bar\nu$       & $-1.27$ & $ 6.66 $  & $5.05$   &  $-0.111$       &  $0.96$ &                $8.3$                   \\ 
$\ell^-\nu\bar\nu$ & $\rho^+ \bar\nu$   & $-0.51$ &  $6.78$   &  $8.32$  &  $-0.037$       &  $1.00$  &                  $16.9$                 \\ \hline \hline
\end{tabular} }
\label{tab:TThatres}
\end{center}
\end{table}

  \vspace{2mm}
\begin{table}[htbp]
\begin{center}
  \caption{Observables $Q^{ij}$ and $\widehat{Q}^{ij}$ at $\sqrt{s}=10.58\, \GeV$ $(N_{\tau\tau} = 4.5 \times 10^{10})$.  } 
  \vspace{1mm}
  {\renewcommand{\arraystretch}{1.2}
\renewcommand{\tabcolsep}{0.2cm}
\begin{tabular}{c c c c c c c c}  \hline \hline
 $\tau^-\to$ & $\tau^+\to$ & $\kappa_{ab}$ & $\sqrt{\langle Q_{33}^2 \rangle_{ab}}$ & $\delta \Im d_\tau$ & ${\tilde\kappa}_{ab}$ & $\sqrt{\langle \widehat{Q}_{33}^2 \rangle_{ab}}$ &  $\delta \Im d_\tau$  \\ 
              &               & $[\GeV^3]$& $  [\GeV^3]$                          & $(\times \, 10^{-19} \ecm)$ &  &    & \ $(\times \, 10^{-19} \ecm)$ \\ \hline\vspace{0.4mm}
$\pi^-\nu$ & $\pi^+ \bar\nu$ &    $-5.26$   &       $6.56$  &  $3.04$     &     $-0.601$    &   $0.59$    &           $2.38$                         \\ 
$\rho^-\nu$ & $\rho^+ \bar\nu$ & $-2.28 $       & $7.01$     &  $3.18$     &   $-0.171$      &  $0.34$  &              $2.05$                     \\
$\pi^-\nu$ & $\rho^+ \bar\nu$  & $-3.77$       & $7.07$ &    $2.11$       &   $-0.386$      & $0.52$ &                  $1.52$                 \\ 
$\ell^-\nu\bar\nu$ & ${\ell'}^+ {\bar\nu}\nu$ & $ 1.40$ & $4.90$  &  $2.64$   &    $0.201$     & $0.64$ &               $2.40$                    \\ 
$\ell^-\nu\bar\nu$ & $\pi^+ \bar\nu$       & $-1.93$ & $7.24$  & $3.61$   &  $-0.200$       &  $0.65$ &                  $3.14$                 \\ 
$\ell^-\nu\bar\nu$ & $\rho^+ \bar\nu$   & $-0.44$ &  $7.32$   & $10.4$   &  $0.015$       &  $0.54$  &                   $22.5$                \\ \hline \hline
\end{tabular} }
\label{tab:QQhatres}
\end{center}
\end{table}
Tables~\ref{tab:TThatres} and~\ref{tab:QQhatres} contain our results for the expectation values (as defined in Eqs.~\eqref{eq:exsimT} 
and~\eqref{eq:exsimQ}) and square roots of the variances of the observables \eqref{Eq.04.01} -- \eqref{Eq.04.04} for several one-prong decays of $\tau^\mp$ where the 
charged particle has a sizable $\tau$-spin analyzing power. Moreover, the resulting 1~s.d. sensitivities to the real and imaginary parts of the $\tau$ EDM form factor 
 are listed.\footnote{The last digit of the expectation values and variances listed in Tables~\ref{tab:TThatres},~\ref{tab:QQhatres}, and~\ref{tab:optresup} is rounded. The 
  sensitivities $\delta \Re d_\tau$ and $\delta \Im d_\tau$ listed in these tables are computed with these rounded numbers.}
 Results for $T^{ij}$ and $\widehat{Q}^{ij}$ were previously given in \cite{Bernreuther:1993nd}
  and agree with those in Tables~\ref{tab:TThatres} and~\ref{tab:QQhatres}. The  accuracies   $\delta\Re d_\tau$ and $\delta\Im d_\tau$ attainable in the various $\tau^\mp$ decay channels
  listed in  Tables~\ref{tab:TThatres} and~\ref{tab:QQhatres} show that the dimensionful observable $T_{33}$ is more sensitive than $\widehat{T}_{33}$ while in the 
  case of  $Q_{33}$ and $\widehat{Q}_{33}$ it is the other way  around -- except for the $\ell \rho$ decay channel which has, 
  in any case, a rather poor sensitivity compared to the 
  other decay modes.
   
 Next we apply the  optimal observables \eqref{Eq.04.10} for measuring $\Re{\hat d}_{\tau}$ and $\Im{\hat d}_{\tau}$ to the reactions of Sec.~\ref{Sec:03}.
 As in Eq.~\eqref{Eq.04.09} and in the following equations, the labels $a,b$ refer here to the decays of $\tau^-$ and/or $\tau^+$
 to one, two, or three measured particles. In particular, we take now the differential decay density matrices for $\tau\to 2 \pi \nu_\tau$
  and $\tau\to 3 \pi \nu_\tau$ given in Appendix~\ref{app:taudec} into account.
  Using \eqref{Eq.03.13}, \eqref{Eq.03.14}, and \eqref{Eq.03.14a} 
    and the respective decay matrices  $\mathcal{D}^a$ and  $\mathcal{D}^{\bar b}$ we define
 \begin{equation}
  \cO_{R}^{a\bar{b}} = \frac{{\rm Tr}[\chi^R_{CP} \mathcal{D}^a  \mathcal{D}^{\bar b}]}{{\rm Tr}[\chi_{SM} \mathcal{D}^a  \mathcal{D}^{\bar b}]} \, , \qquad
  \cO_{I}^{a\bar{b}} = \frac{{\rm Tr}[\chi^I_{CP} \mathcal{D}^a  \mathcal{D}^{\bar b}]}{{\rm Tr}[\chi_{SM} \mathcal{D}^a  \mathcal{D}^{\bar b}]} \, ,
 \label{eq:optORI}
 \end{equation}
 where the trace is taken with respect to the spin indices of $\tau^-$ and $\tau^+$.  Both observables are $CP$-odd
  and $\cO_{R}^{a\bar{b}}$ is also $T_N$-odd while $\cO_{I}^{a\bar{b}}$ is $T_N$-even.    As already 
   emphasized we compute the expectation values by integrating over the whole phase space.
    According to the general theory discussed in Sec.~\ref{Sec:04} and Appendix~\ref{app:exCoCP}
    the covariance matrix for a  decay channel $a \bar{b}$ is given, for zero $\tau$ EDM, by \eqref{Eq.04.14}, \eqref{Eq.04.27}:
    \begin{equation} \label{eqR:V}
     V\bigl(\cO^{\prime a \bar{b}}\bigr)=V\bigl(\cO^{\prime b\bar{a}}\bigr) =
     		\left(\begin{array}{cc}
                  E_0({\cal O}_R^{' a\bar{b}}  {\cal O}_R^{' a\bar{b}}) &   E_0({\cal O}_R^{' a\bar{b}}  {\cal O}_I^{' a\bar{b}})    \\
     E_0({\cal O}_I^{' a\bar{b}}  {\cal O}_R^{' a\bar{b}})                 & E_0({\cal O}_I^{' a\bar{b}}  {\cal O}_I^{' a\bar{b}}) 
			\end{array}\right) \, ,
    \end{equation}
where 
\begin{equation} \label{eqR:defE}
 E_0({\cal O}_R^{' a\bar{b}}  {\cal O}_R^{' a\bar{b}}) \equiv \langle ({\cal O}_R^{' a\bar{b}})^2\rangle_0 \, , \quad 
 E_0({\cal O}_I^{' a\bar{b}}  {\cal O}_I^{' a\bar{b}}) \equiv \langle ({\cal O}_I^{' a\bar{b}})^2\rangle_0 \, ,
\end{equation}
etc.,  denote the expectation values for $d_\tau = 0$. The expectation values for nonzero $\tau$ EDM are given by
 \eqref{eq:68a}:
\begin{equation} \label{eqR:Edtau}
\left(\begin{array}{c}
     E\left(\frac{1}{2} {\cO}_R^{ a \bar{b}}+\frac{1}{2}{\cO}_R^{ b \overline{a}} \right) \\
      E\left(\frac{1}{2} {\cO}_I^{ a \bar{b}}+\frac{1}{2}{\cO}_I^{ b \overline{a}} \right)\end{array}\right)
      \, \equiv \, \left(\begin{array}{c}\langle {\cal O}_R^{a b} \rangle \\
        \langle {\cal O}_I^{a b} \rangle  \end{array}\right) \, = \, V\bigl(\cO^{\prime a \bar{b}}\bigr)
        \left(\begin{array}{c} \Re{\hat d}_{\tau}(s) \\ \Im{\hat d}_{\tau}(s) \end{array} \right) \, .
 \end{equation}   
   We get for the covariance matrix of the optimal estimators of  $\Re{\hat d}_{\tau}(s)$ and $\Im{\hat d}_{\tau}(s)$;
    see \eqref{Eq.04.18}, \eqref{Eq. 04.20}  and \eqref{Eq.04.31}, \eqref{Eq.04.32}:
  \begin{equation} \label{eqR:covG}
  V(\gamma) \, = \, \frac{1}{N_{ab}}  V^{-1}\bigl(\cO^{\prime a \bar{b}}\bigr) \, .  
  \end{equation}
  Here  $N_{ab}$ is the number of events 
      in the diagonal channels $a=b$ whereas for $a\neq b$ it is the sum of the events $a{\bar b}$ 
      and ${\bar b}a$, assuming that their numbers are equal.
    
 However, with the form of the differential cross section used in this paper considerable simplifications occur.  
 In the case where the $\tau$ leptons decay to one measured particle and/or to $2 \pi \nu_\tau$ where both pions
  are measured we have, as shown in Appendix~\ref{app:exCoCP}:
 \begin{equation}\label{eqR:O1p}
 \langle {\cal O}_i^{a\bar{b}}\rangle_0 = 0 \, , \quad  {\cal O}_i^{' a\bar{b}}  = {\cal O}_i^{a\bar{b}} \, , \quad (i=R,I) \, ,\qquad 
 \langle {\cal O}_R^{ a\bar{b}}  {\cal O}_I^{a\bar{b}} \rangle_0 = 0 \, . 
  \end{equation}
That is, for these channels the respective covariance matrix \eqref{eqR:V} is diagonal.

When $\tau^-$, $\tau^+$, or both $\tau$ leptons 
 decay to three measured pions, $\langle {\cal O}_i^{a\bar{b}}\rangle_0 = 0$ $(i=R,I)$ still holds  in the one-photon approximation
  (see Appendix~\ref{app:exCoCP}), but the  covariance matrix is no longer diagonal. 
  Yet we find  for these decay modes that
  $\langle {\cal O}_R^{ a\bar{b}}  {\cal O}_I^{a\bar{b}} \rangle_0 < {\rm a\; few}\times 10^{-4}$ with numerical uncertainties
   below $10^{-3}$. Therefore, within the precision of our numerical analysis the relations \eqref{eqR:O1p} hold also for these
   decay channels, and  \eqref{eqR:Edtau} simplifies to\footnote{The left-hand sides of  \eqref{eq:expoptO} denote
   averages according to  \eqref{Eq.04.06}.}
    \begin{equation} \label{eq:expoptO}
   \langle \cO_{R}^{a b}  \rangle = w_{a\bar b}(s)~\Re {\hat d}_\tau(s) \, , \qquad 
   \langle \cO_{I}^{a b}  \rangle = \omega_{a\bar b}(s)~\Im {\hat d}_\tau(s) \, ,
    \end{equation}
 where we used the abbreviations
 \begin{equation}\label{eqR:abbrev}
   w_{a\bar b} \equiv \langle ({\cal O}_R^{a\bar{b}})^2 \rangle_0 \, , \qquad  \omega_{a\bar b} \equiv \langle ({\cal O}_I^{a\bar{b}})^2 \rangle_0 \, .
 \end{equation}
The resulting  1 s.d.
  errors of the dimensionful EDM 
  couplings $\Re d_\tau$  and  $\Im d_\tau$ are given by
  \begin{equation} \label{eq:1sdsenopt}
  \delta \Re d_\tau(s) = \frac{e}{\sqrt{s}}
   \frac{1}{\sqrt{N_{ab}}} \frac{1}{\sqrt{\langle ({\cal O}_R^{a\bar{b}})^2 \rangle_0}} \, ,  \quad
   \delta \Im d_\tau(s) = \frac{e}{\sqrt{s}}
   \frac{1}{\sqrt{N_{ab}}} \frac{1}{\sqrt{\langle ({\cal O}_I^{a\bar{b}})^2 \rangle_0}} \, .  
  \end{equation}

   \vspace{2mm}
\begin{table}[htbp]
\begin{center}
  \caption{Optimal observables $\cO_{R}^{a{\bar b}}$ and $\cO_{I}^{a{\bar b}}$ at $\sqrt{s}=10.58\, \GeV$ $(N_{\tau\tau} = 4.5 \times 10^{10})$. 
  } 
  \vspace{1mm}
  {\renewcommand{\arraystretch}{1.2}
\renewcommand{\tabcolsep}{0.2cm}  
\begin{tabular}{c c c c c c c c}  \hline \hline
 $\tau^-\to$ & $\tau^+\to$ & $w_{a{\bar b}}$ & $\sqrt{\langle (\cO_{R}^{a{\bar b}})^2 \rangle_0}$ & $\delta \Re d_\tau$ & ${\omega}_{a{\bar b}}$ & $\sqrt{\langle (\cO_{I}^{a{\bar b}})^2 \rangle_0}$ &  $\delta \Im d_\tau$  \\ 
              &               &      &                           & $(\times \, 10^{-19} \ecm)$ &         &                   & \ $(\times\, 10^{-19} \ecm)$ \\ \hline\vspace{0.4mm}
$\pi^-\nu$ & $\pi^+ \bar\nu$ &    $0.111$   &       $0.333$  &  $2.45$     &     $0.352$    &   $0.593$    &     $1.37$                               \\ 
$\pi^-  \pi^0\nu$ & $\pi^+ \pi^0 \bar\nu$ & $0.111$       & $0.333$     &  $1.04 $     &   $0.352$      &  $0.593$  &       $0.58 $                            \\
$\pi^-\pi^-\pi^+\nu$ & $\pi^+ \pi^+\pi^-\bar\nu$ & $0.111$       & $0.333$     &  $2.84$     &   $0.352$      &  $0.593$  &     $1.59$                              \\
$\pi^-\nu$ & $\pi^+ \pi^0 \bar\nu$  & $0.111$       & $0.333$ &   $1.13$        &   $0.352$      & $0.593$ &            $0.63 $                       \\ 
$\pi^-\nu$ & $ \pi^+ \pi^+\pi^-\bar\nu$ & $0.111$       & $0.333$     & $1.86$      &   $0.352$      &  $0.593$  &  $1.05$                                 \\
$\pi^-\pi^0\nu$ & $ \pi^+ \pi^+\pi^- \bar\nu$ & $0.111$    & $0.333$     & $1.21 $      &   $0.352$      &  $0.593$  &    $0.68 $                               \\
$\ell^-\nu\bar\nu$ & ${\ell'}^+ {\bar\nu}\nu$ & $ 0.004 $       & $0.064 $      & $4.04$    &    $0.055 $     & $0.235 $ &  $1.08$                                 \\ 
$\ell^-\nu\bar\nu$ & $\pi^+ \bar\nu$       & $0.020 $        & $0.142 $      & $2.26$    &    $0.162 $      &  $0.402 $ &   $0.80$                                \\ 
$\ell^-\nu\bar\nu$ & $\pi^+\pi^0 \bar\nu$   & $0.020 $ &  $0.142 $   & $1.47$   &  $0.162 $       &  $0.402 $  &       $0.52 $                           \\
$\ell^-\nu\bar\nu$ & $ \pi^+ \pi^+\pi^-\bar\nu$ & $0.020 $       & $0.142 $     & $2.43$      &   $0.162 $      &  $0.402 $  &    $0.86$                               \\ \hline \hline
\end{tabular} }
\label{tab:optresup}
\end{center}
\end{table}
   
Table~\ref{tab:optresup}  contains our results for the expectation values  defined in Eq.~\eqref{eq:expoptO}
 and for the square roots of the variances of the observables \eqref{eq:optORI} for several  $\tau^\mp$ decays to one, two and/or three measured particles.
 The numbers in this table show that taking into account the full kinematic information on the hadronic system in the $\tau\to 2 \pi \nu_\tau$ 
 and $\tau\to 3 \pi \nu_\tau$ decays results in maximal $\tau$-spin analyzing power \cite{Rouge:1990kv,Kuhn:1995nn}, as is the case in the decay $\tau\to  \pi \nu_\tau$. 
 In addition, the resulting 1 s.d. sensitivities to the real and imaginary parts of the $\tau$ EDM form factor 
 are given in Table~\ref{tab:optresup}, assuming  again $4.5 \times 10^{10}$ $\tau$-pair events.   
 The 1 s.d. statistical errors  $\delta\Re d_\tau$ and $\delta\Im d_\tau$ exhibited in Table~\ref{tab:optresup} signify that taking into account the channels where one or both $\tau$ leptons
  decay to two and/or three measured pions yields a significant improvement in the sensitivity to the $\tau$ EDM form factor. 
 Comparing for each channel the  accuracies  $\delta\Re d_\tau$ and $\delta\Im d_\tau$ exhibited in Table~\ref{tab:optresup} with those in 
 Tables~\ref{tab:TThatres} and~\ref{tab:QQhatres} shows that, as expected, the optimal observables  \eqref{eq:optORI} are significantly more sensitive to the $\tau$ EDM
 than the observables $T_{33}$ and $\widehat{Q}_{33}$.

If the measurement errors of the various exclusive $\tau^+ \tau^-$ decay modes are uncorrelated, we may add in quadrature
the statistical errors  of $\Re d_\tau$ and $\Im d_\tau$
attainable for each channel:
\begin{equation} \label{dredab}
 \delta\Re d_\tau  = \left(\sum\limits_{ab} \frac{1}{\left(\delta\Re d_\tau\right)^2_{ab}} \right)^{-1/2} \, ,
\end{equation}
and analogously for $\delta\Im d_\tau$. 
 Performing these quadratures with the uncertainties listed in Tables~\ref{tab:TThatres},~\ref{tab:QQhatres}, 
and~\ref{tab:optresup} yields
the 1 s.d. errors $\delta\Re d_\tau$ and $\delta\Im d_\tau$  given in Table~\ref{tab:EDMerrup}. 
As to the optimal observables  we assumed here for the purpose of comparison  that they are measurable for 
all channels listed in Table~\ref{tab:optresup}. 
 For the leptonic modes this may not be possible in an unambiguous way; see below.
The numbers in Table~\ref{tab:EDMerrup} show that 
 the sensitivity to 
 $\Re d_\tau$  is improved by a factor of about 6 with the optimal observable $\cO_R$ as compared to using  
 the simple ones, whereas the 
  sensitivity to $\Im d_\tau$ is improved by a factor of about 4.

\vspace{2mm}
\begin{table}[htbp]
\begin{center}
  \caption{Ideal 1 s.d. statistical errors on $\Re d_\tau$ and $\Im d_\tau$ that result from adding the 
   respective uncertainties attainable in the various decay channels in quadrature.} 
 \vspace{1mm}
  {\renewcommand{\arraystretch}{1.2}
\renewcommand{\tabcolsep}{0.2cm}
\begin{tabular}{c c c c  c c} \hline \hline
$\delta\Re d_\tau \; [\ecm]$ &  &  & $\delta\Im d_\tau \; [\ecm]$&  &  \\ \hline 
 $\langle T_{33}\rangle_{ab}$ & $\langle\widehat{T}_{33}\rangle_{ab}$ &$\langle \cO_{R}^{a b}  \rangle$ &
 $ \langle Q_{33}\rangle_{ab}$    &$ \langle \widehat{Q}_{33}\rangle_{ab}$ & $\langle \cO_{I}^{a b}  \rangle$ \\
 $ 2.93 \times 10^{-19}$ & $ 4.53 \times 10^{-19}  $ &  $ 5.1 \times 10^{-20} $ &
  $1.23 \times 10^{-19} $ &   $9.4 \times 10^{-20}$ & $2.4 \times 10^{-20} $ \\ \hline \hline
 \end{tabular} }
\label{tab:EDMerrup}
\end{center}
\end{table}

   We briefly discuss the measurability  of the observables used in this section. 
 The KEKB accelerator is an asymmetric $e^+ e^-$ collider; particle momenta measured in
   the laboratory frame can of course  be transformed to the $e^+ e^-$ c.m. frame.
   The simple $CP$ observables \eqref{Eq.04.01} -- \eqref{Eq.04.04} applied to the $\tau^+ \tau^-$ decay channels
    listed in Tables~\ref{tab:TThatres},~\ref{tab:QQhatres} require the momenta of charged mesons and of $ e, \mu$ in the
     $e^+ e^-$ c.m. frame. They can be straightforwardly measured, except for the momentum of $\rho^\pm$ whose 
     determination requires 
      the reconstruction of the decay $\rho^\pm \to \pi^\pm \pi^0$. \\
     The optimal observables involve the momenta of various particles from $\tau^\pm$ decay in the 
     respective $\tau^\pm$ rest frame. 
     This requires
      the knowledge of the $\tau^\pm$ momenta in the $e^+ e^-$ c.m. frame. If both $\tau^+$ and $\tau^-$ decay 
      semihadronically their momenta can be 
      reconstructed in an unambiguous way \cite{Kuhn:1993ra}. If one of the $\tau$ leptons decays semihadronically and 
      the other one to either $e$ or $\mu$, one 
      may discard radiative events in this class such that the  $\tau^+$ and $\tau^-$ in the remaining events are, 
      to good approximation, back to back 
      and carry half of the c.m. energy in the $e^+ e^-$ frame. If the $\tau$ momentum can 
      be reconstructed in the semihadronic decay, e.g. by reconstructing 
       the $\tau$ production and decay vertices, the momentum of the leptonically decaying $\tau$ can be inferred. If both $\tau$ leptons decay
        leptonically the determination of their momenta is not possible in an unambiguous way.
        Therefore, we discard the results for the $\ell \ell'$ channels in Table~\ref{tab:optresup} 
        and add in quadrature  the statistical errors  of $\Re d_\tau$ and $\Im d_\tau$
        attainable with the events listed in  Table~\ref{tab:optresup}  where both $\tau$'s decay semihadronically 
        and for the case where the semihadronic-leptonic decays of $\tau^+ \tau^-$
        are added to the purely semihadronic events. The resulting 1 s.d. errors are given in Table~\ref{tab:EDMerl}.
        The numbers in this table and in Table~\ref{tab:EDMerrup} show that restriction to purely semihadronic 
      $\tau^+ \tau^-$ decays does not lead to a significant decrease in sensitivity to $\Re d_\tau$ and $\Im d_\tau$.

\vspace{2mm}
\begin{table}[htbp]
\begin{center}
  \caption{Ideal 1 s.d. statistical errors on $\Re d_\tau$ and $\Im d_\tau$ that result from adding in quadrature the 
   respective uncertainties attainable with the optimal observables $\cO_{R}^{a b}$ 
     and $\cO_{I}^{a b}$    in the semihadronic decays $( h h)$
   and in the semihadronic and semihadronic-leptonic $( h h + h \ell)$ decays of $\tau^+ \tau^-$.} 
 \vspace{1mm}
  {\renewcommand{\arraystretch}{1.2}
\renewcommand{\tabcolsep}{0.2cm}
\begin{tabular}{c c c } \hline \hline
                               & $\delta\Re d_\tau \; [\ecm]$ &  $\delta\Im d_\tau \; [\ecm]$  \\ \hline 
 $ h h:$    &  $ 5.8 \times 10^{-20}$         &       $3.2 \times 10^{-20} $ \\  
 $ h h + h \ell:$  &  $ 5.1 \times 10^{-20}$   & $2.5 \times 10^{-20} $ \\ \hline \hline
 \end{tabular} }
\label{tab:EDMerl}
\end{center}
\end{table}

Next we investigate the effects of cuts on the  sensitivities to the $\tau$ EDM. A full-fledged Monte Carlo analysis
 with detailed cuts is beyond the scope of this paper. We analyze in the following only the expectation values of the optimal observables 
 in the channels where both $\tau$ leptons decay semihadronically, as these observables and decay 
 modes appear to have the highest sensitivity to $d_\tau$ and allow for an unambiguous reconstruction of the $\tau^\pm$ momenta.
 We apply the following $CP$-invariant phase-space cuts on the final-state pions in the $e^+e^-$ c.m. frame:
 \begin{equation}\label{eq:picuts}
   23^\circ < \theta^* < 157^\circ \, , \qquad p_T > 0.2 \, \GeV \, ,
 \end{equation}
 where $\theta^*$ is the polar angle of a pion with respect to the  $e^+e^-$ beam 
 and $p_T$ its transverse momentum.\footnote{The cut on $\theta^*$ is  inspired by the acceptance of the 
 Belle II detector in the KEKB laboratory frame \cite{Kou:2018nap}.}
 Table~\ref{tab:optcut} contains the resulting
  coefficients $w_{a{\bar b}}$ and ${\omega}_{a{\bar b}}$ of
 the expectation values of   $\cO_{R}^{a{\bar b}}$ and $\cO_{I}^{a{\bar b}}$, respectively, defined in \eqref{eq:expoptO}.
 The event numbers and sensitivities given in Table~\ref{tab:optcut} are 
 estimated by assuming an integrated luminosity of $50~{\rm ab}^{-1}$ that corresponds to
 assuming $N_{\tau\tau}=4.5\times 10^{10}$ in the case of no cuts. The expectation values are somewhat increased by the cuts while 
  the event numbers are, of course, diminished.
 The resulting overall sensitivities are given in Table~\ref{tab:EDMecut}. Comparing these numbers with those of Table~\ref{tab:EDMerl} shows 
 that the cuts \eqref{eq:picuts} lead only to a slight decrease in sensitivity to the $\tau$ EDM.

\begin{table}[htbp]
\begin{center}
  \caption{Optimal observables $\cO_{R}^{a{\bar b}}$ and $\cO_{I}^{a{\bar b}}$ at $\sqrt{s}=10.58\, \GeV$ for the semihadronic $\tau$ decay channels
   with cuts specified in \eqref{eq:picuts}. In the case of nondiagonal channels 
    the event numbers $N_{ab}$ include those of the charge-conjugate mode.
  } 
  \vspace{1mm}
  {\renewcommand{\arraystretch}{1.2}
\renewcommand{\tabcolsep}{0.2cm}  
\begin{tabular}{c c c c c c c }  \hline \hline
 $\tau^-\to$ & $\tau^+\to$ &  $N_{ab}$ & $w_{a{\bar b}}$ & $\delta \Re d_\tau$ & ${\omega}_{a{\bar b}}$ &   $\delta \Im d_\tau$  \\ 
             &             &                     &            & $(\times \, 10^{-19} \ecm)$ &               & \ $(\times\, 10^{-19} \ecm)$ \\ \hline\vspace{0.4mm}
$\pi^-\nu$  & $\pi^+ \bar\nu$ &    $4.21\times 10^8$      &  $0.128$   &       $2.54 $  &  $0.359$     &   $1.52 $   \\ 
$\pi^-  \pi^0\nu$ & $\pi^+ \pi^0 \bar\nu$    &$16.88 \times 10^8$  &  $0.137 $ & $1.23 $     &  $0.390$  &   $0.73 $                             \\
$\pi^-\pi^-\pi^+\nu$ & $\pi^+ \pi^+\pi^-\bar\nu$ & $1.73 \times 10^8$   & $0.139 $ &  $3.81$     &  $0.408 $     & $2.22 $                                 \\
$\pi^-\nu$       & $\pi^+ \pi^0 \bar\nu$      &  $16.53\times 10^8$ & $0.135$ & $1.25$      &   $0.386 $ & $0.74$     \\ 
$\pi^-\nu$       & $ \pi^+ \pi^+\pi^-\bar\nu$ &  $5.18 \times 10^8$  & $0.137 $ & $2.21 $     & $0.401 $      &   $1.29$                                     \\
$\pi^-\pi^0\nu$ & $ \pi^+ \pi^+\pi^- \bar\nu$  & $10.74 \times 10^8$  &  $0.138 $ & $1.53$     & $0.401 $      &   $0.90 $      \\ \hline \hline
\end{tabular} }
\label{tab:optcut}
\end{center}
\end{table}

\begin{table}[htbp]
\begin{center}
  \caption{Ideal 1 s.d. statistical errors on $\Re d_\tau$ and $\Im d_\tau$ that result from adding in quadrature the 
   respective uncertainties attainable with the optimal observables $\cO_{R}^{a b}$ 
     and $\cO_{I}^{a b}$    in the semihadronic decays $( h h)$ of $\tau^+ \tau^-$ given in Table~\ref{tab:optcut}.} 
 \vspace{1mm}
  {\renewcommand{\arraystretch}{1.2}
\renewcommand{\tabcolsep}{0.2cm}
\begin{tabular}{c c c } \hline \hline
                               & $\delta\Re d_\tau \; [\ecm]$ &  $\delta\Im d_\tau \; [\ecm]$  \\ \hline 
 $ h h:$    &  $ 6.8 \times 10^{-20}$         &       $4.0 \times 10^{-20} $ \\   \hline \hline
 \end{tabular} }
\label{tab:EDMecut}
\end{center}
\end{table}
 
 Moreover, the following remark is in order. As already indicated below Eq.~\eqref{eq:fullphpr} our results for the normalized 
 expectation values listed in Tables~\ref{tab:TThatres},~\ref{tab:QQhatres},~\ref{tab:optresup}, and~\ref{tab:optcut} do not depend 
  on the fact that there is a resonance enhancement at $\sqrt{s} = 10.58$ GeV; these numbers hold also for the 
   direct continuum production of $\tau$ pairs.
  In addition, we emphasize again that the event numbers, respectively the integrated luminosity that we use for our 
  sensitivity estimates to the $\tau$ EDM
   are expectations taken from \cite{Kou:2018nap}.

The sensitivity to the $\tau$ EDM that the Belle II experiment may eventually  achieve with 
purely semihadronic  $\tau^+ \tau^-$ decays 
 was investigated also in \cite{Chen:2018cxt}. The authors of this paper use the term proportional to $d_\tau$ of the 
  matrix element for $e^+e^-\to \tau^+ \tau^- \to h \nu_\tau h' {\bar\nu}_\tau$ as optimal observable. It is evaluated with 
   the momenta of the mesons and the reconstructed one of the neutrinos. The real and imaginary parts of the $\tau$ EDM are not separately 
    determined. Assuming the same $\tau^+\tau^-$ event number as we did above, 
    the authors of Ref.~\cite{Chen:2018cxt} find that a  1 s.d. statistical sensitivity $\delta|d_\tau| = 2 \times 10^{-19} \ecm$ 
     can be achieved with their approach.

\section{The $\tau$ EDM form factor in some SM extensions}
\label{sec:FFBSM}

In the SM the EDM  $d_\ell$ of a charged lepton is extremely tiny and generated only at high loop order. 
The dominant short-distance contribution to $d_\ell$ is thought to arise via  Kobayashi-Maskawa  phase induced four-loop 
 contributions that contain, for instance, the induced EDM form factor of the $W$ boson. It can be estimated to be of the order 
 $d_\tau \sim {\cal O}(10^{-42})~\ecm$.
 (One may take, for instance, the estimate of \cite{Pospelov:2013sca} for $d_e$ and apply it to the 
 $\tau$ lepton.) Recently it was pointed out that 
 long-distance hadronic contributions are considerably larger \cite{Yamaguchi:2020eub}. For the $\tau$ EDM is was found that these
  contributions amount to $d_\tau \simeq -7.3\times 10^{-38}~\ecm$  \cite{Yamaguchi:2020eub}. Nevertheless, this is undetectable for the time being.
  
Thus, the detection of a nonzero particle EDM, in particular of the $\tau$ lepton, in a present-day experiment or
one in the foreseeable future would be evidence for a new type of $CP$ violation. 
 In this section we consider three SM extensions with $CP$-violating interactions that generate EDM 
  form factors of fundamental fermions already at one loop. The models we are interested in have 
   $CP$-violating Yukawa couplings. These interactions can  induce a $\tau$ EDM that can be much larger 
  than the electron EDM generated in these models.\footnote{We recall that in models with Higgs-Yukawa-like $CP$-violating couplings
   the dominant contribution to the electron EDM occurs at two loops~\cite{Barr:1990vd}.}
   We compute the $\tau$ EDM at one loop  in a type-II two-Higgs-doublet model and in two scalar leptoquark models and investigate its 
   potential magnitude in the timelike region $q^2 \sim (10~\GeV)^2$, taking into account phenomenological constraints, in particular the 
   tight upper bound \eqref{Eq.01.01} on the electron EDM.
\begin{figure}[h!]
\begin{center}
{\includegraphics[width=0.45\textwidth]{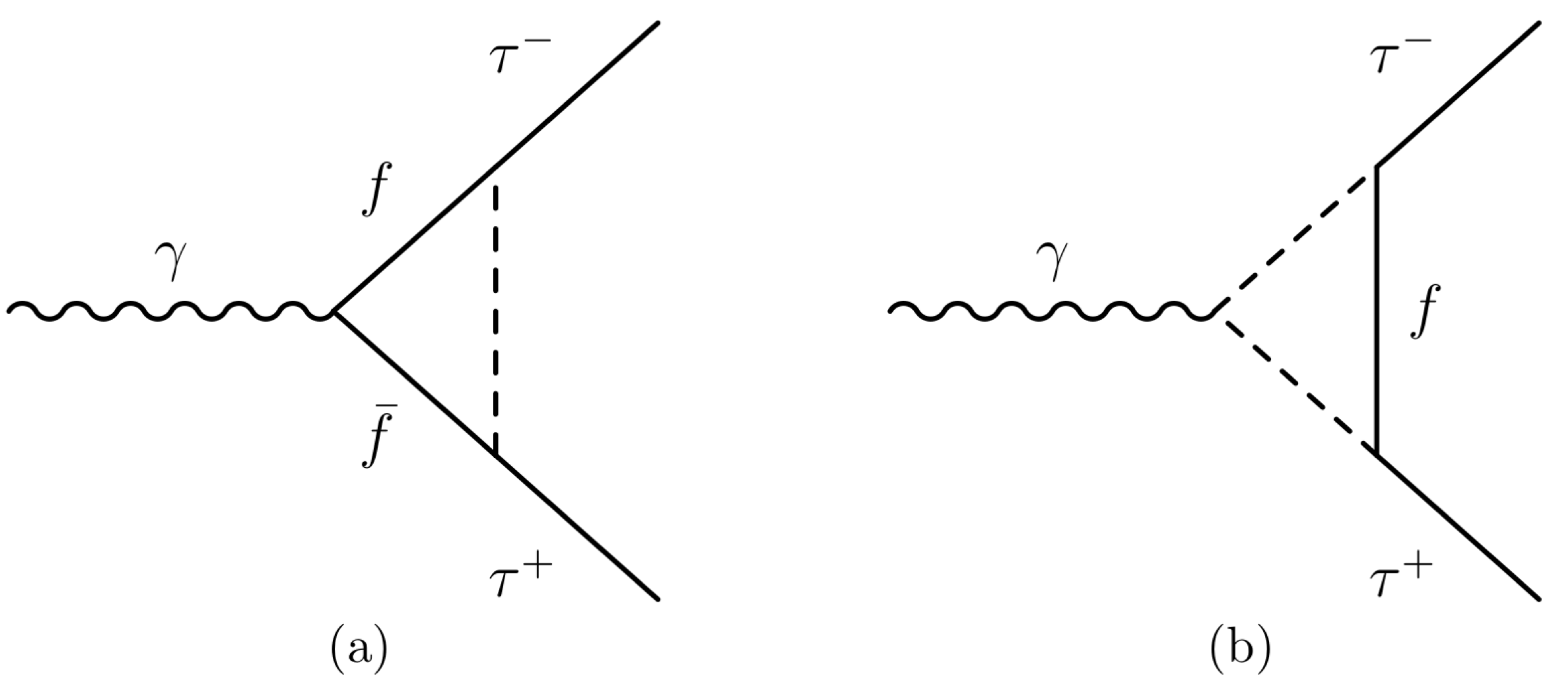}}
\caption{One-loop diagrams that contribute to the $\tau$ EDM form factor in the models considered in Section~\ref{sec:FFBSM}.
In the type-II 2HDM only diagram a) contributes and the dashed and solid internal lines correspond to $h_j$ $(j=1,2,3)$ and $\tau$, respectively.
 In the leptoquark models both diagrams contribute and the dashed and solid internal lines correspond to a spin-zero leptoquark and the top quark, respectively.}
\label{Fig.EDM}
\end{center}
\end{figure}
%

\subsection{Type-II two-Higgs doublet extension} 
\label{suse:2hdm}
In two-Higgs doublet models (2HDM) the field content of the SM is extended by an additional Higgs doublet $H_2$. We consider 
here as an example
 the so-called type-II model. It is defined by its Yukawa coupling structure: the
doublet $H_1$ is coupled to right-chiral down-type quarks and
charged leptons, while $H_2$ is coupled to right-chiral up-type
quarks only.  By construction, flavor-changing neutral currents are absent at tree
level in this model. Assuming a $CP$-violating
Higgs potential $V(H_1, H_2)$ the particle spectrum of the 2HDM contains three neutral
 Higgs bosons $h_j$ $(j=1,2,3)$ that are $CP$ mixtures. In flavor-conserving 2HDM their Yukawa couplings to quarks
 and leptons are of the form 
 \begin{equation}
{\cal L}_{Y,f} \; = \; - (\sqrt{2} G_F)^{1/2} m_f \left[ a_{f,j} {\bar f} f  \, - \, b_{f,j}  {\bar f}i \gamma_5 f \right] h_j \, ,  
\label{yukint}
\end{equation}
where  $f=q,\ell$,  $G_F$ is the Fermi constant, and the reduced Yukawa couplings $ a_{f,j}$ and  $b_{f,j}$ depend on the specific type of 2HDM.
 In the type-II model the reduced couplings of the mass eigenstates $h_j$ to the $\tau$ lepton are 
(we use here the conventions of \cite{Bernreuther:2015fts}):
 \begin{equation} \label{eq:abHiko}
 a_{\tau,j} = R_{j1}/\cos\beta \, , \qquad b_{\tau,j} = R_{j3}\tan\beta \, .
 \end{equation}
 Here $\tan\beta = {\v}_2/{\v}_1$ is the ratio of the vacuum expectation values of the two Higgs doublet fields, and $(R_{ij})$ 
 is a 
  real orthogonal matrix that relates the $CP$ eigenstates and the mass eigenstates of the three  physical neutral Higgs bosons.
  The relations \eqref{eq:abHiko} hold also for the other charged leptons and the down-type quarks. 
   (For up-type quarks, see for instance \cite{Bernreuther:2015fts}.) If $ a_{f,j} b_{f,j} \neq 0$ then \eqref{yukint} violates $CP$.
 
 Here we identify $h_1$ with the $125\, \GeV$ Higgs boson and assume that $h_2$ and $h_3$ are heavier than $400\, \GeV$.
 The exchange of the $h_j$ induces a $\tau$ EDM at one loop shown by the diagram Fig.~\ref{Fig.EDM}a. 
 With the convention of Eq.~\eqref{Eq.02.03} we get\footnote{The real and imaginary parts of the EDM form factor of a fermion were 
 computed for a class of 2HDM including the type-II model in \cite{Bernreuther:1992dz} and evaluated for the top quark.}
 \begin{equation} \label{eqM:dtcomp}
 d_\tau(s) = \sum\limits_{j=1}^3 a_{\tau,j} b_{\tau,j }d^{(j)}_\tau(s) \, ,
 \end{equation}   
 \begin{equation} \label{eqM:d2hdm}
  d^{(j)}_\tau(s) =  - \frac{e \sqrt{2}G_F m_\tau^3}{4 \pi^2 s \beta_\tau^2} 
  \left[B_0(s,m_\tau^2,m_\tau^2) -B_0(m_\tau^2,m_j^2, m_\tau^2)
       + m_j^2 C_0(s,m_\tau^2,m_j^2,m_\tau^2)\right] \, ,
  \end{equation}
 where $\beta_\tau=(1-4 m_\tau^2/s)^{1/2}$ and $m_j$ is the mass of $h_j$.
 The functions $B_0$ and $C_0$ denote the standard scalar one-loop two-point and three-point functions \cite{tHooft:1978jhc}.
  For $s \ge 4 m_\tau^2$ the
  EDM form factor \eqref{eqM:dtcomp}  has both a real and an imaginary part. 
 
 However, apart from the upper bound \eqref{Eq.01.01} on the electron EDM
  existing constraints from experiments at the LHC  preclude a $\tau$ EDM of order $10^{-20}\ecm$ or larger in this model.
  A recent analysis of the decay of the 125 GeV Higgs boson  to $\tau^+\tau^-$ by the CMS experiment 
   restricts the size of a potentially existing pseudoscalar coupling of $h_1$ to the $\tau$ lepton:
   $|b_{\tau,1}/a_{\tau,1}| \le 0.38$ at $68\%$ C.L. \cite{CMS:2020rpr}. Searches for additional neutral Higgs bosons 
   with decays to $\tau^+\tau^-$ exclude Higgs-boson masses of about 400 GeV and below for a large range of Higgs coupling to $\tau$ leptons;
   see, for example, \cite{Caputo:2019bgy,Aad:2020zxo} and references therein.

  We exemplify the order of magnitude of $d_\tau$ that is compatible with these constraints by assuming the 
  masses of the Higgs bosons $h_2$ and $h_3$ to be $m_2= 500$ GeV and $m_3=800$ GeV, respectively. Moreover, we choose
   $\tan\beta = 1$ and the angles of the  mixing matrix $R$, in the parametrization of \cite{Bernreuther:2015fts}, 
    to be $\alpha_1=\alpha_3= 0.785$, $\alpha_2=0.209$. The resulting real and imaginary parts of the $\tau$ EDM \eqref{eqM:dtcomp}
   are given in Table~\ref{tab:D2HDM}
   for several c.m. energies in the energy range considered in this paper. 
 
 \begin{table}[tbh!]
\begin{center}
\caption{\label{tab:D2HDM} Values of the real and imaginary parts of the $\tau$ EDM  form factor  \eqref{eqM:dtcomp} in the type-II
 2HDM, evaluated with  the parameter choice given in the text.}
\vspace{2mm}
 \begin{tabular}{ c c c c c }\hline \hline
    $\sqrt{s}$ [GeV] & 3.6 & 4 & 10.58 & 12 \\ \hline   
    $\Re d_\tau(s)$ $[10^{-24} \ecm]$ & 2.24  & 2.13 & 1.38 & 1.30\\ [1mm]                            
      $\Im d_\tau(s)$ $[10^{-24} \ecm]$ & 0.13 & 0.38 & 0.77 & 0.78 \\[1mm]  \hline \hline                              
 \end{tabular}
 \end{center}
 \end{table}
 
By and large the order of magnitude of the $\tau$ EDM form factor listed in Table~\ref{tab:D2HDM} is characteristic for a large class of Higgs models.
Significantly larger values of $\Re d_\tau(s)$ and $\Im d_\tau(s)$ would be possible if, for instance, Higgs bosons exist 
 with exclusive $CP$-violating couplings to the third generation of quarks and leptons, such that the stringent constraint  \eqref{Eq.01.01}
  on the electron EDM can be evaded.

 \subsection{Spin-zero leptoquarks} 
 \label{suse:lequ}
 Leptoquarks, whose interactions connect a lepton and a quark, occur naturally in unified models of strong and electroweak interactions. 
 In recent years they  have come again into the focus of numerous investigations
  in the context of possible explanations of semileptonic $B$ and $D$ 
   meson decay  and muon $(g-2)$ anomalies; see, for instance, \cite{Bauer:2015knc,ColuccioLeskow:2016dox,Crivellin:2018qmi,Bigaran:2020jil,Crivellin:2020mjs} and references therein.
   Here we are interested in spin-zero leptoquarks with  $CP$-violating Yukawa couplings. 
   They can generate EDMs of the muon and tau lepton that are significantly larger than that of the electron,\footnote{A recent analysis of the effects 
   of spin-zero leptoquarks  on the EDMs of leptons, quarks, and nucleons 
    was made in \cite{Dekens:2018bci}.}   as pointed out some time ago 
   in  \cite{Bernreuther:1991mn,Bernreuther:1996dr} (cf. also \cite{Poulose:1997kt}).   
 
 We consider in the following two different spin-zero leptoquark models, namely the SM extended by 
  a weak ${\rm SU(2)}$ leptoquark doublet $\Phi$ with  ${\rm SU(3)_c\times SU(2)_L \times U_Y(1)}$ quantum numbers
  $\Phi(3,2,7/6)$ (model I) and a SM extension by a weak singlet $S$ with quantum numbers $S(3,1,-1/3) $ (model II).
 The gauge-invariant  interaction Lagrangians are  \cite{Buchmuller:1986zs}
 \begin{equation}
  \label{eq:LIdoub}
  {\cal L}_I = [\overline{L_L} \Lambda_L \epsilon u_R + \overline{e_R}  \Lambda_R Q_L]~\Phi^\dagger \; + \; {\rm H.c.} \,, 
 \end{equation}
 \begin{equation}
  \label{eq:LIsing}
  {\cal L}_{II} = [\overline{L_L^c} Y_L \epsilon Q_L + \overline{e^c_R}  Y_R u_R]~S^\dagger \; + \; {\rm H.c.} \, .
 \end{equation}
 Here $L_L =(\nu_{iL}, e_{i,L})^T$, $Q_L = (u_{i,L}, d_{i,L})^T$, $e_R =(e_{i,R})$, $u_R=(u_{i,R})$, where $i=1,2,3$ is a generation index.
 The label $c$ denotes charge conjugation. The $2\times 2$ matrix $\epsilon= i \tau_2$ acts on the SU(2) indices.
 The electric charge (in units of $e>0$) of $S$ is  $Q_S=-1/3$. For the components of the doublet $\Phi = (\varphi, \varphi')^T$
 we have  $Q_\varphi = 5/3$ and $Q_{\varphi'} = 2/3.$
 The   $\Lambda_L$, $\Lambda_R$ and $Y_L$, $Y_R$  denote complex
 $3\times 3$ matrices in flavor space. Usually the interactions \eqref{eq:LIdoub} and \eqref{eq:LIsing} are defined in the weak basis
  and are rotated, after electroweak symmetry breaking, to the mass basis. We can choose a basis in which the Yukawa matrices of the up-type
   quark and of the charged-lepton couplings to the SM Higgs boson are already diagonal. Then only the down-type quark and neutrino fields must be 
   rotated with their respective 
   mixing matrices when one transforms to the mass basis. The interactions in  \eqref{eq:LIdoub} involving charged leptons and up-type quarks, with 
     which we are concerned here, remain unaffected.  

 We assume that the off-diagonal elements of the matrices $\Lambda_L$, $\Lambda_R$ and $Y_L$, $Y_R$ in generation space are very small and can
  be neglected. Let us denote 
  \begin{equation} \label{eqM:defcop}
  \lambda_J = (\Lambda_J)_{33} \quad \text{and} \quad y_J = (Y_J)_{33} \, , \quad J=L, R \, , 
  \end{equation}
  and 
  \begin{equation}
   f_{\rm I} =  {\rm Im}(\lambda_L^* \lambda_R)    \quad \text{and} \quad f_{\rm II} = {\rm Im}(y_R^* y_L) \, .
  \end{equation}
If $ f_{\rm I}\neq 0$ $(f_{\rm II} \neq 0)$ then the interaction Eq.~\eqref{eq:LIdoub} (Eq.~\eqref{eq:LIsing}) generates a nonzero $\tau$ EDM at one loop.
It is represented by Figs.~\ref{Fig.EDM} a) and ~\ref{Fig.EDM} b) where the internal fermion and boson lines correspond to the top 
 quark $t$ and the $\varphi$ leptoquark in model I and to $t$ and $S$ in model II, respectively.
 The $\tau$ EDM form factor is given by  \cite{Bernreuther:1996dr}  
    \begin{equation}\label{eq:tauEDMdouS}
     d_\tau(s) = e m_t N_c \frac{f_\kappa}{8 \pi^2}
      \frac{1}{s \beta_\tau^2} \left[ Q_t K_t(s) - Q_\chi K_\chi(s) \right] \, , \quad \kappa = {\rm I, II} \, 
   \end{equation}
  where $N_c=3$, $m_t$ is the mass of the top quark which provides the chirality flip, $Q_t=2/3$ and $Q_\chi = 5/3~(-1/3)$ 
  in case of model I (II), where $\chi$ denotes either $\varphi$ or $S$. Moreover
  \begin{eqnarray}
   K_t(s) & = & B_0(s,m_t^2,m_t^2) - B_0(m_\tau^2,m_t^2,m_\chi^2)
   +(m_\chi^2+m_\tau^2 - m_t^2) C_0(s, m_t^2,m_\chi^2,m_t^2) \, , \label{eqM:Kt}\\
   K_\chi(s) & = &  B_0(s,m_\chi^2,m_\chi^2) - B_0(m_\tau^2,m_t^2,m_\chi^2)
   +(s/2 + m_t^2 - m_\chi^2- m_\tau^2) C_0(s, m_\chi^2,m_t^2,m_\chi^2) \, . \label{eqM:Kchi}
  \end{eqnarray}
 Here $m_\chi$ is the mass of $\varphi~ (S)$ in the case of  model I (II). Because 
 $m_t, m_\varphi, m_S \gg \sqrt{s}$ in the kinematic range that we consider here, the $\tau$ EDM form factor \eqref{eq:tauEDMdouS}
  is real. 
 
 In order to estimate the potential size of $d_\tau$ we choose the leptoquark masses $m_\chi=1.5$ TeV ($\chi = \varphi, S$) which are compatible with the 
 experimental  bounds from LHC \cite{Sirunyan:2018ruf,Aaboud:2019bye} and the constraints from the anomalous magnetic moments of the electron 
  and muon \cite{Bigaran:2020jil}. For comparison we evaluate  \eqref{eq:tauEDMdouS} also for $m_\chi =1$ TeV and $2$ TeV.
  With $m_t=172.4$ GeV \cite{Zyla:2020zbs} we get from \eqref{eq:tauEDMdouS} the values listed in
   Table~\ref{tab:Dlepq}.

 \begin{table}[tbh!]
\begin{center}
\caption{\label{tab:Dlepq} Values of the $\tau$ EDM  form factor  \eqref{eq:tauEDMdouS}
 in the doublet (I) and singlet (II) leptoquark model. The numbers in the first, second, and third row of each model
  are obtained with
   $m_\chi =1$, $1.5$, and $2$  TeV ($\chi = \varphi, S$). Moreover, we use $m_t=172.4$ GeV.}
\vspace{2mm}
 \begin{tabular}{ c c c c c }\hline \hline
                                               $\sqrt{s}$ [GeV] & 3.6 & 4 & 10.58 & 12 \\ \hline   
 Model I:   $\Re d_\tau(s)$ $[10^{-20}   f_{\rm I} ~ \ecm]$ &  14.44 & 14.44 & 14.45 & 14.45  \\ [1mm]      
                                                              & 7.89  & 7.89 & 7.89 & 7.89\\ [1mm]          
                                                              & 5.04& 5.04 & 5.04& 5.04 \\ \hline 
 Model II:     $\Re  d_\tau(s)$ $[10^{-20}  f_{\rm II} ~ \ecm]$ &8.85   &8.85  & 8.86   & 8.86   \\[1mm]
                                                         & 5.24 & 5.24 & 5.25 & 5.25 \\[1mm] 
                                                          &3.51 & 3.51 &3.51 & 3.51 \\ \hline \hline                              
 \end{tabular}
 \end{center}
 \end{table}
 
 The numbers in Table~\ref{tab:Dlepq} show that for a given leptoquark mass the form factor $\Re d_\tau(s)$ is essentially flat in the 
 kinematic range considered here. So far, the experimental bounds on the $CP$ parameters $f_{\rm I}$, $f_{\rm II}$ are not stringent.
 Using the experimental bound  \eqref{Eq.01.03} and the numbers given in  Table~\ref{tab:Dlepq}\ for $\sqrt{s}=10.58$ GeV and $m_\chi =1.5$ TeV,
 we get
 \begin{equation} \label{eqM:bound12}
  |f_{\rm I}| < 570  \, , \qquad   |f_{\rm II}| < 857 \quad \text{for} \; \; m_\chi = 1.5~{\rm TeV} \, .
 \end{equation}
 
 If  leptoquark couplings to the $\tau$ lepton and the $c$ quark are taken into account in \eqref{eq:LIdoub} and \eqref{eq:LIsing} then $d_\tau(s)$ develops also 
  an imaginary part for $\sqrt{s} > 2 m_c$. However, away from the charm threshold, the $c$-quark contribution to $d_\tau$ is suppressed in magnitude by  the factor $m_c/m_t\sim 10^{-2}$ as compared 
  to the leading contribution \eqref{eq:tauEDMdouS}, 
 regardless of additional suppression due to small off-diagonal Yukawa couplings.

 One may expect that the Yukawa couplings of the spin-zero leptoquarks are of the Higgs-boson type. Then the (diagonal) couplings of $\varphi$ and $S$ 
   in \eqref{eq:LIdoub} and \eqref{eq:LIsing}   will be proportional to the right-handed fermion involved. That is, 
 \begin{equation} \label{eqM:YukHL}
  \lambda_L \sim m_t/M_{\rm I}\, , \quad \lambda_R \sim m_\tau/M_{\rm I} \, , \qquad y_L \sim m_\tau/M_{\rm II}  \, , \quad y_R \sim m_t/M_{\rm II} \, ,
  \end{equation}
 where $ M_{\rm I}$ and  $ M_{\rm II}$ are mass scales that are expected to be larger than the electroweak symmetry breaking scale $\v=246$ GeV.
 In this case the magnitude of $\Re d_\tau(s)$ will be smaller by a factor of at least $10^{-2}$ than the numbers 
 listed in Table~\ref{tab:Dlepq}.

 \subsection{Box contributions} 
 \label{suse:boxc}
 
\begin{figure}[h!]
\begin{center}
{\includegraphics[width=0.80\textwidth]{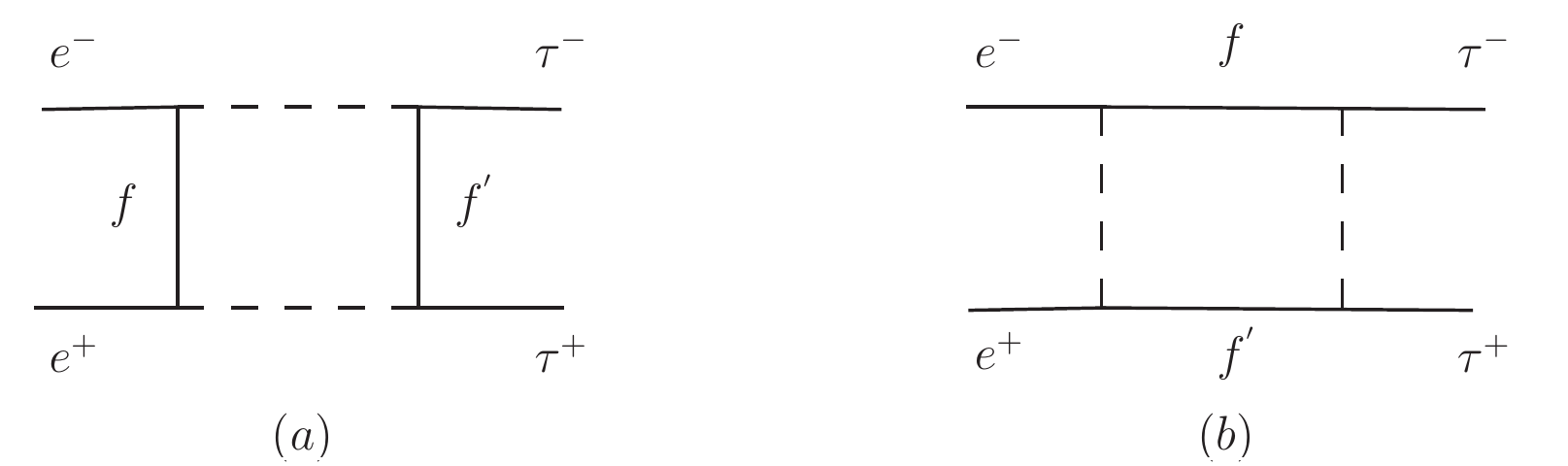}}
\caption{One-loop box diagrams that contribute to the  $S$-matrix element of $e^+e^-\to \tau^+ \tau^-$ in the models considered in 
Sections~\ref{suse:2hdm} and~\ref{suse:lequ}.
In the type-II 2HDM only diagram \ref{Fig.box} a) contributes and the dashed and solid internal lines correspond to $h_j$ $(j=1,2,3)$ 
and $f=e$, $f'=\tau$, respectively.
 In the leptoquark models both diagrams can contribute and the dashed and solid internal lines correspond
 to a spin-zero leptoquark and an up-type quark, respectively. Crossed diagrams are not shown.}
\label{Fig.box}
\end{center}
\end{figure}

 The one-loop $S$-matrix element of $e^+e^-\to \tau^+ \tau^-$ can receive in SM extensions also one-particle 
 irreducible $CP$-violating 
 box contributions that involve Lorentz structures such as $({\bar e} e) ({\bar\tau}i\gamma_5 \tau)$. 
 Here we argue that in the models considered in 
 Sections~\ref{suse:2hdm} and~\ref{suse:lequ} these contributions that are depicted in 
  Fig.~\ref{Fig.box} can be neglected compared to those of the $\tau$ EDM form factors. 
  
 In the type-II 2HDM only diagram a) appears. From the Yukawa interaction \eqref{yukint} one obtains that this 
 contribution is proportional to $G_F m^2_e$.
 Thus this contribution to the $S$-matrix element of $e^+e^-\to \tau^+ \tau^-$ is negligible compared to that 
 of the $\tau$ EDM form factor \eqref{eqM:d2hdm}.
 
 As to the spin-zero leptoquark models: If one considers interactions \eqref{eq:LIdoub}, \eqref{eq:LIsing} that are diagonal 
 in generation space, then only 
 diagram a) contributes with $f=u$, $f'=t$ and this contribution entails a suppression factor $m_u/\sqrt{s}$ where $m_u$
 is the mass of the $u$ quark.
 In the case of  interactions that are nondiagonal in generation space, diagrams \ref{Fig.box} a) and \ref{Fig.box} b) contribute,
 but those contributions that involve leptoquark couplings
  between the electron and the $c$ and $t$ quark contain off-diagonal matrix elements 
  $(\Lambda_J)_{1 j}$ or $(Y_J)_{1j}$ $(J=L,R, j\neq 1)$ that are 
   small due to experimental constraints (see, e.g., \cite{Bigaran:2020jil,Crivellin:2020mjs}).
   
   Thus we conclude that within the above SM extensions the $CP$-violating part of the one-loop $S$-matrix element 
   of $e^+e^-\to \tau^+ \tau^-$ is 
    given to very good approximation by the contribution from the $\tau$ EDM form factor.
    In addition, we remark that the one-loop EDM form factors computed
    in Sections~\ref{suse:2hdm} and~\ref{suse:lequ} are gauge invariant.
   Needless to say, the contribution of the electron EDM form
   factor to this matrix element is completely irrelevant.

\section{Conclusions}
\label{sec:concl}
 The huge data samples of $\tau^+ \tau^-$ production and decay that will eventually be recorded at existing low-energy $e^+ e^-$ colliders will 
 allow, among other investigations, the search for a $\tau$ electric dipole form factor $d_\tau(s)$  with a precision that is significantly 
 higher than existing bounds.
 We reconsidered the issue of using  simple and optimal $CP$ observables  for such measurements. We discussed the general 
 formalism of optimal observables and applied it
  to two $CP$-odd observables based on $CP$-odd 
  $\tau$-spin correlations and polarization asymmetries that are sensitive to the real and imaginary parts of $d_\tau(s)$, respectively. 
  Special emphasis was put on the covariance of these observables.
  In our numerical analysis we computed the expectation values and covariances of the optimal $CP$ observables for 
  $\tau$-pair production in $e^+ e^-$ collisions at the $\Upsilon(4{\rm S})$ resonance with subsequent decays of $\tau^\pm$ to major
  leptonic or semihadronic modes. 
  These results hold also for the continuum production of $\tau$ pairs at  $\sqrt{s}=10.58$ GeV.
  For the $\tau$ decays to two pions and three charged pions we took the full kinematic information
  of the hadronic system into account by
   incorporating the respective differential $\tau^\pm$ decay density matrices into the optimal observables. In this 
   way the maximal $\tau$-spin analyzing power 
    is obtained also with these decay modes.
   Assuming that the Belle II experiment will eventually record and analyze $4.5 \times 10^{10}$  $\tau^+ \tau^-$ events 
   at $\sqrt{s}=10.58$ GeV we found that 
    with purely semihadronic $\tau^+\tau^-$ decays 
   1 s.d. sensitivities $\delta \Re d_\tau = 5.8 \times 10^{-20} \ecm$ and $\delta \Im d_\tau = 3.2 \times 10^{-20} \ecm$   can be 
   obtained with these  optimal observables. 
    For $\Re d_\tau$  this is better than a factor of 5 and for  $\Im d_\tau$ better than  a factor of 3 as the 
    sensitivities attainable with the simple $CP$-odd observables that 
    we analyzed, too. Including events where one (or both) $\tau$ leptons decay leptonically 
  does not lead to a significant increase in sensitivity to $\Re d_\tau$ and $\Im d_\tau$.
  These results were obtained without cuts. We analyzed also the sensitivity of the optimal observables to $\Re d_\tau$ and $\Im d_\tau$
     in the purely semihadronic $\tau^+\tau^-$ decay channels by applying cuts on the final-state pions. Assuming an integrated luminosity 
      of 50 ab$^{-1}$, which corresponds to the above number of   $\tau^+ \tau^-$ events in the case of no cuts, 
     we obtained  $\delta \Re d_\tau = 6.8 \times 10^{-20} \ecm$ and $\delta \Im d_\tau = 4.0 \times 10^{-20} \ecm$.
    That is, the  1 s.d. sensitivities decrease only slightly.

   Furthermore, we  discussed a few SM extensions with nonstandard $CP$ violation that predict an nonzero $\tau$ EDM already 
   at one-loop order. The tight experimental upper bound on the 
    electron EDM, experimental results from the LHC 
    on the $CP$ nature of the 125 GeV Higgs boson, and bounds on the mass and couplings of new particles severely 
    constrain the potential magnitude of $d_\tau$.
    Within the type-II 2HDM, which we consider in this context to be exemplary for a large class of two-Higgs 
    doublet extensions of the SM, the $\tau$ EDM form factor turns out to 
    be too small to be detected in the foreseeable future. However, in scalar leptoquark extensions of 
    the SM $\Re d_\tau(s) \sim 10^{-20} \ecm$ is still possible in the energy range considered 
     in this paper. In any case, future $\tau$ EDM measurements with the Belle II and also the BES III 
     experiment using optimal observables will provide significant 
      information about new sources of $CP$ violation.

\section*{Acknowledgments}
The authors thank M. Diehl, R. Karl, F. M. Krinner, F. Nerling, A. Rostomyan, and A. Szczurek for discussions
 and correspondence, and C. Ewerz for help with one figure. The work of L.C. was supported by the Deutsche Forschungsgemeinschaft under Grant  No.
 396021762-TRR 257.

 \pagebreak
 \newpage
\appendix
\section{$\tau$ decay density matrices}
\label{app:taudec}
%

Here we list the density matrices that describe several major  decays of polarized $\tau^\mp$. 
 Most of them given below  are
 used in section~\ref{sec:results}. The kinematic variables in this
  appendix are defined in the respective $\tau^\pm$ rest frame unless stated otherwise.
  The decay density matrices are computed in the Standard Model; potential $CP$-violating effects
   in $\tau$ decays are not taken into account.
  
  \par
  First we consider $\tau^\mp$ decays into one charged prong, $\tau^\mp\to a^\mp +X$
  with particle multiplicity $\langle n_a\rangle = 1$. The charged particle $a^\mp$ acts as the $\tau^\mp$
   spin analyzer.
 Assuming $CP$ invariance in the decays of $\tau^\mp$ we have in the $\tau$ rest frame:
 \begin{equation} \label{eq:cptaupm}
  \big{\langle}a^-({\bq}), X|{\cal T}|\tau^{-}_\beta\big{\rangle} =
  \eta_a \big{\langle}a^+(-{\bq}), X^{CP}|{\cal T}|\tau^{+}_\beta\big{\rangle}  \, ,
 \end{equation}
  where  $X^{CP}$ is the $CP$ transform of $X$ and $\eta_a =\pm 1$. If $a$ denotes a lepton, we have $\eta_a =1$; for $a=$ meson,
 $\eta_a$ is the product of the intrinsic parity and charge-parity quantum numbers of $a$. Thus, for a pion ($\rho$ meson)
 we get $\eta_\pi= -1$ $(\eta_\rho =+1)$.
 Equation~\eqref{eq:cptaupm} implies for the $\tau^\mp$ decay density matrices 
 \begin{equation} \label{eq:cpDD}
    \mathcal{D}^{a^-}_{\beta'\beta}(\tau^- \to a^-({\bq})+X) 
  = \mathcal{D}^{a^+}_{\beta'\beta}(\tau^+ \to a^+(-{\bq})+X^{CP})  \, . 
 \end{equation}
  The respective decay density 
  matrix $\mathcal{D}^a= (\mathcal{D}^a_{\beta'\beta})$ defined in \eqref{Eq.03.05} and \eqref{Eq.03.07} is of the form 
  \begin{equation} \label{eqA:1prong}
    \frac{d^{3}q_{\mp}}{(2\pi)^{3}2E_{a^\mp}} \,\mathcal{D}^{a^\mp}(\tau^\mp \to a^\mp(q_{\mp})+X) 
    = dE_{a^\mp} \frac{d\Omega_{a^\mp}}{4\pi}~n(E_{a^\mp})\bigl[\one \pm h(E_{a^\mp}){\hbq}_\mp \cdot  \ssig \bigr] \, , 
  \end{equation}
 where $E_{a^\mp}$ is the energy of $a^\mp$ and $d\Omega_{a^\mp}= d\cos\theta_{a^\mp}d\varphi_{a^\mp}$.
 In \eqref{eqA:1prong} the symbol $\one$ denotes the two-dimensional unit matrix and 
 $\ssig = (\sigma_1, \sigma_2, \sigma_3)$ is the 
  vector of Pauli matrices. 
 The function $n(E_a)$ determines the energy spectrum of $\tau\to a$ while $h(E_a)$ encodes the $\tau$-spin analyzing power of the
  charged prong. Equation~\eqref{eqA:1prong} is used in the calculations of Sec.~\ref{sec:results}.
   If the right-hand side of \eqref{eqA:1prong} is integrated over $E_{a^\mp}$, it takes, 
   due to the normalization convention \eqref{Eq.03.07},
   the form
  \begin{equation} \label{eqA:1pronga}
  \frac{d\Omega_{a^\mp}}{4\pi} \int dE_{a^\mp} ~n(E_{a^\mp})\bigl[\one \pm h(E_{a^\mp}){\hbq}_\mp \cdot  \ssig \bigr] \, = \,                     
  \frac{d\Omega_{a^\mp}}{4\pi}~\bigl[\one \pm \alpha_a{\hbq}_\mp \cdot  \ssig \bigr] \, ,
  \end{equation}
 where $\alpha_a$ $(|\alpha_a|\leq 1)$ is a measure of the $\tau$ spin-analyzing power of $a$. 
  
Next we list  the spectral functions $n(E_a)$ and $h(E_a)$ of several
decay density matrices \eqref{eqA:1prong}. 
The functions  $n(E_a)$ have dimension 1/energy while
the functions  $h(E_a)$ are dimensionless.

\subsection*{The decay $\tau^{\mp} \to \ell^{\mp}(q_\mp) + \nu_{\ell} \nu_{\tau}$}

In the leptonic decays $\tau^{\mp} \to \ell^{\mp} \nu_{\ell} \nu_{\tau}$ 
the mass of  $\ell=e, \mu$ can be 
neglected. (Here and below the symbol $\nu$ denotes a neutrino or antineutrino, depending on the case.)
Using $x = 2E_{\ell} / m_{\tau}$, where  $E_{\ell}$ is 
defined in the $\tau$ rest frame, one has \cite{Tsai:1971vv}
\begin{eqnarray}
n_\ell(E_{\ell}) 
& = & 
\frac{4}{m_{\tau}} x^{2}\,\left(3-2x\right)
\, ,
\qquad
h_\ell(E_{\ell})
=
\frac{1-2\, x}{3-2\, x}
\label{eq:lep_nx_bx}
\end{eqnarray}
with $0 \le x \le 1$.
Integrating over the charged lepton energy
in \eqref{eqA:1prong} yields  \eqref{eqA:1pronga} with the $\tau$-spin analyzing power
\begin{equation} \label{eq:sppoel}
 \alpha_\ell = -\frac{1}{3} \, .
\end{equation}
The value of $\alpha_\ell$ can be increased by a suitable cut on $E_\ell$.


\subsection*{The decay $\tau^\mp \to\pi^\mp(q_\mp) +\nu_{\tau}$}

In the two-body decay $\tau\to\pi+\nu_{\tau}$ the energy $E_{\pi}$ 
in the $\tau$ rest frame is fixed and the functions 
$n_{\pi}(E_{\pi})$ and $h_{\pi}(E_{\pi})$ are given by \cite{Tsai:1971vv}:
\begin{eqnarray} \label{eq:1pincpi}
n_{\pi}(E_{\pi}) 
& = & 
\delta\left(E_{\pi} - \frac{m_{\tau}^2 + m_{\pi}^2}{2m_{\tau}}\right)
\, ,
\qquad
h_{\pi}(E_{\pi})\,\,=\,\,1
\, .
\end{eqnarray}
Here the  $\tau$-spin analyzing power is maximal,
\begin{equation} \label{eq:sppopi}
 \alpha_\pi = 1 \, .
\end{equation}

\subsection*{The decay $\tau^{\mp} \to \rho^{\mp}(q_\mp) +  \nu_{\tau}$}

 If the four-momentum of the intermediate $\rho$ meson can be determined  in 
 the decay $\tau^\mp \to \pi^\mp \pi^0\nu_{\tau}$ by measuring the energies and momenta of 
 both $\pi^\mp$ and $\pi^0$, the $\rho$ meson can be used as $\tau$-spin analyzer.
 It is well known that in the two-body decay of a polarized $\tau$ to a transversely or longitudinally
  polarized spin-1 meson and $\nu_{\tau}$ the $\tau$-spin analyzing power of the meson is  maximal \cite{Hagiwara:1989fn}. However, the polarization 
   of the vector meson cannot be determined event by event. Summing over the polarizations of the $\rho$ meson
    and treating it as an on-shell particle, one obtains  $\tau^\mp$ decay density matrices of the form \eqref{eqA:1prong}
    with the spectral functions \cite{Tsai:1971vv,Hagiwara:1989fn}
 \begin{eqnarray} \label{eq:1pincro}
n_{\rho}(E_{\rho}) 
& = & 
\delta\left(E_{\rho} - \frac{m_{\tau}^2 + m_{\rho}^2}{2m_{\tau}}\right)
\, ,
\qquad
h_{\rho}(E_{\rho})\,\,=\,\frac{m_\tau^2 - 2 m_\rho^2}{m_\tau^2 + 2 m_\rho^2}
\, .
\end{eqnarray}
Using $m_\rho = 0.775~\GeV$  we obtain 
\begin{equation} \label{eq:spporo}
 \alpha_\rho = 0.45 \, .
\end{equation}  
We  use this two-body decay mode with \eqref{eqA:1prong} and \eqref{eq:1pincro} in our 
 analysis of the simple $CP$ observables in Sec.~\ref{sec:results}.


 \subsection*{The decay $\tau^{\mp} \to   \pi^{\mp}(q_1) + \pi^{0}(q_2) + \nu_{\tau}$}

The differential rate of the decay of polarized  $\tau$ leptons to a
charged  and neutral pion via a $\rho$ meson was calculated in \cite{Tsai:1971vv}
 in the on-shell approximation for the intermediate $\rho$ meson.
 A more elaborate description of this decay mode takes the $\rho$ and $\rho'$ resonances 
and their finite widths as intermediate states into account \cite{Kuhn:1990ad,Jadach:1990mz,Hagiwara:2012vz}.
We use the matrix element of \cite{Kuhn:1990ad,Jadach:1990mz} for the decay chain 
$\tau \to \rho~(\rho') \to 2 \pi \nu_\tau$. Exact isospin invariance is assumed. In the $\tau^-$ rest frame
 we obtain for the   $\tau^-\to \pi^- \pi^0 \nu_\tau$ decay density matrix
     $\mathcal{D}^{2\pi}$ that is differential in the pion momenta:
\begin{eqnarray}\label{Eq.tau2pidec}
\prod_{i=1}^{2} \frac{d^{3}q_{i}}{(2\pi)^{3}2q_{i}^{0}}
\mathcal{D}^{2 \pi}\bigl(\tau^{-}(k)\rightarrow \pi^{-}(q_{1})\pi^{0}(q_{2})  \nu_{\tau}\bigr) =
\frac{1}{2 m_\tau \Gamma_{2\pi}}  d\Phi_2 |\mathcal{M}_2|^2 \, ,
\end{eqnarray}
where $\Gamma_{2\pi}=\Gamma(\tau^{-}\to\pi^{-}\pi^{0}{\nu}_{\tau})$ and
\begin{equation} \label{eq:M2pi0}
 |\mathcal{M}_2|^2 = G_F^2 |V_{ud}|^2 |F_\pi(Q^2)|^2\left( A_2 \one + \boldsymbol{H}_2 \cdot \ssig \right) \, .
\end{equation}
 Here $G_F$ and $V_{ud}$ denote the Fermi constant and the $ud$ Cabibbo-Kobayashi-Maskawa matrix element, respectively.
 The terms in the squared matrix element are
 \begin{eqnarray}
  A_2 & = & 4 \left[ 2 (k\cdot q)^2 + q^2(Q^2-k\cdot Q) \right] \, ,\label{eq:A2sq} \\
H_2^j & = & 4 m_\tau \left[2 (k \cdot q) q^j  + q^2 Q^j \right] \, , \label{eq:B2sq} 
 \end{eqnarray}
where $j=1,2,3$, $k=(m_\tau, \boldsymbol{0})^T$ in the $\tau$ rest frame,  $Q=q_1 + q_2$, and $q= q_1 -q_2$. \\
The phase-space measure $d\Phi_2$  can be  parametrized as follows:
  \begin{eqnarray}\label{eqrecPhas3}
    d\Phi_2 = &  \frac{1}{ 64~(2\pi)^5} 
   dQ^2 ~\theta(m_\tau^2-Q^2) \theta(Q^2-4 m_\pi^2)
  \displaystyle{ d\Omega_Q \frac{\lambda^{1/2}(m_\tau^2,Q^2,0)}{m_\tau^2}  d\Omega_1^*  \frac{\lambda^{1/2}(Q^2,m_\pi^2\,m_\pi^2)}{Q^2} }\, ,
 \end{eqnarray}
 where $d\Omega_Q=d\cos\theta_Qd\varphi_Q$ is the solid angle element of $Q$, i.e. of $\rho~(\rho')$, in the $\tau$ rest frame and 
 $d\Omega_1^*$ is the solid angle element of the charged pion $\pi^-(q_1)$ in the rest frame of $\rho~(\rho')$.
  Moreover,
 \begin{equation}
  \lambda(x,y,z) = x^2 + y^2 + z^2 - 2xy -2xz -2yz \, .
 \end{equation} 
 The form factor $F_\pi$ in  \eqref{eq:M2pi0} can be parametrized by  \cite{Kuhn:1990ad}
\begin{equation}
\label{eq:deffpion2}
 F_\pi(Q^2) = \frac{B_\rho(Q^2) + \beta_2 B_{\rho'}(Q^2)}{1+\beta_2} \, ,
\end{equation}
where
\begin{equation}
 \label{eq:defBrho2}
 B_\rho(x) =  \frac{m^2_{\rho} }{m^2_{\rho} - x -i m_{\rho} \Gamma_{\rho}(x) }  \; \qquad  \text{and} \; \rho \to \rho' \, .
\end{equation}
The label $\rho$ ($\rho'$) refers to the  $\rho$ ($\rho'$) resonance and $\beta_2$ is a tuning parameter (see below). \\
 We use for the energy-dependent off-shell widths of the $\rho$ and $\rho'$   that are needed in \eqref{eq:defBrho2}:
\begin{equation} 
  \label{eq:widrhopr}
  \Gamma_{\rho}(x) = \Gamma_{\rho}(m_{\rho}^2) \frac{m_{\rho}}{\sqrt{x}}\left(\frac{p(x)}{p(m_{\rho}^2)}\right)^3 
  \theta(x - 4 m_\pi^2) \, ,
\end{equation}
where 
\[ p(x) = \frac{1}{2} \sqrt{x - 4 m_\pi^2} \; , \]
and  $\Gamma_{\rho'}(x)$ is given by the same formula with label $\rho \to \rho'$. 
A value for the on-shell width  $\Gamma_{\rho}(m_{\rho}^2)$ and $\Gamma_{\rho'}(m_{\rho'}^2)$, 
respectively, is given in \eqref{eq:inpar2}. 

  We use the following input values for the computations of the optimal $CP$ observables in  Sec.~\ref{sec:results}:
\begin{eqnarray}
 m_\tau=1.777~{\rm GeV},\quad m_\pi =0.140~{\rm GeV}, & \nonumber \\
 m_\rho =0.775~{\rm GeV}, \quad m_{\rho'} =1.465~{\rm GeV},  & \nonumber \\
 G_F= 1.1664 \times 10^{-5}~({\rm GeV})^{-2}, \quad V_{ud}=0.974,  & \nonumber \\
 \Gamma_\rho = 0.149~{\rm GeV}, \quad
 \Gamma_{\rho'} = 0.400~{\rm GeV} \,  . &
 \label{eq:inpar2}
\end{eqnarray}
 With this input, agreement 
with the experimental width  $\Gamma(\tau^-\to \pi^-\pi^0\nu_\tau)_{\rm exp.} = 5.78 \times 10^{-13}~{\rm GeV}$
is obtained when the tuning parameter $\beta_2$ in 
Eq.~\eqref{eq:deffpion2} is chosen to be
\begin{equation} \label{eq:beta2}   
\beta_2 = -0.175 \, .
\end{equation}
 The differential decay density matrix for the charge-conjugate decay
 \[ \tau^{+}(k) \rightarrow\pi^{+}(q_1) \, \pi^{0} (q_2) \; \bar{\nu}_{\tau} \]
 is of the same form as \eqref{Eq.tau2pidec} with the squared matrix element
 \begin{equation} \label{eq:M2pip}
 |\mathcal{M}'_2|^2 = G_F^2 |V_{ud}|^2 |F_\pi(Q^2)|^2 \left( A_2 \one - \boldsymbol{H}_2 \cdot \ssig \right) \, ,
\end{equation}
 and $A_2$ and $\boldsymbol{H}_2$ are given in Eqs.~\eqref{eq:A2sq} and~\eqref{eq:B2sq}, respectively.

\begin{figure}[h!]
 \begin{center}
 {\includegraphics[width=0.79\textwidth]{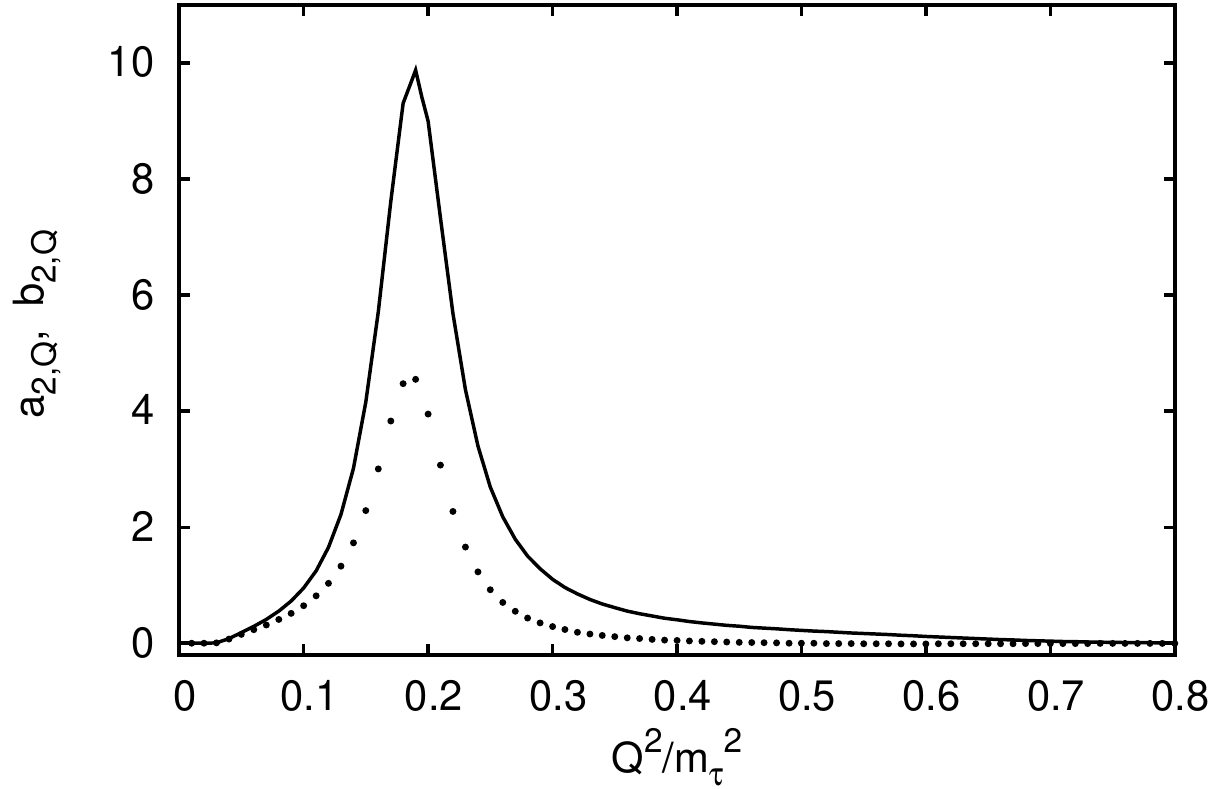}}
  \caption{The spectral functions $a_{2,Q}$ (solid curve) and $b_{2,Q}$ (dotted curve) defined in Eq.~\eqref{eqA:2pipQ}. }
 \label{fig:a2Qb2Q}
 \end{center}
\end{figure}

 One may also determine the $\tau$-spin analyzing power of the ``resonance'' ${Q^\mp}$  
 in the $\tau^\mp\to \pi^\mp\pi^0\nu_\tau$
 decay mode by computing the following decay density matrix:
 \begin{equation} \label{eqA:2pipQ}
     dx~ \frac{d\Omega_Q}{4 \pi} \,\mathcal{D}^{Q^\mp}
    =  dx~ \frac{d\Omega_Q}{4\pi}~\bigl[a_{2,Q}(x)\one \pm b_{2,Q}(x)
    {\boldsymbol{\hat Q}} \cdot  \ssig \bigr]  \, , 
  \end{equation}
    where $4(m_\pi/m_\tau)^2 \leq x  \equiv Q^2/m_\tau^2 \leq 1$
    and $\boldsymbol{\hat Q}=(\boldsymbol{q_1}+ \boldsymbol{q_2})|/|(\boldsymbol{q_1}+ \boldsymbol{q_2})|$.
The spectral functions $a_{2,Q}$ and $b_{2,Q}$ are shown in Fig.~\ref{fig:a2Qb2Q}.
 Integrating the right-hand side of \eqref{eqA:2pipQ} over $x$ the decay density matrix takes the form 
\eqref{eqA:1pronga} with $\alpha_a \to \alpha_{2,Q}$ and 
${\hbq}_\mp \to \boldsymbol{\hat Q}.$
We get for  $\alpha_{2,Q}$:
 \begin{equation} \label{eq:al2Q}
  \alpha_{2,Q} = 0.42 \, .
\end{equation}
Comparison with \eqref{eq:spporo} shows that taking into account the finite widths of the intermediate resonances
and the whole kinematic range of $Q^2$ leads to a slightly smaller $\tau$-spin analyzing power. Nevertheless, we will
 use the value \eqref{eq:spporo} in the computation of the expectation values of the simple $CP$
  observables in Sec.~\ref{sec:results}.

\begin{figure}[h!]
 \begin{center}
 {\includegraphics[width=0.79\textwidth]{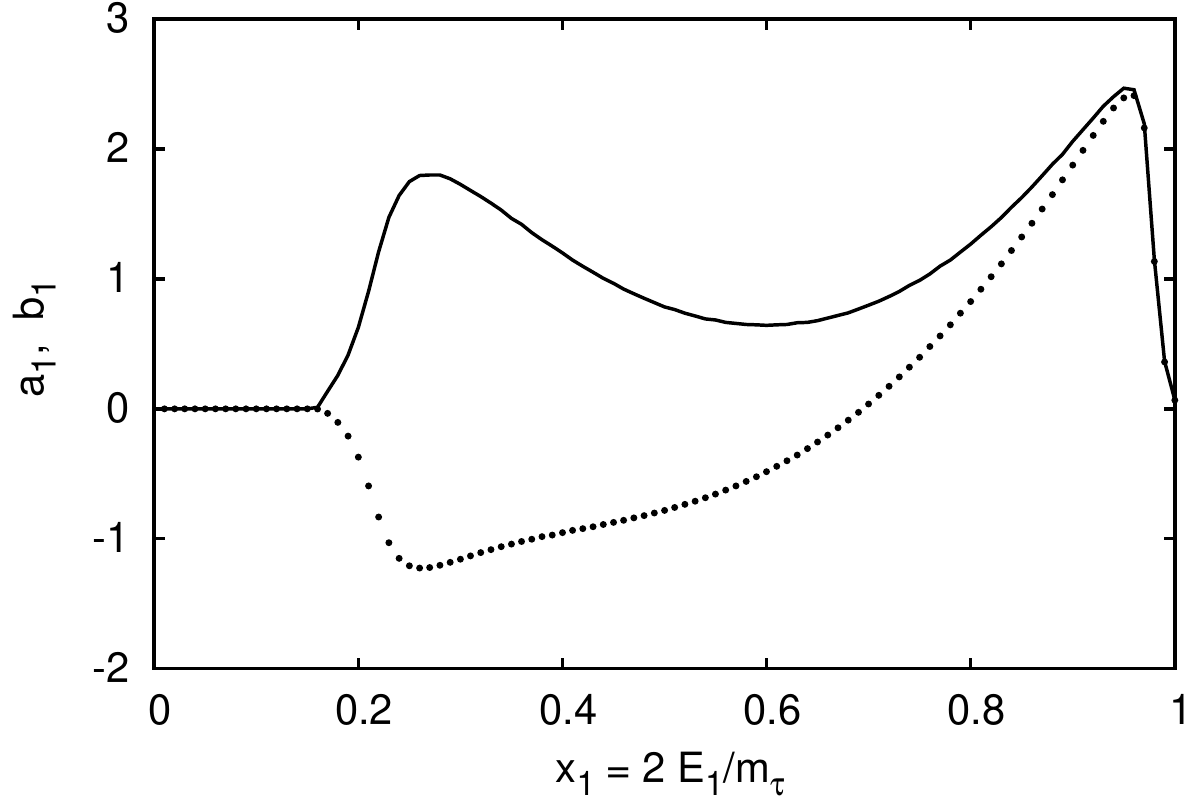}}
  \caption{The spectral functions $a_1$ (solid curve) and $b_1$ (dotted curve) defined in Eq.~\eqref{eqA:pip2p}. }
 \label{fig:a1b1}
 \end{center}
\end{figure}

For completeness we determine also the $\tau$-spin analyzing power of the charged pion 
 in $\tau^\mp \to \pi^\mp(q_1) + \pi^0 \nu_\tau$.
 The respective 1-prong decay density matrix  is given by
 \begin{equation} \label{eqA:pip2p}
    \frac{d^{3}q_1}{(2\pi)^3 2E_1} \,\mathcal{D}^{\pi^\mp}(\tau^\mp\to \pi^\mp(q_1) + \pi^0 \nu_\tau)
    =  dx_1 \frac{d\Omega_1}{4\pi}~\bigl[a_1(x_1)\one \pm b_1(x_1){\hbq}_1 \cdot  \ssig \bigr] \, ,
  \end{equation}
 where $x_1 = 2 E_1/m_\tau$ and  $2 m_\pi/m_\tau \leq x_1 \leq 1$. 
The spectral functions $a_1$ and $b_1$ are shown in Fig.~\ref{fig:a1b1}.
Integrating the right-hand side of \eqref{eqA:pip2p} over $x_1$ the decay density matrix takes the form 
\eqref{eqA:1pronga} with $\alpha_a \to \alpha_1$ and ${\hbq}_\mp \to {\hbq}_1$.
 We get for the $\tau$-spin analyzing power $\alpha_1$ of the charged pion\footnote{This 
  decay mode was analyzed in \cite{Bernreuther:1993nd} using only the intermediate $\rho$ in the narrow-width
   approximation.}
\begin{equation} 
 \label{eq:al1cp}    
 \alpha_1 = -0.036\, .
\end{equation}
Figure~\ref{fig:a1b1} shows that negative and positive contributions cancel to a large extent when $b_1$ is integrated 
over the whole kinematic range, leading to the small value \eqref{eq:al1cp}. The value of $\alpha_1$ can be enhanced by
 a suitable cut on $x_1$. We do not use \eqref{eqA:pip2p} in our analysis of  Sec.~\ref{sec:results}.

 \subsection*{The decay $\tau^{\mp} \to a_1^{\mp} \to 
\pi^{\mp}(q_1) + \pi^{\mp}(q_2) + \pi^{\pm} (q_3) + \nu_{\tau}$}

 The decay mode to three charged prongs proceeds mainly via an intermediate $a_1$ resonance. If one 
  approximates the $\tau \to 3 \pi$ decay mode by $\tau$ decay to an on-shell $a_1$,
  the $\tau$-spin analyzing power of this resonance would be maximal, as stated above,
  if the $a_1$ polarization states can be separated efficiently \cite{Hagiwara:1989fn,Rouge:1990kv}. 
  If one sums over the $a_1$ polarizations the $\tau\to a_1 \nu_\tau$ decay density matrix is of the form 
  \eqref{eqA:1prong} and \eqref{eq:1pincro} with the label $\rho \to a_1$. The $a_1$ mass is not precisely determined
   but, in any case, the $\tau$-spin analyzing power of this resonance is poor in the on-shell approximation. Using  
   \eqref{eq:1pincro} (with $m_\rho \to m_{a_1}$) with the value $m_{a_1} = 1.230~\GeV$ 
   given by  the Particle Data Group \cite{Zyla:2020zbs}  one obtains $\alpha_{a_1} = 0.02$. 
  
   However, maximal sensitivity to the $\tau$ polarization can be obtained with the $3 \pi$ decay mode if the full
    decay dynamics is exploited and the energies and momenta of the three pions are measured. We use the $\tau \to 3 \pi \nu_\tau$
    matrix element given in  \cite{Kuhn:1990ad} (cf. also \cite{Jadach:1990mz,Hagiwara:2012vz}) where this decay is described by
     the decay chain $\tau\to a_1 \to \rho \, (\rho')\,  \pi \to 3 \pi$ with off-shell intermediate resonances.  
     Exact isospin invariance  is assumed.\footnote{The $\tau\to 3\pi \nu_\tau$ decay was analyzed in 
     \cite{Dumm:2009va} within the resonance chiral theory using an elaborate description of the $a_1$
      off-shell width.} We obtain for the  differential $\tau^-\to 2\pi^- \pi^+ \nu_\tau$ decay density matrix
     $\mathcal{D}^{3\pi}$ in the $\tau^-$ rest frame with the normalization conventions \eqref{Eq.03.08} and \eqref{Eq.03.09}:
\begin{eqnarray}\label{Eq.tau3pidec}
\prod_{i=1}^{3} \frac{d^{3}q_{i}}{(2\pi)^{3}2q_{i}^{0}}
\mathcal{D}^{3 \pi}\bigl(\tau^{-}(k)\rightarrow \pi^{-}(q_{1})\pi^{-}(q_{2})\pi^{+}(q_{3})  \nu_{\tau}\bigr) =
\frac{1}{2 m_\tau \Gamma_{3\pi}}  d\Phi_3 |\mathcal{M}_3|^2 \, ,
\end{eqnarray}
where $\Gamma_{3\pi}=\Gamma(\tau^{-}\to\pi^{-}\pi^{-}\pi^{+}{\nu}_{\tau})$ and
 the phase-space measure is given in the recursive phase-space parametrization by
  \begin{eqnarray}\label{eqrecPhase}
  d\Phi_3 =   \frac{1}{ 2^{9}~(2\pi)^8 }
   dQ^2 du~\theta(m_\tau^2 -Q^2)\theta(Q^2- 9 m_\pi^2)
   \theta((\sqrt{Q^2}-m_\pi)^2 -u) \theta(u- 4 m_\pi^2) & \nonumber \\
    \times ~  d\Omega_Q \frac{\lambda^{1/2}(m_\tau^2,Q^2,0)}{m_\tau^2}  d\Omega_3^*  \frac{\lambda^{1/2}(Q^2,u,m_\pi^2)}{Q^2}
    d\Omega_2^{**}  \frac{\lambda^{1/2}(u,m_\pi^2,m_\pi^2)}{u} \, . & 
 \end{eqnarray}
  Here $Q=q_1 + q_2 + q_3$, 
 $u=(q_1+q_2)^2$ and $d\Omega_Q=d\cos\theta_Qd\varphi_Q$ is the solid angle element of $Q$, i.e. $a_1$, in the $\tau$ rest frame,
 $d\Omega_3^*$ is the solid angle element of $\pi^+(q_3)$ in the rest frame of $a_1$, and  
 $d\Omega_2^{**}$ is the solid angle element of $\pi^0(q_2)$ in the rest frame of $\rho$, i.e., the zero-momentum frame of
 $q_1 + q_2$. 
 Note that the statistics factor $1/2$ for two identical particles in the final state is compensated 
 here by the normalization convention \eqref{Eq.03.09}. 
 The squared matrix element is given by
\begin{equation} \label{eq:M3pim}
 |\mathcal{M}_3|^2 = G_F^2 |V_{ud}|^2 \left( A_3 \one + \boldsymbol{H}_3 \cdot \ssig \right) \, ,
\end{equation} 
  where 
  \begin{eqnarray}
 A_3 = & ~|F_1|^2\left[ 4 (k\cdot V_1)^2 - 2 (k\cdot Q - Q^2) V_1^2\right] \nonumber \\
   & + |F_2|^2\left[ 4 (k\cdot V_2)^2 - 2 (k\cdot Q - Q^2) V_2^2\right ] \nonumber \\
   & + {\rm Re}(F_1 F_2^*)\left[8 k\cdot V_1 k\cdot V_2 - 4(k\cdot Q - Q^2) V_1 \cdot V_2\right]\nonumber \\
   & - 2 i(F_2 F_1^* - F_1 F_2^*) \epsilon(k, Q, V_1, V_2) \, ,
   \label{eq:formA}
\end{eqnarray}
\begin{eqnarray}
H_3^j = &  2 m_\tau \left\{ |F_1|^2\left[ 2k\cdot V_1 V_1^j +  V_1^2 Q^j \right] \right. \nonumber \\
        & + |F_2|^2\left[ 2k\cdot V_2 V_2^j +  V_2^2 Q^j \right ] \nonumber \\
        & \left. + 2 {\rm Re}(F_1 F_2^*)\left[ k\cdot V_2 V_1^j  + k\cdot V_1 V_2^j  + V_1 \cdot V_2 Q^j \right] \right\} \nonumber \\
        &  + 2 i m_\tau (F_2 F_1^* - F_1 F_2^*) \epsilon(q', j, V_1, V_2) \, ,
  \label{eq:formHj}      
\end{eqnarray}
and $j=1,2,3$, $k=(m_\tau, \boldsymbol{0})^T$, $q'=k-Q$, 
\begin{equation}
\label{eq:defv12}
V_1^\mu= \left( g^{\mu\nu} - \frac{Q^\mu Q^\nu}{Q^2} \right) (q_1-q_3)_\nu \, , \quad
V_2^\mu=\left( g^{\mu\nu} - \frac{Q^\mu Q^\nu}{Q^2} \right) (q_2-q_3)_\nu \, ,
 \end{equation}
 and $\epsilon(k, Q, V_1, V_2)=\epsilon_{\mu\nu\alpha\beta} k^\mu Q^\nu V_1^\alpha V_2^\beta$,
  $\epsilon(q', j, V_1, V_2) = \epsilon_{\mu j\alpha\beta}q'^{\mu} V_1^\alpha V_2^\beta$
  and we use the convention $\epsilon_{0123} = +1$.
 Moreover,
\begin{equation}
 \label{eq:deff12}
 F_1 = F(Q^2,s)  \, , \qquad F_2 = F(Q^2,t) \, ,
\end{equation}
where $s =(q_1+q_3)^2$ and $t=(q_2+q_3)^2$. The function $F$ is given by \cite{Kuhn:1990ad}:
\begin{equation}
 \label{eq:deffunF}
 F(Q^2,x) = \frac{2\sqrt{2}}{3 f_\pi} B_{a_1}(Q^2) F_\pi(x) \, ,
\end{equation}
where $f_\pi$ is the pion decay constant (in the convention $f_\pi = 0.093$ GeV) and
 $B_{a_1}$ denotes the Breit-Wigner enhancement factor of the $a_1$ meson:
\begin{equation}
 \label{eq:defBa1}
  B_{a_1}(Q^2) = \frac{m^2_{a_1} }{m^2_{a_1} - Q^2 -i m_{a_1} \Gamma_{a_1}(Q^2) } \, .
\end{equation} 
We use as a model for the energy-dependent off-shell width of the $a_1$ meson:
 \begin{equation}
  \label{eq:gama1wid}
  \Gamma_{a_1}(Q^2) = \Gamma_{a_1}(m^2_{a_1}) \frac{g(Q^2)}{g(m^2_{a_1})} \, ,
 \end{equation}
where $\Gamma_{a_1}(m^2_{a_1})$ is the on-shell width (see below) and the function $g$ is given in Eq.~(3.16) of 
 Ref.~\cite{Kuhn:1990ad}.
 Moreover, the pion ``form factor'' $F_\pi(x)$ is given by the formulas \eqref{eq:deffpion2} -- \eqref{eq:widrhopr} above
where now the tuning parameter $\beta_2$ is to be replaced by $\beta_3$ that will be determined below.

 The differential decay density matrix for the charge-conjugate decay
 \[ \tau^{+}(k) \rightarrow\pi^{+}(q_1) \, \pi^{+}(q_2) \, \pi^{-}(q_3) \; \bar{\nu}_{\tau} \]
 is of the same form as  Eqs.~\eqref{Eq.tau3pidec} with the squared matrix element
 \begin{equation} \label{eq:M3pip}
 |\mathcal{M}'_3|^2 = G_F^2 |V_{ud}|^2 \left( A_3 \one - \boldsymbol{H}_3 \cdot \ssig \right) \, ,
\end{equation}
 and $A_3$ and $\boldsymbol{H}_3$ are given in Eqs.~\eqref{eq:formA} and \eqref{eq:formHj}, respectively.
 
 To the best of our knowledge the differential $\tau\to 3\pi \nu$ density matrix \eqref{Eq.tau3pidec} -- \eqref{eq:formHj}
 was so far not given in this explicit form in the literature.

 For our computation of the expectation values of the optimal observables 
 in Sec.~\ref{sec:results} we use the above formulas with the  input values \eqref{eq:inpar2} and 
\begin{equation}
 f_\pi =0.093~{\rm GeV}, \quad m_{a_1} =1.230~{\rm GeV}, \quad \Gamma_{a_1} = 0.483~{\rm GeV} \, . 
 \label{eq:inpar3}
\end{equation}
It remains to fix the tuning parameter $\beta_3$. Using the above squared matrix element and 
 input parameters we find agreement with the experimental width
 $\Gamma(\tau^-\to 2\pi^- \pi^+ \nu_\tau)_{\rm exp.} = 2.11 \times 10^{-13}~{\rm GeV} $
 when the tuning parameter $\beta_3$ is chosen to be
\begin{equation} 
 \label{eq:beta3}    
 \beta_3 = -0.204 \, .
\end{equation} 

 \subsection*{The decay $\tau^{\mp} \to a_1^{\mp} \to 
\pi^{0}(q_1) + \pi^{0}(q_2) + \pi^{\mp} (q_3) + {\nu}_{\tau}$}

For completeness we discuss here also this decay mode, although we do not use it in the
 analysis of Sec.~\ref{sec:results}.  Assuming exact isospin invariance the 
 differential decay density matrices for $\tau^\mp\to 2 \pi^\mp \pi^\pm \nu_\tau$
 derived in the previous subsection can be used also for these decay modes.
 Using the above input parameters  with the exception 
 $m_\pi = m_{\pi^+} = 0.140~\GeV \rightarrow m_\pi = m_{\pi^0} = 0.135~\GeV$,
 agreement with the experimental width $\Gamma(\tau^-\to 2\pi^0 \pi^-\nu_\tau)_{\rm exp.} = 2.10 \times 10^{-13}~{\rm GeV}$
  is obtained with the following value of the tuning parameter, here denoted by $\beta'_3$:
 \begin{equation} 
 \label{eq:betpr}    
 \beta'_3 = -0.190 \, .
\end{equation} 

\begin{figure}[h!]
 \begin{center}
 {\includegraphics[width=0.89\textwidth]{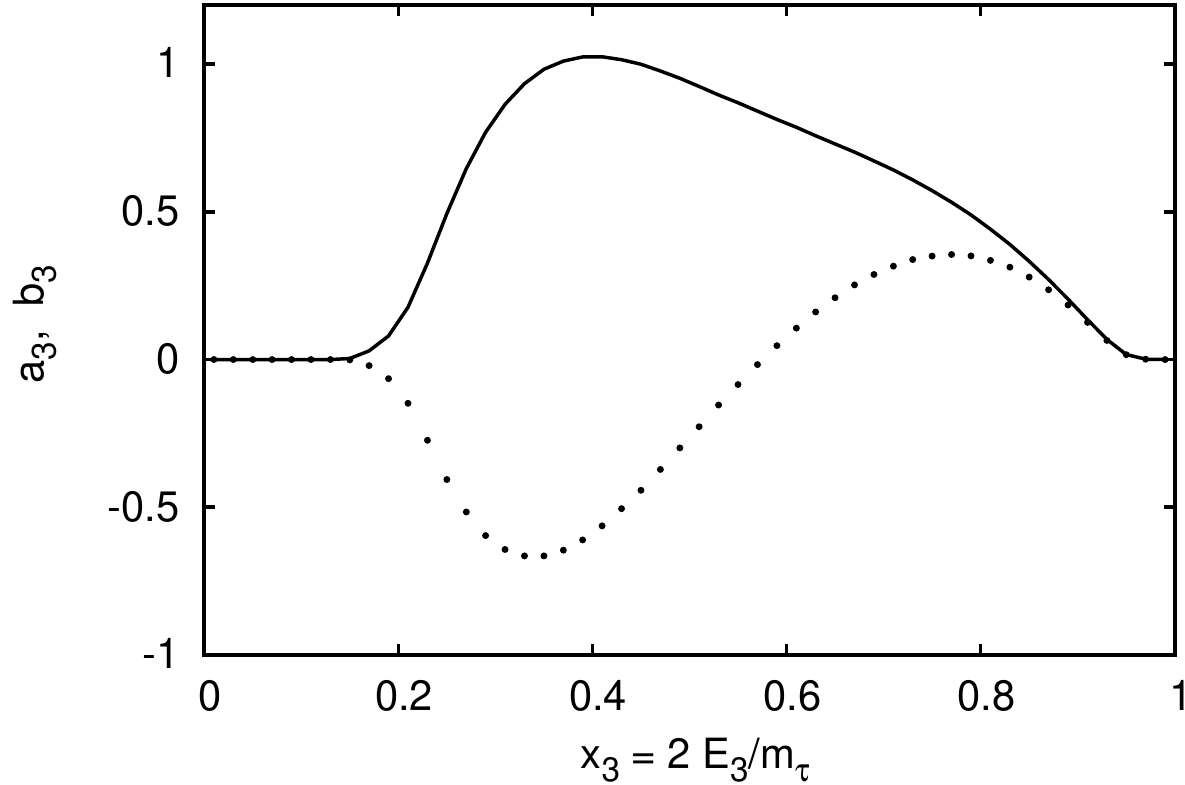}}
  \caption{The spectral functions $a_3$ (solid curve) and $b_3$ (dotted curve) defined in Eq.~\eqref{eqA:pip3p}. }
 \label{fig:a3b30}
 \end{center}
\end{figure}

 Moreover,  the $\tau$-spin analyzing power of the charged pion in this decay mode is also of interest.
 The 1-prong decay density matrix for $\tau^\mp \to \pi^\mp(q_3) + 2 \pi^0 \nu_\tau$, normalized to the charged
  particle multiplicity $n_{\pi^\pm}=1$, is given by
 \begin{equation} \label{eqA:pip3p}
    \frac{d^{3}q_3}{(2\pi)^3 2E_3} \,\mathcal{D}^{\pi^\mp}(\tau^\mp \to \pi^\mp(q_3) + 2 \pi^0 \nu_\tau)
    = dx_3 \frac{d\Omega_3}{4\pi}~\bigl[a_3(x_3)\one \pm b_3(x_3){\hbq}_3 \cdot  \ssig \bigr]  \, , 
  \end{equation}
 where $x_3 = 2 E_3/m_\tau$ and  $2 m_\pi/m_\tau \leq x_3 \leq 1 - 3 (m_\pi/m_\tau)^2$. 
The spectral functions $a_3$ and $b_3$ are shown in Fig.~\ref{fig:a3b30}.
 Integrating the right-hand side of \eqref{eqA:pip3p} over $x_3$ the decay density matrix takes the form 
\eqref{eqA:1pronga} with $\alpha_a \to \alpha_3$ and 
${\hbq}_\mp \to {\hbq}_3.$
We get for the $\tau$-spin analyzing power $\alpha_3$ of the charged pion:\footnote{A simpler description of this 
  decay mode was used in \cite{Bernreuther:1993nd} and the value $\alpha_\pi = -0.18$ was obtained.}
\begin{equation} 
 \label{eq:al3cp}    
 \alpha_3 = -0.144\, .
\end{equation}
This number is rather small because $b_3$ has both negative and positive contributions that cancel  
 to a large extent when integrated over $x_3$.
The analyzing power can be enhanced by a suitable cut on $x_3$. 

\section{Expectation values and covariances of $CP$-odd observables}
\label{app:exCoCP}
%
In this appendix we discuss general properties of expectation values and covariances
 of the $CP$-odd observables introduced in Sec.~\ref{Sec:04} and computed in Sec.~\ref{sec:results}. 
 We treat first  case i) of Sec.~\ref{Sec:03} where only one charged particle is measured from
 $\tau^-$ and $\tau^+$ decays, respectively.
 The differential cross section of the two-particle inclusive reaction \eqref{eq:2pincl} and \eqref{eq:1prdec}
  as used in this paper is given by  \eqref{Eq.03.10}:
\begin{eqnarray} \label{Eq.apBdsig}
d\sigma_{a\bar{b}} & = & \dfrac{\sqrt{1-4m_{\tau}^{2}/s}}{16\pi s} 
{\rm Br}(\tau^{-}\rightarrow A) \, {\rm Br}(\tau^{+}\rightarrow \overline{B}) 
\nonumber \\
& & \times ~{\rm Tr}\left[R {\cal D}^a {\cal D}^{\bar b}\right] 
\frac{|\qm^*|}{(2\pi)^2}\frac{|\qp^*|}{(2\pi)^2} dE_-^* dE_+^*
\dfrac{d\Omega_{k_+}}{4\pi} \frac{d\Omega_-^*}{4\pi} \frac{d\Omega_+^*}{4\pi} \, ,
\end{eqnarray}
where we have used in \eqref{Eq.03.10} the momenta of the charged particles
 $a$ and ${\bar b}$ and the corresponding phase-space measures in the respective $\tau^-$ and $\tau^+$ rest frame.
 The one-particle inclusive decay density matrices in the  $\tau^\mp$ rest frames are given in \eqref{eqA:1prong}.
 We recall the relation between the respective rest-frame momenta $q_\mp^*$ and $k_\mp^*=(m_\tau,\boldsymbol{0})^T$
  and the momenta $q_\mp$ and $k_\mp$ in the $e^+ e^-$ c.m. frame. With the Lorentz boost	
 \begin{equation}\label{eq:Lorboost}
		\Lambda_{\kk}	=
		\left(\begin{array}{cc}
	\frac{k^0}{m_\tau} & \frac{ k^j}{m_\tau}    \\
  \frac{k^i}{m_\tau} & \delta^{ij} + {\hat k}^i {\hat k}^j \left(\frac{k^0 - m_\tau}{m_\tau}\right)
			\end{array}\right) \, ,
\end{equation}
where $\kk$ is the three-momentum of $\tau^+$ in the $e^+ e^-$ c.m. frame, we have
\begin{equation} \label{eq:rebocm}
\Lambda_{\pm \kk}~k_\mp = k_\mp^* \, , \qquad \Lambda_{\pm \kk}~q_\mp = q^*_\mp  \, .
\end{equation}

   Next we decompose \eqref{Eq.apBdsig} according to  \eqref{Eq.04.08} and \eqref{Eq.04.09}, neglecting terms
   quadratic in ${\hat d}_\tau$. Here our phase-space variables are
   \begin{equation} \label{eq:phaphi}
    \phi =  (E^*_-, E^*_+, \hk, \hqm^*, \hqp^*) \, ,
   \end{equation}
and the measure is
 \begin{equation}\label{eq:phamea}
 d\phi = dE_-^* dE_+^*
\dfrac{d\Omega_{k}}{4\pi} \frac{d\Omega_-^*}{4\pi} \frac{d\Omega_+^*}{4\pi} \, .
  \end{equation} 
 We get
 \begin{equation} \label{eq:decsab}
  d\sigma_{a\bar{b}} =\left\{S_{\rm SM}^{a\bar{b}}(\phi)+
  S_{CP,R}^{a\bar{b}}(\phi)\Re \,{\hat d}_{\tau}  + S_{CP,I}^{a\bar{b}}(\phi) \Im \, {\hat d}_{\tau} \right\} d\phi \, ,
 \end{equation}
where, using  \eqref{eqA:1prong},
\begin{eqnarray} \label{eq:chrSMa}
 S_{\rm SM}^{a\bar{b}}(\phi) & = & \dfrac{\sqrt{1-4m_{\tau}^{2}/s}}{16\pi s} 
{\rm Br}(\tau^{-}\rightarrow A) \, {\rm Br}(\tau^{+}\rightarrow \overline{B})  
\frac{\chi_{\rm SM,\alpha\alpha'\beta\beta'}}{| 1 + e^2 \Pi_c(s)|^2}  \nonumber \\
& & \times~ n_a(E_-^*)\bigl[\delta_{\beta'\beta} + h_a(E_-^*){\hbq}_-^* \cdot  \ssig_{\beta'\beta} \bigr] ~
n_b(E_+^*)\bigl[\delta_{\alpha'\alpha} - h_b(E_+^*){\hbq}_+^* \cdot  \ssig_{\alpha'\alpha} \bigr] \, .
\end{eqnarray}
 The quantities $S_{CP,R}^{a\bar{b}}$ and $S_{CP,I}^{a\bar{b}}$ are obtained from \eqref{eq:chrSMa} by the replacements
 \begin{equation} \label{eq:Brepl}
 \chi_{\rm SM} \rightarrow \chi_{CP}^R \qquad \text{and} \qquad  \chi_{\rm SM} \rightarrow \chi_{CP}^I \, ,
   \end{equation}
respectively; see \eqref{Eq.03.13} -- \eqref{Eq.03.14a}.

We can now perform the traces in \eqref{eq:chrSMa}. With \eqref{Eq.03.16} we see that this amounts to make the following
 replacements in \eqref{Eq.03.13} -- \eqref{Eq.03.14a}:
 \begin{eqnarray} \label{eq:BTrrepl}
  \one & \rightarrow & 4 ~n_b ~n_a \, , \nonumber \\
  \sip  & \rightarrow & -4 ~n_b h_b ~n_a ~{\hbq}_+^*  \, , \nonumber \\
  \ssim  & \rightarrow & 4 ~n_b ~n_a h_a  ~ {\hbq}_-^*\, , \nonumber \\
  \sigma_+^r  \sigma_-^s  & \rightarrow & - 4 ~ n_b h_b ~n_a h_a ~{\hat q}_+^{* r} {\hat q}_-^{* s}  \, . 
 \end{eqnarray}
 Next, we consider the transformation 
 \begin{equation} \label{eq:tracms}
  \kk \rightarrow -\kk \, , \qquad \qp \rightarrow -\qp \, ,  \qquad \qm \rightarrow -\qm \, ,
 \end{equation}
which, using  \eqref{eq:rebocm}, implies
 \begin{equation} \label{eq:traresf}
  \hk \rightarrow -\hk \, , \qquad \hqp^* \rightarrow -\hqp^* \, ,  \qquad \hqm^* \rightarrow -\hqm^* \, ,
 \end{equation}
 and vice versa. These transformations correspond to the naive ``time reversal'' transformation\footnote{One may also transform 
 $\pp \to - \pp$, but this is irrelevant here.}
 $T_N$ referred to in
  Sec.~\ref{Sec:03}.
  Inspection of $\chi_{\rm SM}$, $\chi_{CP}^R$, and $\chi_{CP}^I$, i.e., of Eqs.~\eqref{Eq.03.13} -- \eqref{Eq.03.14a}  with 
 the replacements  \eqref{eq:BTrrepl}, shows that applying \eqref{eq:traresf} we have
 \begin{equation} \label{eq:traSSS}
 S_{\rm SM}^{a\bar{b}}(\phi) \rightarrow    S_{\rm SM}^{a\bar{b}}(\phi) \, , \qquad
 S_{CP,R}^{a\bar{b}}(\phi) \rightarrow   - S_{CP,R}^{a\bar{b}}(\phi) \, , 
 \qquad S_{CP,I}^{a\bar{b}}(\phi) \rightarrow    S_{CP,I}^{a\bar{b}}(\phi) \, .
   \end{equation}

 We turn to the simple and optimal observables of Sec.~\ref{Sec:04}. We assume integration over the whole phase space or,
  if cuts are applied, we assume the cuts to be $CP$-symmetric. In addition we assume the cuts to be invariant under 
   \eqref{eq:tracms}. The tensors ${\widehat T}^{ij}$ and ${T}^{ij}$ of \eqref{Eq.04.01} and \eqref{Eq.04.02}
    are odd  whereas 
    ${\widehat Q}^{ij}$ and ${Q}^{ij}$ of \eqref{Eq.04.03} and \eqref{Eq.04.04} are even under the transformation \eqref{eq:tracms}.
  
 Let us first consider the case $a=b$. The $CP$ properties of the observables $T$ and $Q$ imply
 \begin{eqnarray} \label{eqB:CPTQ}
 E_0(T^{ij}) = 0 \, , & \qquad & E_0( {\widehat T}^{ij}) = 0 \, , \nonumber \\
  E_0(Q^{ij}) = 0 \, , & \qquad & E_0( {\widehat Q}^{ij}) = 0 \, .
 \end{eqnarray}
 Moreover, turning to the covariance matrix of one of the $T$ and one of the $Q$ variables,
 the transformation \eqref{eq:tracms} implies 
 \begin{equation} \label{eqB:covS}
 E_0(T^{ij}Q^{k l}) = 0 
   \end{equation}
and likewise for the other $T Q$ correlations. That is, the covariance matrix of the $T, Q$ variables
 is diagonal. 
 
 The optimal $CP$ observables are in the case $a = b$:
 \begin{equation} \label{eqB:optO}
 {\cal O}_R^{a\bar{a}}(\phi) = \frac{S_{CP,R}^{a\bar{a}}(\phi)}{S_{\rm SM}^{a\bar{a}}(\phi) } \, ,\qquad
 {\cal O}_I^{a\bar{a}}(\phi) = \frac{S_{CP,I}^{a\bar{a}}(\phi)}{S_{\rm SM}^{a\bar{a}}(\phi) } \, .
  \end{equation}
 The $CP$ transformation properties of these observables imply
 \begin{equation} \label{eqB:E0optO}
  E_0({\cal O}_R^{a\bar{a}}) = 0 \, , \qquad  E_0( {\cal O}_I^{a\bar{a}}) = 0 \, ,
  \end{equation}
 and applying the $T_N$ transformation \eqref{eq:tracms}  it follows that
 \begin{equation} \label{eqB:covoO}
 E_0({\cal O}_R^{a\bar{a}}  {\cal O}_I^{a\bar{a}}) = 0 \, .
  \end{equation}
 Thus, the covariance matrix is diagonal in this case:
 \begin{equation}\label{eqB:covaa}
		V({\cal O}^{a\bar{a}})	=
		\left(\begin{array}{cc}
                  E_0({\cal O}_R^{a\bar{a}}  {\cal O}_R^{a\bar{a}}) & 0    \\
  0 & E_0({\cal O}_I^{a\bar{a}}  {\cal O}_I^{a\bar{a}}) 
			\end{array}\right) \, .
\end{equation}
 
 Next we turn to the case $a\neq b$. As this final state is no longer $CP$-symmetric, the $CP$ transformation properties
  of the observables are no longer of immediate use. Let us first consider the simple observables, for instance, 
$ T^{ij}$ and ${\widehat Q}^{ij}$. Applying the $T_N$ transformation  \eqref{eq:tracms}  we get
 \begin{equation} \label{eqB:Tmat0}
 E_0^{a\bar{b}}(T^{ij}) \equiv \langle T^{ij} \rangle_{0, a\bar{b}} = 0.
 \end{equation}
The transformation  \eqref{eq:tracms}  implies also that expectation values of the form \eqref{eqB:covS} 
vanish in the nondiagonal case.
  As to the SM expectation value
  \begin{equation} \label{eqB:Qhmat0}
 E_0^{a\bar{b}}({\widehat Q}^{ij}) \equiv \langle {\widehat Q}^{ij} \rangle_{0, a\bar{b}} 
 \end{equation}
 there is, however,  in the case $a\neq b$  no symmetry argument implying that it vanishes, too. 
  Therefore, one should use in this case 
    in general the observables 
    \begin{equation} \label{eqB:QprO}
     {\widehat Q}^{~' ij} = {\widehat Q}^{ij} - \langle {\widehat Q}^{ij} \rangle_{0, a\bar{b}} \, .
 \end{equation}
 From the $CP$ property of ${\widehat Q}^{ij}$   one gets, of course,
  \begin{equation} \label{eqB:Qhimag}
 \langle {\widehat Q}^{ij} \rangle_{0, a\bar{b}} + \langle {\widehat Q}^{ij} \rangle_{0, b \bar{a}} = 0 \, .
  \end{equation}
  Thus, the respective quantity to probe for $CP$ violation is \eqref{Eq.04.06}:
  \begin{equation}
   \frac{1}{2}\left\{\langle {\widehat Q}^{ij} \rangle_{a\bar{b}} + \langle {\widehat Q}^{ij} \rangle_{b \bar{a}} \right\} \, .
  \end{equation}
  But the corresponding variance in the SM, for instance of the $i=j=3$ components, has to be calculated in general  as
  \begin{equation} \label{eqB:QQcov}
  \langle {\widehat Q}^{~' 33}{\widehat Q}^{~' 33} \rangle_{0, a\bar{b}} \, =  \, 
  \langle {\widehat Q}^{~' 33}{\widehat Q}^{~' 33} \rangle_{0, b\bar{a}}  \, .
  \end{equation}
   The above  statements apply, of course, also to $Q^{ij}$. 
   Yet in our analysis where we use the SM matrix element of the form \eqref{eq:chrSMa} and integrate over the whole phase space 
   we find that \eqref{eqB:Qhmat0} vanishes within our numerical uncertainties of order $10^{-4}$. This holds also for 
   the respective expectation values of $Q^{ij}$.

  The optimal observables are in the case $a\neq b$ (cf. \eqref{eq:optORI}):
  \begin{equation} \label{eqB:OORI}
   \cO_{R}^{a\bar{b}}(\phi) = \frac{S_{CP,R}^{a\bar{b}}(\phi)}{ S_{\rm SM}^{a\bar{b}}(\phi) }   \, ,
   \qquad  \cO_{I}^{a\bar{b}} (\phi) = \frac{S_{CP,I}^{a\bar{b}}(\phi)}{ S_{\rm SM}^{a\bar{b}}(\phi)}    \, ,
  \end{equation}
 where  $\cO_{R}^{a\bar{b}}(\phi)$ and $\cO_{I}^{a\bar{b}} (\phi)$ are odd and even under the transformation 
 \eqref{eq:traresf}, respectively; see \eqref{eq:traSSS}. Therefore, we have
 \begin{equation} \label{eqB:E0OR}
  E_0(\cO_{R}^{a\bar{b}}) \, = \, 0 \, .
  \end{equation}
  For analyzing  $E_0(\cO_{I}^{a\bar{b}})$ we perform in $d\sigma_{a\bar{b}}$, Eq.~\eqref{eq:decsab},
  the variable transformation 
  \begin{equation} \label{eq:ApBvark}
   \kk \rightarrow -\kk  \, .
   \end{equation}
The term  $S_{\rm SM}^{a\bar{b}}(\phi)$ remains invariant, while $\cO_{i}^{a\bar{b}} (\phi)$ $(i=R,I)$  change sign; 
 see  \eqref{Eq.03.13}  -- \eqref{Eq.03.14a}      and \eqref{eq:chrSMa}, \eqref{eq:Brepl}.
Thus 
\begin{equation} \label{eqB:E0OI1p}
  E_0(\cO_{I}^{a\bar{b}}) \, = \, 0 
  \end{equation}
  if one integrates over the whole angular range of $\kk$ or over a range that is symmetric with respect to  \eqref{eq:ApBvark}.
   Beyond the one-photon approximation 
    $E_0(\cO_{i}^{a\bar{b}})$ $(i=R,I)$ will in general be nonzero. Therefore, one should use in general (cf. \eqref{Eq.04.12}):
  \begin{eqnarray}\label{eqB:ExORI}
   \cO_{R}^{' a\bar{b}}(\phi) & = & \cO_{R}^{ a\bar{b}}(\phi) - E_0(\cO_{R}^{a\bar{b}}) \, , \nonumber \\
   \cO_{I}^{' a\bar{b}}(\phi) & = & \cO_{I}^{ a\bar{b}}(\phi) - E_0(\cO_{I}^{a\bar{b}}) \, .
     \end{eqnarray}
 In our case here the transformations  \eqref{eq:traresf} and \eqref{eq:ApBvark} imply that the  
 covariance matrix $V(\cO^{' a\bar{b}})$ 
  is still diagonal and is given by 
 \begin{equation}\label{eqB:covab}
		V({\cal O}^{' a\bar{b}})	= V({\cal O}^{ a\bar{b}}) =
		\left(\begin{array}{cc}
                  E_0({\cal O}_R^{ a\bar{b}}  {\cal O}_R^{ a\bar{b}}) & 0    \\
  0 & E_0({\cal O}_I^{ a\bar{b}}  {\cal O}_I^{ a\bar{b}}) 
			\end{array}\right) \, .
\end{equation}
 This covariance matrix is then used in \eqref{Eq.04.31}
    and   \eqref{Eq.04.32} for the estimators $\gamma_i$ and their covariance matrix $V(\gamma)$.

 We come now to the final states of case ii) of  Sec.~\ref{Sec:03} to which we apply  the optimal $CP$ observables.
 The channels where $\tau\to 3 \pi \nu_\tau$
 is involved require a more detailed discussion. 
  In our models for the hadronic $\tau$ decays outlined in Appendix~\ref{app:taudec} the squared matrix element
  of this decay mode given in \eqref{eq:M3pim} -- \eqref{eq:formHj} differs from the 
   respective squared matrix element of $\tau\to 2 \pi \nu_\tau$ and those of
   the one-particle inclusive decays in that it contains contributions from absorptive parts caused by  the finite widths of
    the intermediate resonances. This implies that a $T_N$ transformation can no longer be used to discriminate between
     the optimal observables $\cO_R$ and $\cO_I$.  In order to see this explicitly let us for definiteness consider
      the case where $\tau^-$ decays to three observed pions (labeled by the symbol $A$), while in the decay of $\tau^+$
       only one charged particle is measured (label ${\bar b}$). 
       The differential cross section is obtained by inserting the respective 
         decay density matrices into  \eqref{Eq.03.10}, taking into account  \eqref{Eq.03.08} -- \eqref{Eq. 03.11}.
       Using  \eqref{eq:M3pim} the matrix element $S_{\rm SM}^{A\bar{b}}$ is obtained from \eqref{eq:chrSMa}, up to an overall factor,
        by replacing 
       \begin{equation} \label{eqB:repl}
      n_a(E_-^*)\bigl[\delta_{\beta'\beta} + h_a(E_-^*){\hbq}_-^* \cdot  \ssig_{\beta'\beta} \bigr]   
        \rightarrow \bigl[A_3 \delta_{\beta'\beta} +  \boldsymbol{H}_3 \cdot  \ssig_{\beta'\beta} \bigr]  \, ,
       \end{equation}
 The matrix elements $S_{CP,R}^{A\bar{b}}$ and $S_{CP,I}^{A\bar{b}}$ are obtained in the same fashion.
  Inspection of the functions $A_3$ and $H_3^j$ shows that neither has a definite behavior under the 
   following transformation that is analogous to \eqref{eq:traresf}:
   \begin{equation}\label{eqB:tra123}
   \kk \rightarrow -\kk \, , \qquad \bq^*_i \rightarrow  -\bq^*_i \quad (i=1,2,3) \, .
       \end{equation}
  The dispersive terms in $A_3$ and $H_3^j$ are even under \eqref{eqB:tra123} whereas the absorptive terms are odd.
  Therefore, the matrix elements   $S_{\rm SM}^{A\bar{b}}$,   $S_{CP,R}^{A\bar{b}}$, and $S_{CP,I}^{A\bar{b}}$ 
   do not have a definite transformation behavior under \eqref{eqB:tra123}, too.  Hence we expect  that
   \begin{equation} \label{eqB:corRI3}
   E_0(\cO_{CP,R}^{A{\bar b}}\cO_{CP,I}^{A{\bar b}}) \neq 0 \, .
   \end{equation} 
  Thus, the covariance matrix $V(\cO^{' A {\bar b}})$ can have nondiagonal elements that are nonvanishing.  
   On the other hand, applying the transformation \eqref{eq:ApBvark}, $\kk \to - \kk$, to $S_{\rm SM}^{A\bar{b}}$ and
   to $S_{CP,i}^{A\bar{b}}$ $(i=R,I)$
   shows that the first term remains invariant while the two others change sign.
   Therefore, with our matrix elements we have 
    \begin{equation} \label{eqB:E0nze}
   E_0(\cO_{CP,R}^{A{\bar b}}) = 0 \, , \qquad  E_0(\cO_{CP,I}^{A{\bar b}}) = 0 \, .
     \end{equation}
  Beyond the one-photon approximation \eqref{eqB:E0nze} will no longer hold. 
  Thus when $\tau$-pair
   decays to final states $A{\bar b} + b{\bar A}$, $A{\bar A}$, and $A {\bar B}+ B {\bar A}$ are considered, 
   one should in general apply
   -- especially in experimental analyses -- the full formalism of the optimal observable method as 
   explained in Sec.~\ref{Sec:04}.
  The nondiagonal elements of the respective covariance matrix of the optimal $CP$ observables computed 
   with the matrix elements for $\tau$-pair production and decay used in this paper
    at $\sqrt{s}=10.58$ GeV are very small and can be neglected in view of our numerical uncertainties; see
    Sec.~\ref{sec:results}.

   At last a remark that applies if the full formalism of Sec.~\ref{Sec:04} has to be used. Suppose the parameters $\Re d_\tau$
    and $\Im d_\tau$ have been measured in $k$ decay channels. Let us denote the results for their mean values in the channel $\kappa$  by
 \begin{equation}\label{eqB:meanvd}
		\overline{\boldsymbol{X}}^{(\kappa)}	=
		\left(\begin{array}{c}
                     \Re {\bar d}_\tau^{(\kappa)} \\
                       \Im {\bar d}_\tau^{(\kappa)}
			\end{array}\right) \, ,  \qquad (\kappa =1, \cdots,k) \, ,
\end{equation}   
 and for the respective covariance matrix by $V^{(\kappa)}$. Furthermore, we assume these $k$ measurements to be independent, i.e.,
  uncorrelated. We define the matrix
  \begin{equation} \label{eqB:MatVm1}
   V^{-1} = \sum\limits_{\kappa=1}^k \left(V^{(\kappa)}\right)^{-1}
   \end{equation}
 and the overall mean
 \begin{equation} \label{eqB:ovmea}
 \overline{\boldsymbol{X}} = V  \sum\limits_{\kappa=1}^k \left(V^{(\kappa)}\right)^{-1} \overline{\boldsymbol{X}}^{(\kappa)} \, .
   \end{equation} 
 The covariance matrix is then given by $V$.
 That is, the 1 s.d. error ellipse for the mean values \eqref{eqB:ovmea}
 in the  $\Re d_\tau  - \Im d_\tau$ plane is given by
 \begin{equation} \label{eqB:erell}
 \left({\boldsymbol{X}} - \overline{\boldsymbol{X}}\right)^T V^{-1} \left({\boldsymbol{X}} - \overline{\boldsymbol{X}}\right) = 1 \, .
  \end{equation}   


\end{document}